\documentclass[twocolumn]{aastex631}
\usepackage{multirow}
\usepackage{xcolor}
\hypersetup{linkcolor=red,citecolor=blue,filecolor=cyan,urlcolor=magenta}
\usepackage{longtable}
\usepackage{array}
\usepackage{amsmath}

\DeclareUnicodeCharacter{2212}{-}
\begin{document}

\title{Probing the Low Radio Frequency Emission in PG Quasars with the uGMRT - II}
\correspondingauthor{Sanna Gulati}
\email{sanna@ncra.tifr.res.in}
\author[0000-0003-3785-1725]{Sanna Gulati}
\affiliation{National Centre for Radio Astrophysics - Tata Institute of Fundamental Research, S. P. Pune University Campus, Pune 411007, India}
\author[0000-0003-0667-7074]{Silpa S. }
\affiliation{Departamento de Astronomía at Universidad de Concepción, Casilla 160-C, Concepción, Chile}
\author[0000-0003-3203-1613]{P. Kharb}
\affiliation{National Centre for Radio Astrophysics - Tata Institute of Fundamental Research, S. P. Pune University Campus, Pune 411007, India}
\author[0000-0001-6947-5846]{Luis C. Ho}
\affiliation{Kavli Institute for Astronomy and Astrophysics, Peking University, Beijing 100871, China}
\affiliation{Department of Astronomy, School of Physics, Peking University, Beijing 100871, China}
\author[0009-0000-1447-5419]{S. Ghosh}
\affiliation{National Centre for Radio Astrophysics - Tata Institute of Fundamental Research, S. P. Pune University Campus, Pune 411007, India}
\author[0000-0002-0367-812X]{J. Baghel}
\affiliation{National Centre for Radio Astrophysics - Tata Institute of Fundamental Research, S. P. Pune University Campus, Pune 411007, India}

\begin{abstract}
We present results from uGMRT 685 MHz observations of 87 QSOs belonging to the Palomar Green (PG) quasar sample with $z<0.5$. Radio emission is detected in all sources except for 3 radio-quiet (RQ) sources, viz., PG~0043+039, PG~1121+422, and PG~1552+085. 
The radio-loud (RL) – RQ dichotomy persists at 685~MHz with only 1 source, PG~1216+069, changing its classification from RQ to RL. Approximately 1/3 of the detected RQ quasars display AGN-dominated radio emission while the rest may show additional contributions from stellar-related processes. 
Consistent with this, the RL and RQ quasars occupy distinct tracks on the `fundamental plane' of black hole activity. We find that RL quasars have $\log_{10}(L_{685\,\mathrm{MHz}}/\mathrm{W\,Hz}^{-1}) > 25.5$, while RQ quasars have ${\log_{10}(L_{685\,\mathrm{MHz}}/\mathrm{W\,Hz}^{-1})} <23.5$. Furthermore, the radio sizes display the RQ$-$RL divide as well with RQ sources typically having sizes $\lesssim30$~kpc, with only 16 ($\sim22$\%) RQ sources having sizes between 30 and 100 kpc where there is an overlap with RL quasar sizes. A strong correlation exists between 685~MHz radio luminosity and black hole mass which is tightened when accretion rate is considered, highlighting the important role played by the accretion rate and accretion disk structure in jet production. We found no difference in the minimum-energy magnetic field strengths of the radio cores of RL and RQ quasars; however, different assumptions of source volume and volume filling factors may apply. High-resolution X-ray observations and radio-X-ray flux comparisons are needed to independently test the `magnetic flux paradigm'.
\end{abstract}

\section{Introduction} \label{sec:intro}
Active galactic nuclei (AGN) are powered by mass accretion on to supermassive black holes (SMBHs) in the centers of galaxies \citep{Rees84}. The majority of AGN ($\sim$90\%) comprise weak outflows that extend to sub-kpc scales and are often diffuse or wind-like \citep[e.g.,][]{Giroletti2009}. These are the radio-quiet (RQ) AGN. Radio-loud (RL) AGN, on the other hand, exhibit powerful relativistic jets that extend to hundreds of kpc and megaparsec scales. The radio-loudness parameter, $R$, defined as the ratio of the 5 GHz flux density to the optical (B band) flux density, is used to distinguish between these two classes \citep{Kellermann89}, with RL AGN having $R>10$ and RQ AGN having $R\leq10$. 

Multiple explanations have been proposed for the observed RL-RQ dichotomy including differences in SMBH masses \citep[e.g.,][]{Laor2000, McLure2001, Wu2001, Gopal2008, Richings2011}, SMBH spins \citep[e.g.,][]{Tchekhovskoy10, Sikora07}, accretion rates \citep[e.g.,][]{Ho02, Sikora07, Best2012} and environments \citep[e.g.,][]{Best2005, Croft2007, Kauffman2008,Wylezalek2013}. One hypothesis is that quasars experience intermittent radio activity; they remain RQ most of their life and become RL during active phases \citep[e.g.,][]{Coziol2017, Silpa2021}. This latter idea can be connected to the `magnetic flux paradigm' by \citet{Sikora2013} which has been invoked to explain the RL-RQ divide. According to the `magnetic flux paradigm', RQ AGN must have much lower magnetic field strengths close to the central black holes compared to RL AGN \citep[e.g.,][]{Chamani2021}. Alternatively, the `spin paradigm' suggests that SMBH spin controls jet power with rapidly spinning black holes launching strong relativistic jets (producing radio-loud AGN), while slowly spinning black holes produce only weak radio emission \citep[e.g.][]{Moderski1998, Sikora07, Tchekhovskoy10}. 

Low-frequency surveys do not show a significant change in the relative fraction of RL to RQ quasars \citep{Ennis1982, Robson1985}. Studies suggest that geometry and relativistic beaming play a role, with radio emission appearing stronger when jets are oriented toward us \citep{Urry1995, Fan2004, Xiao2015, Pei2016}. However, beaming alone is insufficient to explain the dichotomy \citep{Kellermann89, Kellermann2004}. The host galaxy may also influence radio emission, as low-luminosity radio emission have been linked to star formation and supernovae activity \citep{Kellermann2016}. 


In RL AGN, jets are thought to originate from the central engine via two key mechanisms: the Blandford–Znajek process, which extracts rotational energy from a spinning supermassive black hole \citep{BlandfordZnajek77}, and the Blandford–Payne mechanism, an MHD process in which the rotating accretion disk launches outflows by accelerating material along large-scale magnetic field lines anchored in the disk \citep{BlandfordPayne82}. The radio emission in AGN jets is produced via the synchrotron process. However, the origin of radio emission in RQ AGN remains unclear \citep{Panessa19}, with proposed mechanisms including star formation \citep{Terzian65, Gurkan2019} and coronal activity \citep{LaorBehar08}, starburst and/or AGN-driven winds \citep{Condon13, IrwinSaikia03, HotaSaikia06, Mizumoto19}, weak radio jets \citep{Falcke00, Jarvis19, Kharb19}, and free-free emission from disk or torus winds \citep{BlundellKuncic07, Ho99, LalHo10}.

This paper is the second in the series after \citet[][henceforth, referred to as Paper I]{Silpa2020}, where we examine the origin of radio emission in RQ AGN and investigate the RL-RQ dichotomy using low-frequency data at 685~MHz from the upgraded Giant Metrewave Radio Telescope (GMRT), in the Palomar-Green \citep[PG;][]{Green86} sample of quasars. In this paper, we present the results on 65 PG quasars, in addition to the 22 sources presented in Paper I. The GMRT 685~MHz angular resolution of $\sim4\arcsec$, translates to spatial scales of a few kpc for the PG quasars.  

The paper is organized as follows. The PG sample is described in Section~2. Section~3 describes the uGMRT data reduction and analysis. The results are presented in Section~4, discussion in Section~5, and the conclusions of this work are given in Section~6. Throughout this paper, we have assumed $\Lambda$ cold dark matter cosmology with $H_0$ = 73~km~s$^{-1}$~Mpc$^{-1}$, $\Omega_{m}=0.27$ and $\Omega_{v}=0.73$. The spectral index $\alpha_{R}$, derived using the 685~MHz peak flux densities (this work) and the 5~GHz peak flux densities reported by \citet{Kellermann94}, is defined such that flux density at frequency $\nu$, $S_\nu\propto\nu^{\alpha_{R}}$.

\section{The Sample}
The PG catalog encompasses $\sim$1800 UV-excess objects (i.e., U$-$B$<-$0.44), identified through an optical survey covering an area of approximately $\sim$10,714~deg$^2$ at absolute galactic latitudes exceeding 30 degrees. This survey utilized 266 double U and B exposures from the Palomar 18-inch Schmidt Telescope \citep{Green86}. The Palomar Bright Quasar Survey (BQS), a subset of the PG survey, selected objects based on specific criteria: (1) morphological criteria indicating a dominant star-like appearance and (2) spectroscopic criteria revealing the presence of broad emission lines. The BQS sample comprised 114 objects, including 92 quasars with $M_B < -23$ and 22 Seyferts or low-luminosity quasars with $M_B > -23$. Our focus is on the PG quasar sample, consisting of objects from the BQS with redshifts $z<0.5$. This subset includes 87 sources, encompassing both quasars and Seyfert type 1 galaxies \citep{BorosonGreen92}. Nearly 80\% (71/87) of the sample is RQ, while the remaining 20\% (16/87) is RL \citep{Kellermann89}.

\begin{figure*}
\centering
\includegraphics[height=9cm]{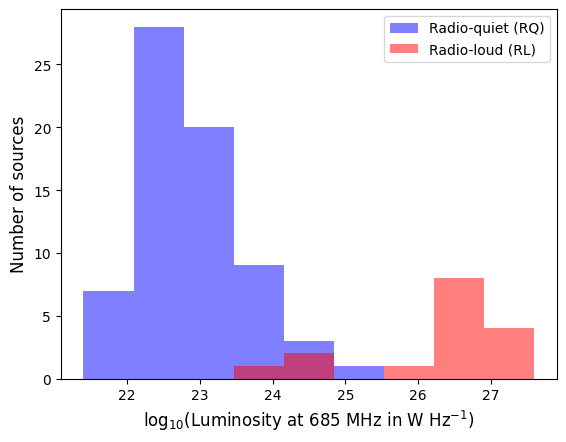}
\caption{\small Distribution of the uGMRT 685 MHz luminosities for the PG quasar sample. The RL-RQ dichotomy persists in the low-frequency uGMRT data.}
\label{RLhist}
\end{figure*}

The PG quasar sample stands out as one of the most extensively studied samples of low-redshift AGN. It boasts a wealth of data, including accurate black hole masses obtained from reverberation mapping \citep{Kaspi00} and single-epoch spectroscopy data \citep{VestergaardPeterson06}, detailed host galaxy morphologies and bulge/disc decompositions from Hubble Space Telescope (HST) imaging data \citep{Kim08, Kim17}, comprehensive broad-band spectral energy distributions (SEDs) and accurate bolometric luminosities across various wavelengths \citep{Shang11}, observations elucidating dust \citep{Petric15, Shangguan18} and gas properties \citep{Evans06, Shangguan20}, enabling analyses of the interstellar medium of host galaxies, and infrared (IR) data facilitating investigations into torus properties \citep{Zhuang18} and star formation rates \citep[SFR;][]{Shi14}. There however, remains a dearth of high-sensitivity low-frequency radio data on this sample.

Paper I discusses the results from our pilot study conducted for a sample of 22 PG quasars (20 RQ and 2 RL) with the uGMRT at 685 MHz. These sources were chosen on the basis of availability of Atacama Large Millimeter/submillimeter Array (ALMA) data \citep{Shangguan20} to study the low-frequency radio emission in tandem with the CO(2–1) molecular emission. In the current paper, we present results from observations carried out at 685 MHz on the 65 remaining PG quasars with the uGMRT, thereby completing the first GMRT low-frequency survey of the PG quasar sample. 

\section{Observations and Data Reduction}
The GMRT 685 MHz observations of the PG quasars were carried out from 2019 December to 2020 January (Project ID: 37\_042; PI: Silpa S.). Data were reduced and analyzed using the standard procedures in CASA\footnote{Common Astronomy Software Applications; \citet{Mullin07}}. To optimize the {\it uv}-coverage, each target was observed for at least two scans of $\sim$20 min each. The flux calibrators 3C~48, 3C~286, 3C~138, and/or 3C~147, were observed for 5 min at the start and end of the observations. The observations of the phase calibrators (of 4 min) and targets were interspersed, allowing the time-dependent amplitude and phase calibration solutions to be interpolated onto the time intervals of the target observations in between the calibrators. The GMRT observations are summarized in Table 1. These observations can measure the largest angular scale of $\sim25\arcmin$, with the FWHM of the synthesized beam being $\sim3–5\arcsec$. The basic calibration strategy is described below.

\begin{figure*}
\centering
\includegraphics[height=9cm]{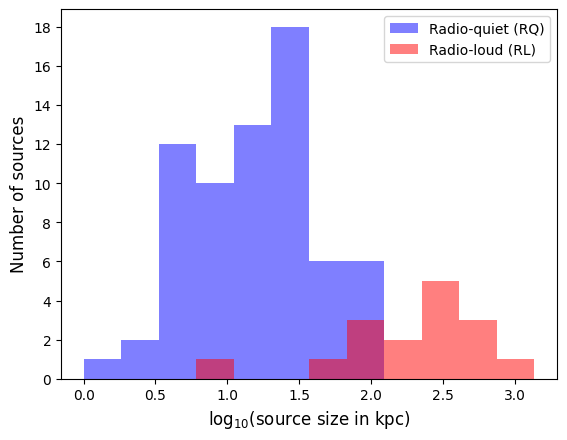}
\caption{\small Distribution of 685 MHz projected radio source sizes in kpc for the PG quasar sample.}
\label{hist}
\end{figure*}

The GMRT produces data is in the Long-Term Archive (LTA) format, which needs to be first converted to the Flexible Image Transport System (FITS) format. This step was carried out using the $\tt{LISTSCAN}$ and $\tt{GVFITS}$ utilities. 4096 channels spanning 560$-$810 MHz were used in these observations; beyond 810 MHz, the sensitivity drops significantly. The non-usable frequency range (810$-$950 MHz) and non-working antennas were omitted from the log file generated by $\tt{LISTSCAN}$, resulting in the final FITS file covering the frequency range 560–810 MHz. Using the $\tt{CASA}$ task $\tt{IMPORTUVFITS}$, the FITS file was converted to a measurement set (MS), which was then provided as the input file to the respective pipelines described below. Out of the 65 sources, 25 were analysed using our $\tt{CASA}$-based pipeline available at \url{https://sites.google.com/view/silpasasikumar/}, while the remaining 40 sources were reduced using the CAsa Pipeline-cum-Toolkit for Upgraded GMRT data REduction (CAPTURE) pipeline \citep{Kale-Ishwara21}, available at \url{https://github.com/ruta-k/CAPTURE-CASA6}. 

We performed four iterations of phase-only self-calibration and four iterations of amplitude and phase self-calibration, except for PG 2251+113, where four iterations of phase-only self-calibration and only two rounds of amplitude and phase self-calibration were performed, as further rounds did not improve the image quality. For all the sources, the \texttt{ROBUST} parameter was chosen to be 0.5, except for PG 0003+158 where a value of \texttt{ROBUST} = 1 was adopted in order to mitigate the imaging artefacts and improve sensitivity, while still maintaining adequate resolution.
The basic calibration and imaging steps have been discussed in detail in Paper I. 

The 685 MHz peak and total flux density of individual sources was calculated using the {\tt AIPS} task \texttt{JMFIT}. For the sources where \texttt{JMFIT} did not converge and/or the sources that have multiple components, tasks \texttt{TVMAXF} and \texttt{TVSTAT} were used to calculate the peak flux and total flux densities, respectively. All flux density measurements are reported in Table~\ref{Table1}. The peak flux densities range from 0.3 mJy~beam$^{-1}$ to $\sim$~$3\times10^4$ mJy~beam$^{-1}$ (Table~\ref{Table1}). The rms noise for the sample, measured locally from blank sky regions around each quasar, is typically around $\sim 30-40$ $\mu$Jy~beam$^{-1}$ (except for a few sources where it is explicitly mentioned). 

The errors in the flux densities were calculated by considering a 10\% amplitude calibration uncertainty (for the uGMRT) and the rms noise of individual sources. The size of the sources reported in this paper were measured using the Gaussian-fitting AIPS task \texttt{JMFIT} (for unresolved sources) and \texttt{TVDIST} (for extended sources). The typical errors in sizes are $\sim3\%$.

The results of our analysis are presented in the following section. Linear fits ($y = mx + c$) to all correlation plots were obtained using the {\tt numpy.polyfit} routine in Python. The corresponding correlation coefficients and p-values are summarized in Table~\ref{Table2}.

\section{Results}\label{morph-alpha}
We detect 685 MHz radio emission in all sources except for the RQ quasars PG~0043+039, PG~1121+422, and PG~1552+085, at the $3\sigma$ level. 
Figure~\ref{RLhist} presents the 685~MHz luminosities of the PG sample with sources classified RL and RQ following \citet{Kellermann89}. Figures~\ref{fig13} to \ref{fig23} present the 685~MHz images of the PG quasars. RL sources in the sample exhibit large-scale jets on hundreds of kiloparsec scales, except for PG~2209+184 with $\sim7$~kpc size and PG~1425+267 with a total extent of $\sim1.3$ Mpc. 

Of the 16 RL sources, seven exhibit morphological features such as X-shaped lobes, hybrid morphology, and highly bent or knotty jets, which could arise either due to environmental effects or due to restarted AGN jet activity.
Five sources (see Table~\ref{Table1}), remain unresolved at the uGMRT resolution of $\sim4\arcsec$. Below we discuss the RL-RQ luminosity and size divide followed by results on the core radio spectral indices and the possible origins of the low-frequency radio emission.

\subsection{The RL-RQ Divide: 685~MHz Luminosity, Sizes \& Equipartition Magnetic Field Strengths}
The RQ and RL populations display a bimodal distribution in the 685~MHz luminosities,  $(L_{685~\mathrm{MHz}})$, with a small overlap (Figure~\ref{RLhist}). Based on this distribution, we find the following empirical division: RQ quasars typically show $\log_{10}(L_{685\,\mathrm{MHz}}/\mathrm{W\,Hz^{-1}}) < 23.5$, while RL quasars mostly show $\log_{10}(L_{685\,\mathrm{MHz}}/\mathrm{W\,Hz^{-1}}) > 25.5$. `Intermediate' sources have $23.5 \leq \log_{10}(L_{685\,\mathrm{MHz}}/\mathrm{W\,Hz^{-1}}) \leq 25.5$. 

\begin{figure*}
\centering
\includegraphics[height=6cm,width=8cm]{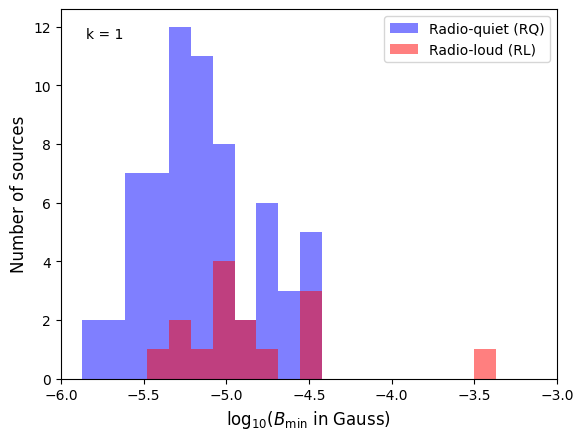}
\includegraphics[height=6cm,width=8cm]{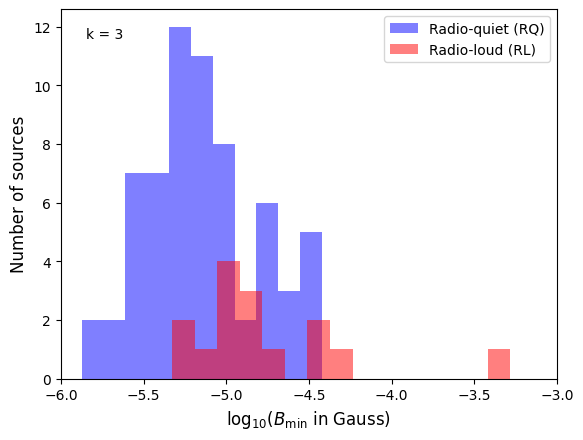}
\caption{\small Distribution of `minimum energy' magnetic field strength for RQ and RL sources for $k=1$ (left panel) and $k=3$ (right panel). The outlier source with the largest ${B_\mathrm{min}}$ value is PG~1226+023 or 3C~273.}
\label{Bmin_hist}
\end{figure*}

The RL sources with $L_{685~\mathrm{MHz}}<10^{25.3}$~W\,Hz$^{-1}$ are: PG~0007+106 (III~Zw~2), PG~1309+355, PG~1425+267, and PG~2209+184. Of these, PG~2209+184 is a flat-spectrum radio quasar (FSRQ) without any signature of kpc-scale jets \citep[see][]{Silpa2020}. PG~1425+267 is a large FRII radio galaxy without a dominant core or hotspots. The following RQ sources have $L_{685~\mathrm{MHz}}>10^{24}$~W\,Hz$^{-1}$: PG~0157+001, PG~1216+069, PG~1543+489, PG~1612+261, PG~1700+518, and PG~2112+059. All these sources are core-dominant sources without substantial extended emission. 

We also examined the distribution of the source sizes in kpc (Table~\ref{Table1}) for the PG sample. As shown in Figure~\ref{hist}, the radio sizes display the RQ$-$RL divide as well, with RQ sources typically having sizes $\lesssim30$~kpc, with only 16 ($\sim22$\%) RQ sources having sizes between 30 and 100 kpc where there is an overlap with RL quasar sizes. For a source size of 30 kpc with a jet speed equal to the speed of light (the actual jet speed would indeed be a fraction of this) would imply a lower limit to the source age to be $\sim10^5$~yrs. The RL quasars, PG~1226+023, PG~1302-102, PG~1309+355, PG~2251+113, and PG~2209+184 have sizes $<100$~kpc. Among these sources, PG~1226+023, PG~1302-102, PG~1309+355, and PG~2209+184 are FSRQs, while PG~2251+113 is a NLSy1/SSRQ. Hence, their small size is likely due to projection effects caused by small viewing angle. The RQ source PG~1216+069 has a large extent of $\sim98$~kpc. Our image shows the source to be a regular radio galaxy albeit with weak diffuse emission. This source appears to be misclassified RL AGN; its 685~MHz luminosity also places it into the RL category. The radio size appears to be a better discriminant of the radio-loudness/radio-quietness of a quasar compared to the radio luminosity. 

As differences in magnetic field strengths close to the central black holes has been suggested to explain the RL-RQ divide \citep[e.g.][]{Tchekhovskoy10, Chamani2021, Lopez-Rodriguez2023}, we examined the distribution of `minimum-energy' magnetic field strengths ($B_{\rm min}$) for the RQ and RL sources in our sample, following the relations in \citet{Odea1987}. We adopted a proton-to-electron energy ratio ($k$) of unity and assumed a volume filling factor ($\phi$) of 1. The radio spectrum was taken to span a frequency range from 10 MHz ($\nu_l$) to 100 GHz ($\nu_u$). For each source, the integrated flux density of the core (listed in Table~\ref{Table1} was used along with a fixed flat spectral index of $\alpha = -0.2$.\footnote The theoretical value of the dimensionless parameter $C_{12}$, that incorporates the frequency integration over the synchrotron spectrum and depends on the adopted spectral index and frequency limits, is taken for $\alpha = -0.2$ to be $8.3\times10^6$ \citep[see][]{Pacholczyk1970}. A spherical geometry was assumed for the emitting volume, with the radius derived from the core sizes measured using the task \texttt{JMFIT} in AIPS. 

Figure~\ref{Bmin_hist} shows the histogram of $\log_{10}(B_{\rm min})$ for the RQ and RL sources. We do not observe any bimodality in the distribution, and the $B_{\rm min}$ values for the two populations show an overlap, with the exception of the RL source, PG~1226+023, a.k.a. 3C~273. While on the surface, this may suggest no difference in the magnetic field strengths, it is important to note that the assumptions made in the magnetic field estimation are likely not valid for both AGN classes. For instance, differences in parameters such as the source volume and volume filling factors, as well as the source relativistic proton-to-electron energy ratio ($k$), are likely to be different between the AGN sub-classes which may result in the net magnetic field strengths being lower in RQ sources \citep[e.g.,][]{Chamani2021}. 
It is interesting to note that the RQ and RL AGN differ statistically significantly at the $\sim3\sigma$ level when $k=3$ in the RL sources. Independent estimates of magnetic field strengths need to be made using core radio and X-ray flux density, the former being due to synchrotron emission and the latter due to inverse-Compton. As of now, high resolution Chandra X-ray observations that can isolate the core emission, do not exist for the PG quasar sample to test the above.

\subsection{Radio Core Spectral Indices \& Origin of Radio Emission}
The radio core spectral index values ($\alpha_R$) derived using the 685~MHz peak flux densities (this work) and the 5~GHz peak flux densities reported by \citet{Kellermann94} have been noted in Table~\ref{Table3}. 
For the sources studied in Paper I, the values were recalculated using the same methodology and are provided in the same Table with source names in bold.  
The cores of RQ sources, on average, exhibit steeper spectra (mean $\alpha_R$: $-$0.7) than the cores of RL sources (mean $\alpha_R$: $-$0.2). All RL sources in our sample, except PG~1100+772 and PG~1704+608 have flat or inverted spectrum cores, indicative of synchrotron self-absorbed bases of jets. PG~1100+772 and PG~1704+608 exhibit very steep spectrum cores ($\alpha_R\sim -1$). The spectral index for PG~2251+113, when calculated using the 5~GHz flux density from \citet{Kellermann89}, yielded an unrealistic value of approximately $-3$. Therefore, a spectral index of $-0.7$ has been adopted here \citep[consistent with][]{Baghel24}. 

The RQ cores exhibit a range of spectral indices, varying from flat to steep, in agreement with the previous studies \citep{Silpa2020, Chiaraluce20}. PG~1351+640 and PG~1435-067 have flat spectrum cores, while PG~1216+069 has an inverted spectrum core. A flat/inverted spectrum can be interpreted as arising from unresolved synchrotron self-absorbed bases of small-scaled radio jets, `frustrated' radio jets in dense environments \citep{O'Dea91}, or coronal emission \citep{LaorBehar08, RaginskiLaor16}. PG~1626+554 has a spectral index of $-0.1$; therefore, thermal free-free emission from the accretion disk, torus, or HII regions cannot be ruled out as alternative explanations for this source. PG~0804+761 and PG~1416$-$129 lie at the cusp of flat-steep division. 

The remaining RQ sources have steep spectrum cores. A steep radio spectrum can arise from unresolved or barely resolved jet/lobe emission on even smaller spatial scales \citep[such as on a few- to 10-kpc scales;][]{Falcke00, Jarvis19, Kharb19} or optically thin synchrotron emission from AGN or starburst-driven winds \citep{Cecil01, IrwinSaikia03, HotaSaikia06, Hwang18}. 

\begin{figure}
\centering
\includegraphics[height=6.5cm]{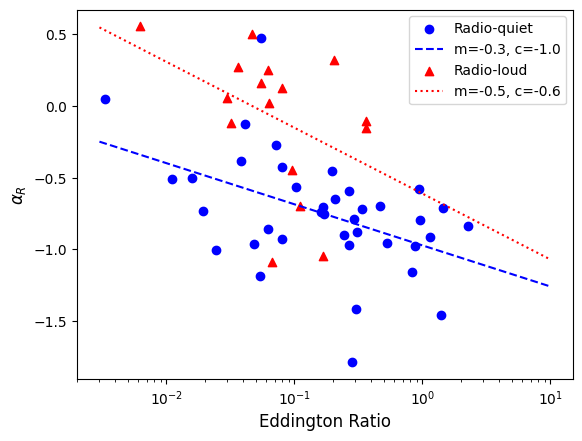}
\caption{\small The 685~MHz - 5~GHz spectral index versus Eddington ratios for the PQ sample. In this and following plots, red triangles denote the RL QSOs, the blue circles denote RQ QSOs, the blue-dashed and red-dotted lines represent the best-fit lines for RQ and RL sources, respectively. }
\label{fig1}
\end{figure}

To understand the origin of the observed spectral indices, $\alpha_R$, we examined the same with respect to the Eddington ratios (see Figure~\ref{fig1}). We found an anti-correlation between the two for both the RL and RQ quasars, suggesting a link between accretion rate and the presence of optically thin steep-spectrum synchrotron emission arising from either unresolved jets or AGN-driven winds. 


\subsection{The Radio-IR \& $\mathrm{\alpha_R - q_{IR}}$ Correlations}
\label{radio-IR}
\begin{figure*}
\centering
\includegraphics[height=6.5cm]{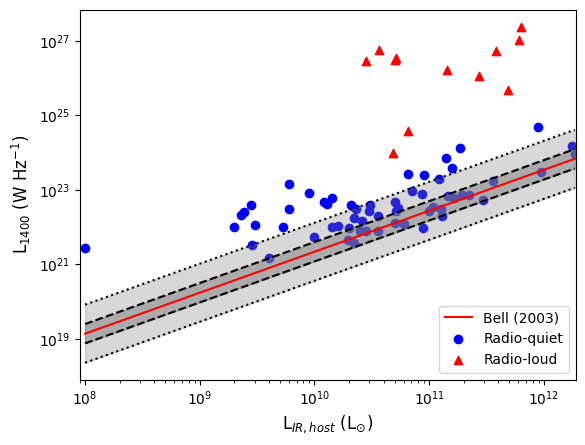}
\includegraphics[height=6.5cm]{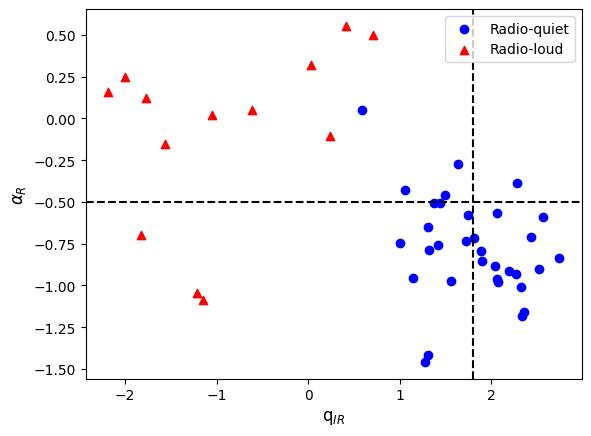}
\caption{\small (Left) The radio-IR correlation for PG sources. The solid black line is the radio−IR correlation for star-forming galaxies \citep{Bell03}. 
The black dashed lines mark the $1\sigma$ regions and the black dotted lines mark the $3\sigma$ regions. (Right) The mean core spectral index as a function of $q_{\mathrm{IR}}$. The black dashed vertical line represents $q_{\mathrm{IR}} = 1.8$ (which is about $2.1\sigma$ below the \citep{Bell03} relation), which discriminates the AGN versus star formation contributions. The black horizontal line is $\alpha_R=-0.5$, which we use to discriminate between steep ($\alpha_R<-0.5$) and flat ($\alpha_R > -0.5$) spectrum cores.}
\label{fig4}
\end{figure*}

Continuing our exploration on the origin of radio emission, we looked at the radio-infrared correlation, which is one of the best-studied and robust correlations in astrophysics. Figure~\ref{fig4} shows $L_{1400}$ as a function of $L_{\mathrm{IR,host}}$. The straight line in the plot represents the radio-IR correlation from \citet{Bell03}, as given by \citet{Silpa2020}:
\begin{equation}
    \log L_{1400}~\mbox{W Hz}^{-1} = 1.10 \log L_{\mathrm{IR,host}} + 10.34
\end{equation}
The $1\sigma$ and $3\sigma$ limits are derived using a scatter of 0.26 dex in the radio-IR correlation \citep{Bell03}. We find that all RL PG sources and $\sim1/3^{\mathrm{rd}}$ (20/59) of the RQ PG sources (with known IR luminosities and $q_{\mathrm{IR}}$ values) lie above the $3\sigma$ limit of the radio-IR correlation. 

Additionally, \citet{Bell03} have defined the parameter, $q_{\mathrm{IR}}$, to quantify the radio-IR correlation, which is given by: 
\begin{equation} 
    q_{\mathrm{IR}} = \log(L_{\mathrm{IR}}/3.75 \times 10^{12}~\mbox{W}) -  \log(L_{1400}/~\mbox{W Hz}^{-1})
\end{equation}
A low $q_{\mathrm{IR}}$ value, along with a significant positional offset in the radio-IR correlation plane, would imply that AGN likely dominates the radio emission in the source {\citep[e.g. see][]{White2017}. A value of $q_{\mathrm{IR}}<1.8$ classifies the source as being ``radio-excess'' due to dominant AGN contribution \citep{Condon02}. 

\begin{figure*}
\centering
\includegraphics[height=6cm,width=8cm]{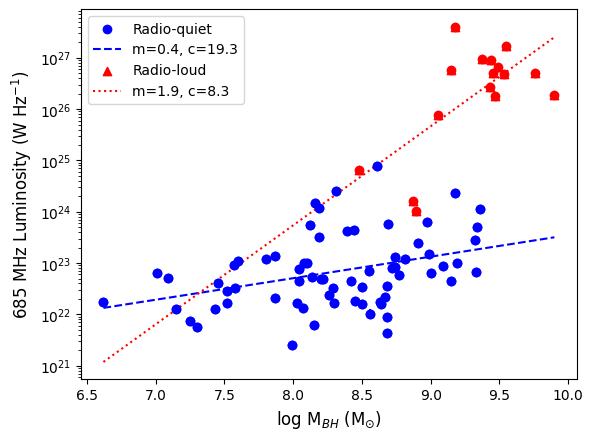}
\includegraphics[height=6cm,width=8cm]{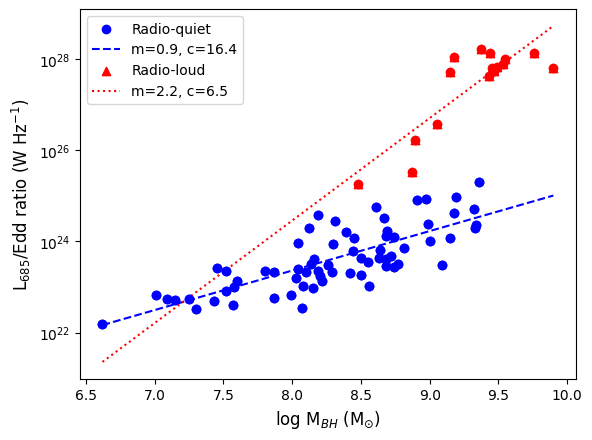}
\caption{\small (Left) Total 685~MHz radio luminosities from the uGMRT versus logarithm of black hole masses for the PG quasars. The errors in 685 MHz radio luminosity are smaller than or comparable to the symbol size. The RL sources do not show any significant correlation. (Right) Ratio of 685~MHz luminosities to Eddington ratios versus black hole masses.}
\label{fig7}
\end{figure*}

To derive $q_{\mathrm{IR}}$ for our sample, we followed the approach of \citet{Silpa2020}, first estimating the 1400 MHz flux-density ($S_{1400}$) from the observed total 685 MHz flux-density ($S_{685}$) using: 

\begin{equation}
    S_{1400} = S_{685}(1400/685)^{\alpha_R}
\end{equation}
where, $\alpha_R$ is the radio spectral index. The value for $\alpha_R$ is taken to be $-0.7$ for all sources. Using $S_{1400}$, we calculated the 1400 MHz luminosity for the sample as $L_{1400} = 4 \pi D_L^2S_{1400}(1+z)^{(-\alpha_R - 1)}$. To calculate the total IR luminosities, \citet{Lyu17} performed SED decomposition using AGN \citep{Elvis1994, Xu2015}, stellar, and star-forming galaxy templates \citep{Rieke2009}, and integrated the best-fit model over the $8–1000~\mu \rm{m}$ range. We used the $8-1000~\mu$m luminosities ($L_{\mathrm{IR,host}}$) from \citet{Lyu17}, which account only for the host galaxy’s contribution minus the AGN, to calculate $q_{\mathrm{IR}}$. 

The sources that lie above the $3\sigma$ limit of the radio-IR correlation also have $q_{\mathrm{IR}}<1.8$ (see Table~\ref{Table3}). This suggests an AGN dominance in their radio emission. While the remaining 39 RQ sources lie on the correlation and within the $3\sigma$ limit, this does not rule out an AGN contribution in them \citep[e.g.][]{Wong16}. The coexistence of AGN and stellar emission is likely, as also suggested in Paper I for the smaller PG sub-sample. In the following section, we look at the relation between AGN jet power and accretion. 

\subsection{Relation between Radio Power \& Accretion}
In the current paper, the black hole masses (${M_{\mathrm{BH}}}$) and AGN bolometric luminosities ($L_\mathrm{bol}$) have been obtained from \citet{Shangguan18} and \citet{Lyu17}, respectively. The ${M_{\mathrm{BH}}}$ values are single-epoch virial estimates calculated using Equation 4 of \citet{HoKim15}, while the bolometric luminosities are derived as $L_\mathrm{bol}$ = 5.29 $L_\mathrm{IR,AGN}$, with $L_\mathrm{IR,AGN}$ being the AGN infrared luminosity from \citet{Lyu17}. 
The prefactor of 5.29 is derived from the intrinsic AGN SED template, calibrated by \citet{Xu2015} and adopted by \citet{Lyu17} in defining the bolometric correction.
The Eddington luminosity is computed using $L_\mathrm{Edd}$ ($L_{\odot}) = 3.2\times10^4$(${M_{\mathrm{BH}}}$/$M_\odot$). The Eddington ratios $L_\mathrm{bol}$/$L_\mathrm{Edd}$) range from $0.01-3.59$ for the PG sample.

We find a significant ($\sim3\sigma$) correlation ($r_s=0.6$) between 685 MHz luminosity and black hole mass (left panel of Fig.~\ref{fig7}). This could be interpreted as the radio jet power being closely related to the black hole mass, with more massive black holes producing the most powerful radio outflows, consistent with \citet{Best2005}, and others. This correlation becomes tighter on invoking the dependence of the Eddington ratio, a proxy for the accretion rate (right panel of Fig.~\ref{fig7}). We also note that RL AGN generally have higher ${M_{\mathrm{BH}}}$ and RQ AGN have lower ${M_{\mathrm{BH}}}$, although with some overlap between the two populations, which suggests that the sources with low values of ${M_{\mathrm{BH}}}$ are less likely to possess powerful jets. This could favor the hypothesis that the jets are powered by black hole spins and that the observed correlation is resulting from a strong connection between ${M_{\mathrm{BH}}}$ and black hole spin \citep{Laor2000}. However, there have been hints of a more slowly rotating population emerging at higher SMBH masses \citep{Reynolds2013, Reynolds2014}. Our correlation becomes tighter on invoking the accretion rate dependence, which suggests that the RL/RQ divide could have both ${M_{\mathrm{BH}}}$ and accretion rate dependence, consistent with the findings of \citet{Lacy01}. 

We carried out a partial correlation test between 685 MHz luminosity and Eddington ratio, controlling the influence of the black hole mass. Interestingly, a significant positive correlation is observed for the entire sample according to the Spearman rank correlation test ($r_s = 0.66$, $p_s = 1.8\times10^{-11}$). The RQ and RL sources separately also exhibit positive correlations (RQ: $r_s = 0.73$, $p_s = 3.3\times10^{-12}$; RL: $r_s = 0.59$, $p_s = 0.02$). The weaker correlation seen in RL quasars may arise from additional contributions of extended jet/lobe emission, whereas in RQ quasars the radio output predominantly traces the nuclear processes, leading to a comparatively tighter correlation. 

Our analysis of the PG quasar sample at 685 MHz reveals a radio luminosity-${M_{\mathrm{BH}}}$ correlation with a slope of 1.4, consistent with the slope of $1.4 \pm 0.2$ reported by \citet{Lacy01} at 5\,GHz. Including the Eddington ratio yields the best-fit relation, $\log L_{685\,\mathrm{MHz}} = 1.9 \log M_{\mathrm{BH}} + 1.0 \log (L/L_{\mathrm{Edd}}) + 8.6$, which is also consistent with the slope of $1.9\pm0.2$ in \citet{Lacy01}.

Furthermore, we estimated the time-averaged jet kinetic power ($\overline Q$) using the relation given by \citet{Punsly2018} 
\begin{equation}
   \overline Q = 3.8 \times 10^{45} f L_{151}^{6/7} ~\ \text{ergs s}^{-1}
\end{equation}
where, $L_{151}$ is the radio luminosity at 151 MHz in units of $10^{28}$ W Hz$^{-1}$ sr$^{-1}$ and $f = 15$ \citep{Blundell2000}. We derived $L_{151}$ using the following relations:

\begin{align}
    S_{151} &=  S_{685} (151/685)^{\alpha_R}\\
    L_{151} &=  4\pi D_L^2S_{151}(1+z)^{(-\alpha_R-1)}
\end{align}
where, $\alpha_R$ is taken to be $-0.7$ for all sources and $S_{685}$ is the total radio flux density listed in Table~\ref{Table1}. The derived $\overline Q$ are presented in Table~\ref{Table3}. 

We find a significant positive correlation between jet power and ${M_{\mathrm{BH}}}$ for the entire sample (Figure~\ref{fig8}, Table~\ref{Table2}). Interestingly, the RQ and RL sources do not show different slopes here, compared to the 685~MHz luminosities versus black hole masses (left panel of Fig.~\ref{fig7}). The jet powers of RQ and RL sources individually show a positive correlation with Eddington ratios (right panel of Fig.~\ref{fig8}). However, the split between the RQ and RL sources is clearer here. This may be consistent with the RQ and RL sources following different relationships between jet power and accretion rates (see Section~\ref{discuss}).

\begin{figure*}
\centering
\includegraphics[height=6cm,width=8cm]{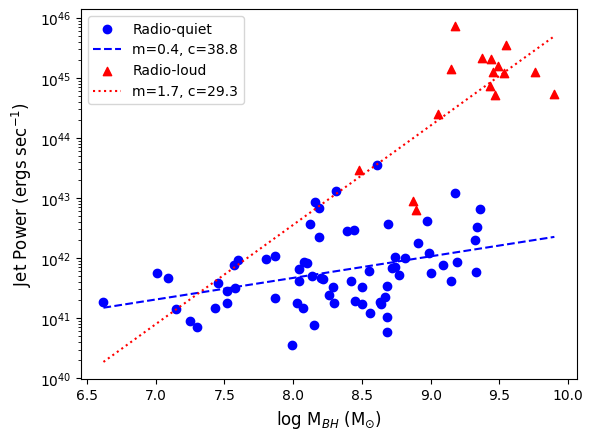}
\includegraphics[height=6cm,width=8cm]{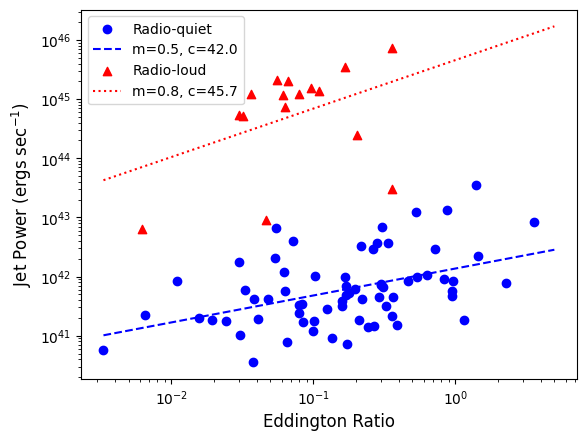}
\caption{\small Jet kinetic power versus black hole masses (left) and Eddington ratios (right) for the PG sample. The RL sources in both the plots do not show any significant correlation (see Table~\ref{Table2}).}
\label{fig8}
\end{figure*}

\subsection{The Radio `Fundamental Plane'}
The radio luminosity (${L_\mathrm{R}}$) is found to have a direct dependence on the mass of the black hole and X-ray luminosity (${L_\mathrm{X}}$), which is characterized as the black hole `fundamental plane' \citep[FP;][]{Merloni03}. We used the $0.2–20$~keV X-ray luminosities provided in \citet{Brandt00, LaorBehar08} taken from the ROSAT All-Sky Survey (RASS) with an effective angular resolution of $\sim1.8\arcmin$.
Figure~\ref{fig11} shows the correlation between ${L_\mathrm{R}}$, ${L_\mathrm{X}}$, and ${M_{\mathrm{BH}}}$ for the PG sample. 

To model the dependence of ${L_\mathrm{R}}$ on ${L_\mathrm{X}}$ and ${M_{\mathrm{BH}}}$, we performed linear regression using the {\tt LinearRegression} routine from the {\tt scikit-learn} Python package. The PG sample studied in this work defines the FP as: log $L_{\mathrm{R}}$ = 0.6 log $L_{\mathrm{X}}$ + 1.1 log ${M_{\mathrm{BH}}}$ - 11.3. The RQ and RL sources show a dichotomy in the FP. The set of best-fit equations for RQ and RL sources are log ${L_\mathrm{R}}$ = 0.3 log ${L_\mathrm{X}}$ + 0.3 log ${M_{\mathrm{BH}}}$ + 8.5 and log ${L_\mathrm{R}}$ = 0.8 log ${L_\mathrm{X}}$ + 1.4 log ${M_{\mathrm{BH}}}-22.5$, respectively (see Figure~\ref{fig11}). We discuss the radio `FP' for the PG sources in the following section.

\section{Discussion}\label{discuss}
The radio-quiet PG quasars exhibit a significant correlation between 685~MHz luminosity and Eddington ratios. However, no such correlation is observed for the entire PG sample (see Table~\ref{Table2}). Similarly, no correlation is found between radio-loudness and Eddington ratios for the PG quasar sample (see Table~\ref{Table2}). \citet{Ho02} had reported an anti-correlation between the radio-loudness and the Eddington ratios for a sample of active galaxies. Since the Eddington ratio varies as mass accretion rate ($\dot M$), this anti-correlation implied that relative radio power increases with decreasing $\dot M$. Their study indicated that most of the weakly active nuclei in nearby galaxies are likely undergoing advection-dominated accretion. A similar anti-correlation was also reported by \citet{Sikora07} for a sample comprising broad-line radio galaxies (BLRG), RL quasars, Seyferts, LINERS, FR I radio galaxies, and PG quasars.

The sample studied in \citet{Ho02} included sources spanning a much broader range of activity, our work focuses on the PG quasars, which are optically selected and, therefore, are unlikely to host an advection-dominated accretion flow (ADAF) disk \citep[e.g.,][]{Narayan1994}. When comparing our results with those of \citet{Sikora07}, we observe that their sample includes 43 PG sources, which, when considered separately, also show a negative correlation between radio-loudness and Eddington ratios (at the $2\sigma$ level). However, when all the 87 PG sources are considered, we find no such correlation. This could likely be because the already weak and scattered trends are further diluted by the large dispersion introduced by the remaining sources.


\begin{figure*}
\centering
\includegraphics[height=6cm,width=8cm]{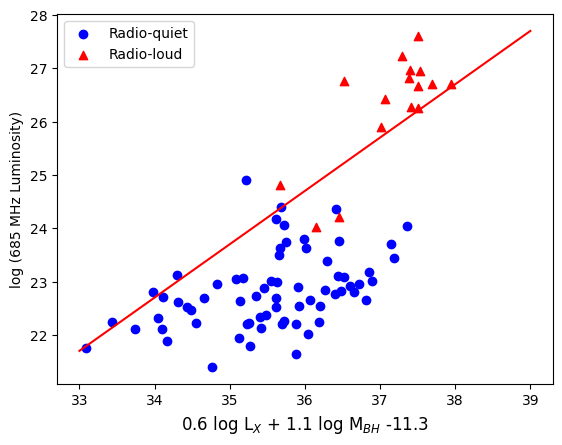}\\
\includegraphics[height=6cm,width=8cm]{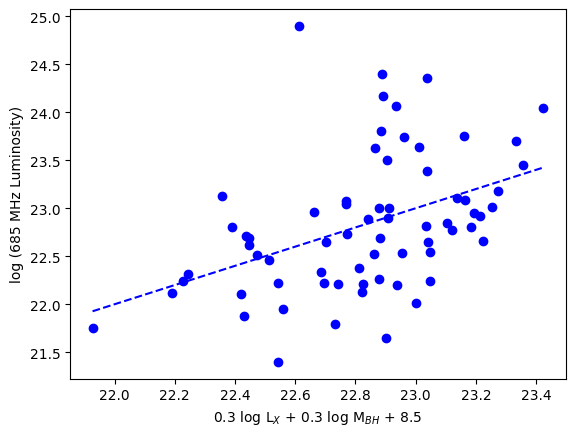}
\includegraphics[height=6cm,width=8cm]{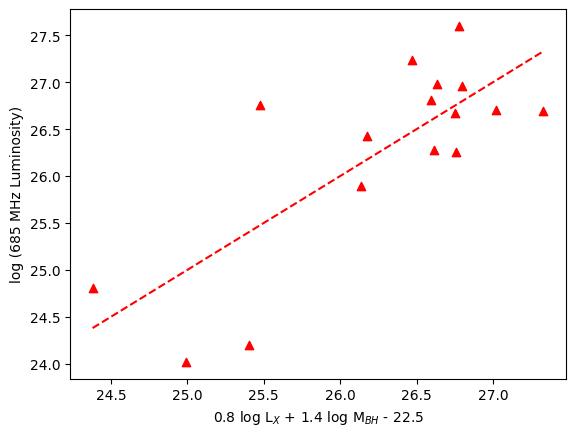}
\caption{\small The `fundamental plane' of black hole activity for the complete sample (top), RQ (left), and RL (right) sources separately. The x-axis is defined following the best-fit parameters respectively. }
\label{fig11}
\end{figure*}

\citet{Franceschini1998} found a positive correlation between ${M_{\mathrm{BH}}}$ and 5~GHz radio luminosity for 13 nearby weakly active galaxies, with the functional form of the correlation arising naturally due to an ADAF. \citet{Laor2000} derived the black hole masses from the H$\beta$ FWHM values and the optical continuum luminosity for all 87 $z<0.5$ PG sources \citep{BorosonGreen92} and found an increasing trend of 5~GHz radio luminosity with ${M_{\mathrm{BH}}}$. However, when they included 29 nearby galaxies in their sample, they found that the tight relation of \citet{Franceschini1998} was not supported by their data. 
\citet{Lacy01} combined the quasars from the FIRST Bright Quasar Survey (FBQS) with the PG sample and studied the correlation between ${M_{\mathrm{BH}}}$ and the 5~GHz radio luminosity. They found a continuous variation of radio luminosity with ${M_{\mathrm{BH}}}$. They reported that the correlation becomes marginally tighter on invoking the dependence on the accretion rate, which is based on the radio-optical correlation studies \citep[e.g.,][]{Serjeant1998, Willott1999}. They suggested a new scheme to ``unify'' RL and RQ sources in which radio luminosity would depend on ${M_{\mathrm{BH}}}$ and accretion rate. Our uGMRT study of the PG quasar sample is consistent with the findings of \citet{Laor2000} and \citet{Lacy01}.

\subsection{Signatures of Episodic AGN Activity}
7 of the 16 RL sources exhibit morphological signatures such as X-shaped lobes, hybrid morphology, or strongly bent jets, which may be consistent with either environmental effects or episodes of restarted AGN activity \citep[also see][]{Baghel23, Baghel24}. In particular, the misaligned northern radio lobe in PG~1704+608 suggests a change in the direction of the jet, as reported by \citet{Vaddi2019, Baghel23}. A clear discontinuity in the surface brightness between the inner jets and the larger lobes in PG~1004+130 is indicative of restarted AGN activity with two distinct episodes, as proposed by \citet{Ghosh23} and \citet{Baghel23} based on spectral index imaging and polarization. Reactivation is suggested to be one of the possible explanations for the peculiar morphology of PG~1100+772 \citep{Marecki12, Fernini07, Baghel23}. 

All RL sources in our sample, except PG~1100+772, PG~1704+608, and PG~2251+113, exhibit flat or inverted spectrum cores, consistent with synchrotron self-absorbed jet bases. PG~1100+772 and PG~1704+608 show very steep spectrum cores ($\alpha \sim -1$), while PG~2251+113 yields an ultra-steep spectral index of $\sim -3$ when using the 5 GHz flux from \citet{Kellermann89}. Consequently, we adopt a spectral index of $-0.7$ for this source, as reported by \citet{Baghel24}. The presence of steep-spectrum cores in these RL sources suggests that restarted activity on small spatial scales cannot be ruled out. All three sources suggest `relic' emission from past episodes of AGN activity \citep{Roettiger94, Kharb16}. Additionally, three radio-quiet sources: PG~1612+261, PG~1613+658, and PG~1700+518, also exhibit very steep-spectrum cores ($\alpha_R < -1$), further suggesting that small-scale restarted activity may not be exclusive to RL quasars.

\subsection{Origin of radio emission in RQ quasars} 
In radio-loud sources, the radio emission is primarily powered by relativistic jets \citep[e.g.,][]{Bridle1984}. In contrast, the origin of radio emission in RQ sources remains an open question, with possibilities ranging from weak jets and AGN winds to star formation and coronal activity \citep[e.g.,][]{Kellermann89, Kellermann94, White2017, Zakamska16, Hwang18, Silpa2020, Baldi2022, Chen2023}.

The PG quasars occupy four different quadrants in the $\alpha_R$ versus q$_{IR}$ plot (see right panel of Figure~\ref{fig4}). The top-right quadrant denotes flat/inverted spectrum radio emission of stellar origin, most likely arising from thermal free-free emission around HII regions or other stellar-related thermal processes. The bottom-right quadrant represents steep-spectrum radio emission of stellar origin, most likely produced by starburst-driven winds and/or star formation. The top-left quadrant corresponds to flat/inverted spectrum radio emission of AGN origin, arising from synchrotron self-absorbed jet bases, coronal emission, or thermal free-free emission from the accretion disk or torus. The bottom-left quadrant represents steep-spectrum AGN-related radio emission, produced by jets or AGN winds.

All RL sources lie in the left quadrants, consistent with the presence of jets. All except PG~1100+772, PG~1704+608, and PG~2251+113 lie in the top-left quadrant. These three sources, which lie in the bottom-left quadrant, are candidate restarted AGN (see Section~\ref{morph-alpha}). Of the 33 RQ sources, 17 lie in the right quadrants (indicative of stellar origin) and 16 in the left quadrants (indicative of an AGN origin), as also observed in the radio-IR correlation plot (Figure~\ref{fig4}). Of the 17, one source (PG~1426+015) exhibits a flat/inverted spectrum, most likely arising from stellar-related thermal processes, while the rest show steep-spectrum stellar-origin emission, likely driven by starburst winds. Four of the 16 AGN-origin sources exhibit a flat/inverted spectrum, likely from the bases of small-scale or frustrated jets or coronal activity. 
The remaining 12 sources have steep-spectrum emission arising from AGN jets or winds. Additionally, two sources (PG~0804+761 and PG~1416$-$129) lie on the flat-steep division and belong to the AGN dominant category.

Therefore, the radio emission in $\simeq48\%$ (16/33) of the RQ sample, for which we obtained $\alpha_R$ values, is AGN-dominated, while the remaining $\simeq52\%$ (17/33) shows contributions from both AGN and stellar-related activity. A plausible breakdown of the physical processes at play in these sources is outlined here: $\simeq75\%$ (12/16) of the AGN-dominated sources host jets or AGN winds, and $\simeq25\%$ (4/16) are likely powered by either jet/wind bases, coronal activity, or AGN thermal processes. $\simeq94\%$ (16/17) of the star-formation-dominated sources host starburst-driven winds, while one source (1/17 $\simeq6\%$) is likely driven by stellar thermal processes, with AGN activities potentially co-existing \citep[also see][]{Silpa2020}.

\subsection{Potential causes of the RL-RQ divide}
In the 685 MHz luminosity versus black hole mass planes (left panel of Figure~\ref{fig7}), RL and RQ sources appear to form a continuous distribution, with RL sources occupying the higher end and RQ sources the lower end, suggesting a possible unification of these classes in terms of black hole activity \citep[e.g.,][]{Franceschini1998, Laor2000, Lacy01}. Based on the partial correlation test (Section~4.4), when the influence of black hole mass is taken into account, the positive correlation observed between 685 MHz luminosity and Eddington ratio strengthens for both RQ and RL quasars separately, indicating that the Eddington ratio, which serves as a proxy for accretion rate, plays a dominant role in driving the RL$-$RQ dichotomy \citep[e.g.,][]{Lacy01}. 

It is worth considering the possibility that the RL-RQ transition in quasars may be linked to changes in the accretion state of the disk. In a previous study of the changing-look quasar PG~0007+106, a.k.a. III~Zw~2 \citep{Silpa2021}, it was suggested that shifts in accretion mode could lead to variations in jet activity, potentially turning jets on or off \citep[e.g.,][]{Baghel23, Baghel24}. If this analogy holds, RQ AGN might correspond to the soft state, characterized by suppressed jets and possibly strong winds, while RL AGN could be in the hard state, where jet launching is more efficient and powerful outflows are produced \citep{Fender2004}. The correlation we observe between 685 MHz radio luminosities and Eddington ratios, especially after controlling for black hole masses, further suggests that accretion state transitions could play a role in driving the RL$–$RQ dichotomy \citep[e.g.,][]{Kording06, Sikora07, Silpa2022}. This would also be consistent with the different relations observed between jet power and Eddington ratios for the RQ and RL sources seen in Figure~\ref{fig8}. Additionally, the `magnetic flux paradigm' that suggests the accumulation of large-scale magnetic flux near the black hole in magnetically arrested disks (MADs) can lead to the launching of powerful relativistic jets \citep{Tchekhovskoy2015, Sikora2016, Velzen2013}, could hold true.

Using a sample of 208 radio AGN, of which 141 were RQ and 67 RL, derived from multi-wavelength surveys of GOODS-N, GOODS-S, and COSMOS/UltraVISTA, \citet{Wang2024} have reported that the `fundamental plane' differs significantly between RQ and RL AGN, indicating a strong dependence on radio-loudness. Our findings are consistent with \citet{Wang2024}. In our work, the RL and RQ PG sources follow distinct FP relations with significantly different coefficients, reinforcing the dichotomy. The radio emission in RQ sources may have contributions from both star-formation-related processes and AGN-driven activity, potentially affecting their position on the FP.

The observed radio-X-ray correlation in this work, also reported by \citet{Brinkman2000} using ROSAT and FIRST observations for a sample of 843 AGN, suggests that both radio and X-ray emissions are likely driven by the same underlying physical mechanisms, primarily involving the central black hole and the accretion disk. The anti-correlation found between the spectral index and the Eddington ratio indicates that sources with higher Eddington ratios have a steeper slope, while those with lower Eddington ratios exhibit a flatter slope \citep[e.g.,][]{Laor19}. This suggests the dominance of optically thin synchrotron emission in the former, which can be explained by the presence of strong radio outflows \citep{Ghosh2025}. In sources with lower Eddington ratios, either coronal emission is dominant \citep[e.g.,][]{Behar2018} or the synchrotron self-absorbed bases of jets, which require very long baseline interferometry (VLBI) to resolve them. 

\section{Conclusions}
Results from uGMRT 685~MHz observations of 87 PG quasars are presented in this paper. Low-frequency radio emission is detected in all sources except for the RQ quasars, PG~0043+039, PG~1121+422, and PG~1552+085. Below are our primary findings. 

\begin{enumerate}
\item Morphological signs of jet disruption, misalignment, and steep spectrum cores observed at 685~MHz in some RL quasars point to episodic or restarted jet activity. The dichotomy observed between RL and RQ PG quasars with the VLA at 5~GHz persists at 685~MHz observations with the uGMRT. The detection of additional diffuse radio emission, including potential relics from previous episodes of AGN activity, often missed in higher-frequency observations due to sensitivity limitations, does not significantly alter the RL$-$RQ division. 

\item We find an empirical division in the 685~MHz luminosities: sources with $\log_{10}(L_{685\,\mathrm{MHz}}/\mathrm{W\,Hz^{-1}}) < 23.5$ can be classified as RQ, while those with $\log_{10}(L_{685\,\mathrm{MHz}}/\mathrm{W\,Hz^{-1}}) > 25.5$ can be classified as RL. Sources with luminosities, $23.5 \leq \log_{10}(L_{685\,\mathrm{MHz}}/\mathrm{W\,Hz^{-1}}) \leq 25.5$, are intermediate. The radio sizes display the RQ$-$RL divide as well with RQ sources typically having sizes $\lesssim30$~kpc, with only 16 ($\sim22$\%) RQ sources having sizes between 30 and 100 kpc where there is an overlap with RL quasar sizes. One RQ quasar with a size of $\sim98$~kpc, PG~1216+069, is clearly a misclassified RL quasar with an appropriately large 685~MHz luminosity and an inverted core spectral index. 

\item We did not find a difference in the minimum-energy magnetic field ($B_{\rm min}$) strengths of the radio cores of RL and RQ quasars, with the exception of the RL quasar PG~1226+023 or 3C~273. However, different assumptions of source volume and volume filling factors may be valid for the two AGN classes. High resolution X-ray observations that can resolve the X-ray core and a radio-to-X-ray flux density comparison are needed to obtain independent magnetic field strengths and thereby test the `magnetic flux paradigm' for RL and RQ AGN.

\item The distribution of PG quasars in the (685 MHz $-$ 5 GHz) $\alpha_R$–$q_{\rm IR}$ plane reveals four distinct quadrants. RL quasars occupy the AGN-dominated quadrants consistent with jet activity, while RQ sources are split nearly evenly between AGN-related and star-formation-related quadrants. This is also consistent with the findings in the 685~MHz radio$–$IR correlation. The AGN-related emission in RQ sources may arise in weak jets or coronal emission. 

\item The RL and RQ PG quasars follow distinct `fundamental plane' relations constructed using 685~MHz radio luminosity, with significantly different coefficients for the two populations. This supports the idea that the coupling between radio emission, black hole mass, and accretion differs between the two populations. This is also observed in the relation between jet power and Eddington ratios of the RQ and RL classes. The deviation of RQ sources from the RL-defined plane might reflect additional radio contribution from star formation, or differences in the accretion disk states.

\item Overall, our results suggest that accretion rates and supermassive black hole masses are likely to be influential in the RL-RQ divide. Additionally, magnetic field strengths at the black holes can be important and need to be estimated. 
\end{enumerate}

\begin{table*}
\begin{flushleft}
\setlength\extrarowheight{-0.02pt}
{\tiny
\caption{PG sample properties and uGMRT observation details for 65 sources. The peak flux density corresponds to the compact core component and is therefore equivalent to the core flux density.}
\label{Table1}
{\begin{tabular}{ccccccccccc}
\hline
Quasar & Redshift & Host galaxy{$^{*}$} & radio-loudness$^{**}$  & Classification & uGMRT         & 685 MHz peak  & 685 MHz total & $\it{rms}$ noise   & Angular & Linear\\
       &          & morphology         & (R)              &   by Radio      &  Phase cal     &  flux density &   flux density  &($\mu$Jy~beam$^{-1}$)  & size & size    \\ 
       &          &  (from NED)        &                &                 & IAU name       &  S$_\mathrm{685}^{peak}$ &  S$_\mathrm{685}$ &  &\\
       &          &                    &                &                 &          &  (mJy~beam$^{-1}$)                 &  (mJy)                     &  & ($\arcsec$) & (kpc)\\
       
\hline
PG 0003+158    & 0.45090 &   -    &     180      & RL   &  0022+002   & 102.2 & $1192.6\pm10.3$ & 864.8 &   47.3 &   264.2  \\
{\bf PG 0003+199}    &    0.02578     &  S0/a         &    0.27  & RQ      &  2330+110    & 11     & 12$\pm$1       & 31.7   &  5.2 & 2.6  \\
{\bf PG 0007+106}    &    0.08934     &  -            &    200   & RL      &  2330+110    & 56     & 95$\pm$6       &  33.5  & 84.4  & 135.3 \\
PG 0026+129    & 0.14520 &   -          &     1.1      & RQ & 0022+002  &  7.0 & $12.0\pm0.7$  & 67.0   &  11.4 &   27.9  \\
{\bf PG 0049+171}    &    0.06400     &  -            &    0.32  & RQ      &  0119+321    & 1.8    & 2.0$\pm$0.2    & 64.2   &  4.6 & 5.4     \\
{\bf PG 0050+124}    &    0.05890     &  Sa;Sbrst     &    0.33  & RQ      &  0059+001    & 9.5    & 11.0$\pm$1.0   & 34.5  &  4.3 & 4.7   \\
PG 0052+251    & 0.15445 &   Sb        &     0.24     & RQ &  0119+321  &  2.3 &  $2.9\pm0.2$ & 53.0 &  10.4 &  26.8  \\
PG 0157+001    & 0.16311 &  Bulge/disc &     2.1      & RQ & 0059+001   &  39.2 & $42.5\pm3.9$  & 27.3 & 6.4 &  17.1  \\
PG 0804+761    & 0.10000 &  E3        &     0.6      & RQ &  0410+769   &  2.4 &  $3.1\pm0.2$ & 25.5 & 27.7  &   49.0  \\
PG 0838+770    & 0.13230 &  S         &   $\leq$0.11 & RQ & 0410+769   &  0.55 &  $0.80\pm0.06$ & 19.2 & 16.5 & 37.4   \\
PG 0844+349    & 0.06400 &  Sp(d)      &     0.027    & RQ &  0741+312  &  0.24 & $1.70\pm0.04$  & 36.2 & 17.7 &    21.0    \\
PG 0921+525    & 0.03529 &   -         &    1.5       & RQ & 0834+555    &  8.2 & $13.9\pm0.8$  & 30.1 & 17.9 &     12.1   \\
{\bf PG 0923+129}    &    0.02915     &  S0?          &    2.1   & RQ      &  0842+185    & 12     & 14$\pm$1       & 27.8   &  10.1 & 5.7   \\
PG 0923+201    & 0.19270 &  E1         &    0.14      & RQ & 0842+185    &  0.35 & $0.79\pm0.04$ & 22.9 & 17.0 &  52.5   \\
{\bf PG 0934+013}    &    0.05034     &  SBab         &    0.38  & RQ      &  0943-083    & 1.2    & 2.1$\pm$0.1    & 23.3   &  8.5 & 8.0   \\
PG 0947+396    & 0.20554 &  S          &    0.25      & RQ   & 1035+564    & 0.98  & $1.3\pm0.1$ & 19.8 & 9.1 &    29.4    \\
PG 0953+414    & 0.23410 &  -          &    0.44      & RQ & 0834+555   &  0.66 & $0.69\pm0.07$  & 20.0 & 5.0 &     18.0    \\
PG 1001+054    & 0.16012 &  -          &    0.50      & RQ & 0943-083    & 0.66  & $2.3\pm0.1$ & 73.3 & 7.5  &    20.0   \\
PG 1004+130    & 0.24074 &  E2         &    230       & RL & 1120+143  &  24.9   & $2106.9\pm2.5$ & 61.3 & 143.5 &    526.0    \\
{\bf PG 1011-040}    &    0.05831     &  SB(r)b: pec  &    0.097 & RQ      &  0943-083    & 1.2    & 1.6$\pm$0.1    & 22.7   &  8.7 & 9.4    \\
PG 1012+008    & 0.18674 &  E          &    0.50      & RQ & 0943-083   &  5.1 & $5.4\pm0.5$  & 21.6 & 23.3  &     70.1   \\
PG 1022+519    & 0.04459 &  SB0/a      &    0.23      & RQ  & 0834+555    &  1.3 & $1.6\pm0.1$  & 19.2 & 5.9 &    5.0    \\
PG 1048+342    & 0.16707 &  E          &  $\leq$0.1   & RQ & 1227+365   &  0.33 & $0.25\pm0.04$ & 27.9 & 3.7 &     10.2$^a$    \\
PG 1048$-$090  & 0.34530 &   -       &    380       & RL & 1130-148    & 38.7  &  $3692.7\pm3.9$  & 220.8 & 91.9 &    434.6   \\
PG 1049$-$005  & 0.35895 &   -         &   0.25       & RQ &  1130-148  &  1.5 & $1.8\pm0.2$  & 72.1 &  13.3 &    64.5   \\
PG 1100+772 & 0.31150    &      &    320     &  RL  &  1313+675   &  659.4  & $4325.7\pm65.9$ & 124.3 & 32.3 &     142.5     \\
PG 1103-006 & 0.42374   &   -   &   270     & RL   & 1130-148    & 183.3   & $1720.9\pm18.3$ & 77.4 & 57.4 &     309.0     \\
PG 1114+445      &  0.14373   &   S          &    0.13    &  RQ  &  1227+365   &  0.90  & $1.70\pm0.09$ & 24.3 & 11.8 &     28.6     \\
PG 1115+407      &  0.15481   &   -          &     0.17   &  RQ  &  1227+365   &  0.98 &  $2.2\pm0.1$ & 25.5 & 11.2 &     28.8    \\
PG 1116+215      &  0.17564    &    E2       &    0.72    &  RQ  &  1120+143  &  8.1  & $8.2\pm0.8$ & 36.9 & 4.5 &    12.9   \\
{\bf PG 1119+120}    &    0.05020     &  SABa         &    0.15  & RQ      &  1120+143    & 3.2    & 5.4$\pm$0.3    & 29.1   &  8.1 & 7.7 \\
PG 1126-041      &  0.05980    &  S           &    0.17   & RQ  &  1130-148   &  1.9  & $2.4\pm0.2$ & 26.2 & 5.3 &    5.9   \\
PG 1149-110      &  0.04870    &  -           &     0.88  &  RQ  &  1130-148  3 & 6.8   & $7.7\pm0.7$ & 30.1 & 9.1 &    4.4   \\
PG 1151+117      & 0.17565    &   -          &    $\leq$0.07   &  RQ  &  1120+143   & 0.45  & $0.51\pm0.06$ & 44.4 & 5.4 &    15.6$^a$  \\
PG 1202+281    &  0.16501       &  E1         &    0.19    &  RQ  & 3C286    &  2.0  & $2.1\pm0.2$ & 33.5 & 10.1 &    27.5   \\
{\bf PG 1211+143}    &    0.08090     &  -            &    0.13  & RQ      &  1254+116    & 4.7    & 6.5$\pm$0.5    & 155.3   &  5.3 & 7.8  \\
PG 1216+069    &  0.33130       &   E          &    1.7          &  RQ  &  1150-003   &  1.94  & $4.7\pm0.2$ & 33.1 & 21.3 &    97.8      \\
PG 1226+023 &   0.15834      &   -          &  1100            &  RL  & 1150-003   & 35878   & $70105.0\pm3587.8$ & 3214 &  36.6 &     96.2 \\
{\bf PG 1229+204}    &    0.06301     &  SB0          &    0.11  & RQ      &  1254+116    & 1.9    & 2.5$\pm$0.2    & 28.71   & 5.2  & 6.0  \\
{\bf PG 1244+026}    &    0.04818     &  E/S0         &    0.53  & RQ      &  1254+116    & 2.9    & 3.1$\pm$0.3    & 26.48   & 5.7  & 5.1  \\
PG 1259+593    &  0.47619      &   -      &    $\leq$0.1      & RQ   & 1400+621    &  0.16  & $0.19\pm0.03$ & 27.1 & 5.6 &    32.3$^a$  \\
PG 1302-102 &  0.27840  &  E4(?)    &   190    &  RL  &  1351-148  &  410.5  & $461.7\pm41.0$ & 96.7 & 19.4 &    79.2   \\
PG 1307+085    & 0.15384     &    E1?         &    0.1        &  RQ  & 1330+251    &  0.87  & $1.20\pm0.09$ & 36.8 & 8.4 &    21.5    \\
PG 1309+355  &  0.18295       &   Sab          &   180           & RL   & 3C286  &  62.8  & $86.4\pm6.3$ & 47.8 & 16.8 &     50.0   \\
{\bf PG 1310$-$108}  &    0.03427     &  -            &    0.095 & RQ      &  1248-199    & 0.74   & 0.87$\pm$0.08  & 30.64   & 4.9  & 3.2  \\
PG 1322+659  &  0.16800       &   S      &   0.098           & RQ   & 1400+621   & 0.63  & $0.71\pm0.07$ & 29.4 & 18.4 &     50.8    \\
{\bf PG 1341+258}    &    0.08656     &  -            &    0.12  & RQ      &  1330+251    & 0.30   & 0.36$\pm$0.04  & 23.9   &  4.3 & 6.7  \\
{\bf PG 1351+236}    &    0.05500     &  -            &    0.26  & RQ      &  1330+251    & 2.1    & 3.0$\pm$0.2    & 23.4   &  4.4  & 4.6  \\
PG 1351+640   &  0.08820       &  -      &  0.26    &  RQ  & 1400+621    &  34.2  & $35.6\pm3.4$ & 33.2 &  15.7 &    24.9   \\
PG 1352+183   &  0.15147       &  E        &   0.11      & RQ   &  3C286   &  0.26  & $0.20\pm0.04$ & 31.0 & 5.2 &     13.3$^a$\\
PG 1354+213   &  0.30046       &   E          &  $\leq$0.08      &  RQ  & 3C286    &  0.35  & $0.30\pm0.05$ & 34.0 &  5.2  &  22.6$^a$    \\
PG 1402+261   &  0.16400       &   SBb          &    0.23          &  RQ  & 3C286   & 1.63   & $1.7\pm0.2$ & 33.3 &  9.0  & 24.5$^a$    \\
{\bf PG 1404+226}    &    0.09800     &  -            &    0.47  & RQ      &  1330+251    & 2.8    & 2.9$\pm$0.3    & 20.2   &  3.9 & 6.8  \\
PG 1411+442   &  0.08960       &  E4           &   0.13           & RQ   & 3C286   &  2.50  & $2.7\pm0.3$ & 31.8 & 8.2 &     13.2   \\
PG 1415+451  &  0.11371       &   -          &    0.17          &  RQ  & 3C286    &   1.57 & $1.8\pm0.2$ & 32.5 & 6.3 &     12.6    \\
PG 1416-129  &   0.1289      &   -          &   1.1           & RQ   &  1351-148  & 2.2  & $2.7\pm0.2$ & 34.3 &  14.3 &    31.5   \\
PG 1425+267 &   0.36361      &   -          & 540       & RL   & 3C286   &  36.8  & $675.4\pm3.7$ & 36.8 & 279.9 &     1369.4  \\
{\bf PG 1426+015}    &    0.08657     &  E/S0         &    0.28  & RQ      &  1445+099    & 2.0    & 2.6$\pm$0.2    & 24.3   &  4.9 & 7.6  \\
PG 1427+480  &  0.22063       &   E          &   $\leq$0.16   &  RQ  &  1400+621   &  0.37  & $0.46\pm0.05$ & 30.1 & 13.9 &    47.8    \\
PG 1435-067  & 0.12600        &   -          &  0.069        & RQ   &  1445+099 1 & 0.47   & $1.0\pm0.7$ & 56.9 &  11.0 &    23.9    \\
PG 1440+356  &  0.07906       &  S           &   0.37           &  RQ  &  3C286   &  7.31  & $7.7\pm0.7$ & 54.1 &  14.0 &    20.1     \\
PG 1444+407  &  0.26791       &   E1(?)          &   $\leq$0.08           &  RQ  &  3C286  &  2.79  & $2.8\pm0.3$ & 34.5 & 6.4 &    25.4$^a$    \\
{\bf PG 1448+273}    &    0.06500     &  -            &    0.23  & RQ      &  3C 286      & 5.3    & 5.3$\pm$0.5    & 27.0   & 5.2  & 6.2  \\
{\bf PG 1501+106}    &    0.03642     &  E            &    0.36  & RQ      &  1445+099    & 3.7    & 5.1$\pm$0.4    & 204.3   & 5.4  & 3.7  \\
PG 1512+370 &  0.37053  &   -          &    190          & RL   & 1609+266   &  37.5  & $1627.3\pm3.8$ & 77.6 & 79.2 &    329.2   \\
PG 1519+226     & 0.13609 &   -          &   0.9           & RQ   & 1609+266    &  0.30  & $0.32\pm0.05$ & 36.9 & 6.3 &     14.6    \\
PG 1534+580     & 0.03023 &   E1?, S0    &    0.7          &  RQ  &  1400+621   &  7.7  & $7.9\pm0.8$ & 28.5 &   21.6 &    12.5   \\   
PG 1535+547     & 0.03893 &             &     0.14         &  RQ  &  1400+621   &  1.6  & $1.6\pm0.2$ & 33.4 &  18.0 &    13.3    \\
PG 1543+489     & 0.39982 &   S          &      0.63        &  RQ  &  1400+621   &  4.3  & $4.4\pm0.4$ & 39.6 &  14.8  &    76.7   \\
PG 1545+210 &  0.26430  &    -         &    420          &  RL  &  1609+266   &  40.5  & $1186.1\pm4.0$ & 145.0 & 81.8 &    321.2   \\
PG 1612+261     & 0.13095 &  -           &      2.8        &  RQ  &  1609+266   & 27.8   &  $30.3\pm2.8$ & 33.7 & 15.9  &    35.7   \\
PG 1613+658     & 0.12109 &  E2          &     0.94         & RQ   & 1400+621    & 8.2  & $8.5\pm0.8$ & 87.5 &  34.9  &    73.2 \\
PG 1617+175     & 0.11460 &  -           &     0.39         & RQ   &  1609+266   &   8.1 & $8.2\pm0.8$ & 34.7 &  2.7  &    5.5$^a$  \\
PG 1626+554     & 0.13360 &   -          &     0.11         &  RQ  &  1400+621   &  0.41  & $0.44\pm0.05$ & 28.5 & 11.3  &    25.8$^a$ \\
PG 1700+518    & 0.28900 &   Sbrst    &  2.4         & RQ & 1634+627    & 37.2  & $44.0\pm3.7$  & 40.2 &  18.7  &    78.2  \\
PG 1704+608    & 0.37152 &            &   620        & RL & 1634+627   & 63.8  &  $5844.9\pm6.4$ & 137.7 & 73.0 &    362.4   \\
{\bf PG 2130+099}    &    0.06298     &  (R)Sa        &    0.32  & RQ      &  2148+069    & 7.5    & 8.7$\pm$0.7    & 24.44   & 4.2  & 4.9   \\
PG 2112+059    & 0.45900 &  -          &   0.32       & RQ &  2130+050  & 5.1  & $5.2\pm0.5$  & 56.3 & 5.3  &     30.1  \\
{\bf PG 2209+184}    &    0.07000     &  -            &    54    & RL      &  2148+069    & 93     & 97$\pm$9       & 95.5   & 5.5  & 7.0   \\
{\bf PG 2214+139}    &    0.06576     &  -            &    0.052 & RQ      &  2148+069    & 0.94   & 0.93$\pm$0.10  & 25.5   &  0.8 & 1.0$^{a}$    \\
PG 2233+134    & 0.32571 &  E          &   0.28       & RQ & 2130+050   & 1.2  & $1.4\pm0.1$ & 22.8 &   18.5  &    84.2  \\
PG 2251+113    & 0.32550 &   -        &   370        & RL &  2212+018   & 1228  &  $2548.3\pm122.8$ & 246.3 &  19.3 &    87.6    \\
{\bf PG 2304+042}    &    0.04200     &  -            &    0.25  & RQ      &  2330+110    & 1.0   & 1.1$\pm$0.1     & 32.1   & 5.8  & 4.6    \\  
PG 2308+098    & 0.43330 &   -          &    190       & RL& 2212+018   &  51.1 &  $1278.6\pm5.1$ & 93.4 &  117.0 &    638.0  \\
\hline
\end{tabular}}

$^{*}$ E=elliptical; S0=lenticular; S=spiral; Sbrst=starburst; pec=peculiar; ?=uncertainty; a,b represent the tightness of the spiral arms; A,B represent bars; (r) represents inner ring. 
$^{**}$ Radio-loudness parameter taken from \citep{Kellermann94}.
$a$: unresolved sources.  
Undetected sources: PG 0043+039, PG 1121+422, PG 1552+085
}
\end{flushleft}
\end{table*}

\begin{table*}
\begin{center}
{\scriptsize
\caption{Results on the correlation studies of 685 MHz luminosity with other physical properties. The values represent the correlation coefficients with respective p-values in brackets. A correlation is considered significant at the $\sim$99.7\% confidence level ($p < 0.001$, $\simeq 3\sigma$).
}
\label{Table2}
{\begin{tabular}{ccccccccc}
\hline
\multirow{3}{*}{Parameters} & \multicolumn{3}{c}{Spearmann Rank}  & \multicolumn{3}{c}{Kendall-tau} \\
    & Full sample   &   RQ &  RL  & Full Sample & RQ  &  RL  \\     
\hline
Spectral index - Edd ratio &  -0.4 (0.0006) & -0.4 (0.008)  & -0.5 (0.05)  & -0.3 (0.0009) & -0.3 (0.008) &  -0.4 (0.06)  \\
&   &   &   & &  &    \\
Radio-loudness - Edd ratio & 0.09 (0.4)  & 0.2 (0.2) &  -0.2 (0.5) & 0.05 (0.6) & 0.1 (0.1) & -0.1 (0.4)  \\
685 MHz total luminosity - Edd ratio & 0.1 (0.3) & 0.4 (0.0005)  & 0.4 (0.1) & 0.1 (0.2)  &  0.3 (0.0005) & 0.3 (0.1) \\
685 MHz total luminosity - $M_{\mathrm{BH}}$ &  0.6 ($1.8\times10^{-10}$) &  0.4 (0.001) &  0.4 (0.1) & 0.4 ($1.9\times 10^{-9}$) & 0.3 (0.002) & 0.2 (0.2) \\
685 MHz total luminosity/Edd. ratio - $M_{\mathrm{BH}}$ &  0.9 ($2.2\times10^{-25}$) &  0.8 ($3.9\times10^{-15}$) & 0.6 (0.008)  & 0.7 ($2.3\times10^{-19}$) & 0.6 ($2.7\times10^{-12}$) & 0.5 (0.003)\\
685 MHz total luminosity - $L_{\mathrm{X}}$ & 0.6 ($9.2\times10^{-10}$) & 0.5 ($1.2\times10^{-5}$)  &  0.6 (0.02) & 0.4 ($3.9\times10^{-9}$)  &  0.4 ($1.8\times10^{-5}$) & 0.4 (0.03) \\
685 MHz total luminosity - $L_{\mathrm{IR}}$ & 0.6 ($4.5\times10^{-9}$) & 0.6 ($8.9\times10^{-8}$)  &  0.5 (0.1) & 0.4 ($6.4\times10^{-8}$) &  0.5 ($3.7\times10^{-7}$) &  0.2 (0.2)  \\
&   &   &   & &  &    \\
Jet Power - $M_{\mathrm{BH}}$ &  0.6 ($1.8\times10^{-10}$) & 0.4 (0.001) & 0.4 (0.1)  & 0.4 ($1.9\times10^{-9}$) &  0.3 (0.002) & 0.2 (0.2) \\
Jet Power - Eddington Ratio &  0.1 (0.3) & 0.4 (0.0004) &  0.4 (0.1) & 0.1 (0.2) & 0.3 (0.0005)   &  0.3 (0.1) \\
\hline
\end{tabular}}
}
\end{center}
\end{table*}

\begin{table*}
\begin{center}
\setlength\extrarowheight{-0.02pt}
{\tiny
\caption{\small {Summary of physical properties of PG quasars. The spectral index ($\alpha_R$) has been calculated using the peak flux densities at 685 MHz (this work) and 5 GHz \citep{Kellermann94}.}}
\label{Table3}
{\begin{tabular}{ccccccccccc}
\hline
Quasar & L$_\mathrm{IR,host}${$^{a}$}  & log L$_\mathrm{X}^{b}$ & log ${M_{\mathrm{BH}}}${$^{c}$} & Edd. {$^{h}$} & L$_{5000}$  & $\alpha_R $ &  L$_{1400}$ & L$_{151}$ & $\overline Q$ & $q_{\mathrm{IR}}$\\  
&  &    &  & ratio & (total) & & ($\times 10^{22}$ & ($\times 10^{22}$ & ($\times10^{41}$ &\\
&   (10$^\mathrm{11}$ L$_{\sun}$)  & (erg s$^\mathrm{-1}$) & (M$_{\sun}$) & $\lambda$ & ($10^{28}$ ergs s$^{-1}$ Hz$^{-1}$) & & W Hz$^{-1}$) &   W Hz$^{-1}$ sr$^{-1}$) & ergs s$^{-1}$) & \\
\hline
PG 0003+158    &   0.50      &    45.55      &     9.45     &  0.08 &    $1.4\times10^5$ & 0.12 &  3.0$\times10^4$ & $1.2\times10^4$ & $1.2\times10^4$ & -1.8  \\  
{\bf PG 0003+199}     &  0.02     &  43.80    &  7.52    &   0.21    &  5.5     &  -0.65     &  1.1    &  0.4     &  1.8     &  1.3\\
{\bf PG 0007+106}     &   0.48    &  44.49    &  8.87    &   0.05    &  $5.4\times10^3$    &  0.50      &  96.8    &   36.6    & 89.8  & 0.7 \\ 
PG 0026+129    &    -         &  44.70     & 8.12        & 0.28 &   236.8 & -1.8 &  33.8 &  12.8 & 36.4 & - \\  
{\bf PG 0049+171}     &   0.03    &  44.05    &  8.45    &   0.02    &  5.9     &  -0.50     &  1.1   &  0.4     &   2.0    & 1.4\\
{\bf PG 0050+124}     &  2.93     &  44.18    &  7.57    &   2.26    &  21.5     &  -0.84     &  5.5    &   2.1    &   7.7    & 2.7\\
PG 0052+251    &      0.71    &     44.93  &  8.99       & 0.06 &   38.9 & -0.9 &  9.2 &  3.5 & 12.0 & 1.9 \\
PG 0157+001    &   17.49    &   44.23    &    8.31       & 0.87 & 468.8 & -0.98  & 151.0 & 57.1  & 131.5  & 2.1\\   
PG 0804+761    &      0.13    &  44.78     &     8.55    & 0.20 &   54.0 & -0.5 & 4.3 &  1.6 & 6.2 & 1.5 \\   
PG 0838+770    &      1.30    &   44.16    &     8.29    & 0.16 &  $*$ & - & 2.0 & 0.8  & 3.2 &  2.8 \\   
PG 0844+349    &       0.14   &   43.99    &     8.03    & 0.10 &  - & - & 1.0 & 0.4  & 1.8 & 2.2 \\   
PG 0921+525    &      0.02    &    43.52   &     7.45    & 0.16 &   11.3 & -0.7 & 2.5 & 1.5  & 3.9 & 1.0 \\  
{\bf PG 0923+129}     &  0.22     &  43.69    &  7.52    &   0.12    &  20.6     &   -        &  1.7    &    0.6      &   2.9    & 2.1 \\
PG 0923+201    &       0.31   &   43.69    &     9.33    & 0.03 &   20.7 & - & 4.0 &  1.5 & 5.8 & 1.9 \\   
{\bf PG 0934+013}     &  0.25     &  43.73    &  7.15    &   0.24    &  3.2     &   -0.90    &  0.8   &   0.3     &   1.4    & 2.5 \\
PG 0947+396    &     1.94     &   44.72    &      8.81   & 0.17 &   29.1 & -0.7 & 7.4 &  2.8 & 9.9 & 2.4 \\   
PG 0953+414    &     -        &   44.98    &     8.74    & 0.30 &   229.4 & - & 5.1 & 1.9  & 7.1 & -\\ 
PG 1001+054    &   0.09       &   42.74    &     7.87    & 0.63 &    46.4 &  - & 8.2 &  3.1 & 10.8 & 1.1\\   
PG 1004+130    &      1.42    &   44.48    &     9.43    & 0.06 &  $5.6\times10^4$ & 0.02 & $1.6\times10^4$ &  $6.2\times10^3$ & $7.3\times10^3$ & -1.0 \\   
{\bf PG 1011–040}     &  0.28     &  42.60    &  7.43    &   0.27    &  2.3     &   -0.59    &  0.8   &  0.3    &   1.5          & 2.6 \\
PG 1012+008    &      0.90    &   44.06    &    8.39     & 0.26 &   78.2  & -0.97 & 25.6 & 9.7  & 28.7 & 1.6 \\ 
PG 1022+519    &      0.20    &   43.66    &     7.25    & 0.14 &   1.7   & - & 0.5 &  0.2 & 0.9 & 2.6\\   
PG 1048+342    &     0.88     &    44.22   &     8.50    & 0.08 &    0.6 & - &  1.0 & 0.4  & 1.7 & 3.0 \\   
PG 1048$-$090  &     0.36     &   45.16    &   9.37      & 0.06 &  $1.7\times10^5$ & 0.2 & $5.7\times10^4$  & $2.2\times10^4$ & $2.1\times10^4$ & -2.2 \\   
PG 1049$-$005  &     0.45     &    44.78   &     9.34    & 0.22 &   132.4 & - & 30.1 &  11.4 & 33.0 & 2.5\\   
PG 1100+772 	&    3.80     &    45.24	&       9.44    &	0.07 &	$1.4 \times 10^5$  & -1.08 & $5.5\times10^4$ & $2.1\times10^4$  & $2.1\times10^4$& -1.1 \\
PG 1103-006   &              -       &       44.90	&       9.49    &	0.10 &	$1.8 \times 10^5$ & -0.4  & $4.0\times10^4 $ & $1.5\times10^4$  & $1.6\times10^4$& -  \\
PG 1114+445   &         0.12        &       43.87	&       8.72    &	0.17 &	10.3 & -0.8 & 4.8 &  1.8 & 6.9 & 1.4\\ 
PG 1115+407   &         2.22        &       44.33	&       7.80    &	0.54 &	- & - & 7.2 & 2.7  & 9.7 & 2.5\\ 
PG 1116+215   &              -        &       44.83	&       8.69    &	0.34 &	194.2 & -0.7 & 34.5 &  13.0 & 37.1 & -  \\ 
{\bf PG 1119+120}     &  0.36     &  43.48    &  7.58    &   0.32    &  5.7     &   -        &  2.0   &  0.8     &   3.2    & 2.3 \\ 
PG 1126-041   &         0.60        &       42.32	&       7.87    &	0.36 &	4.4 & - & 1.2 & 0.5  & 2.1 & 2.7\\ 
PG 1149-110   &         0.30        &       43.83	&       8.04    &	0.05 &	14.9 & -0.96 & 2.7 & 0.9  & 3.8 & 2.1 \\ 
PG 1151+117   &              -        &       44.43	&       8.68    &	0.08 &	$**$ & - & 2.1 &  1.0 & 4.1 & -\\ 
PG 1202+281    &         0.87        &       44.70	&       8.74    &	0.10 &	50.9 & -0.6 & 7.8 &  2.9 & 10.4 & 2.1 \\ 
{\bf PG 1211+143}     &  0.09     &  44.54    &  8.10    &   0.46    &  $2.4\times10^3$     &   -0.70    &  6.0    &  2.3    &  8.3     & - \\
PG 1216+069    &              -      &       45.10	&       9.36    &	0.06 &	945.6 & 0.47 & 67.9 & 25.7  & 66.3 & - \\
PG 1226+023 &         6.32          &       45.67	&       9.18    &	0.36 &	$2.1\times10^6$ & -0.2 & $2.4\times10^5$ &  $9.1\times10^4$ & $7.3\times10^4$& -1.6\\
{\bf PG 1229+204}     &  0.26     &  44.00    &  8.26    &   0.08    &   6.4    &   -0.93    &  1.4    &  0.5     &  2.4     & 2.3 \\
{\bf PG 1244+026}     &  0.16     &  43.58    &  6.62     &  1.15     &  4.6    &  -0.92     &  1.0    &  0.4     &   1.9    & 2.2 \\
PG 1259+593    &              -        &       44.53	&       9.09    &	0.30 &	14.1 & - & 5.4 & 2.1  & 7.6 & -\\ 
PG 1302-102 &         4.89        &       45.09	&       9.05    &	0.20 &	$1.3\times 10^5$ & +0.3 & $4.7\times10^3$ & $1.8\times0^3$ & $2.5\times10^3$ &  0.03\\ 
PG 1307+085    &         0.03        &       44.59	&       9.00    &	0.06 &	18.7 & - & 4.0 &  1.5 & 5.7 & 0.9\\ 
PG 1309+355 &         0.66        &       43.89	&       8.48    &	0.36 &	$4.0\times10^3$ & -0.1 & 391.9 & 148.3  & 297.8 & 0.2 \\ 
{\bf PG 1310–108}     &  0.04     &  43.30    &  7.99     &  0.04     &  0.7     &  -         &  0.1    &    0.06   &   0.4    & 2.4 \\
PG 1322+659    &         0.99        &       44.68	&       8.42    &	0.22 &	12.6 & - & 2.7 &  1.0 &  4.2 & 2.6 \\ 
{\bf PG 1341+258}     &  0.22     &  43.83    &  8.15     &  0.06     &  0.9     &   -        &  0.4    &  0.1     &   0.8    & 2.8 \\
{\bf PG 1351+236}     &  0.50     &  43.11     &  8.67     & 0.006    & 3.7      &  -         &   1.3   &   0.4    &   1.7    & 2.6 \\
PG 1351+640    &         1.59        &       43.53	&       8.97    &	0.07 &	236.6 & -0.3 & 38.4 &  14.5 & 40.7 & 1.6 \\ 
PG 1352+183    &              -        &       44.38	&       8.56    &	0.10 &	13.0 & - &  0.6 & 0.2  & 1.2 &-\\ 
PG 1354+213    &         1.07        &       44.60	&       8.77    &	0.18 &	$**$ & - & 3.6 &  1.3 & 5.3 & 2.5\\ 
PG 1402+261    &         1.63        &       44.44	&       8.08    &	0.97 &	37.2 & -& 6.1 & 2.3  & 8.4 & 2.4\\ 
{\bf PG 1404+226}     &  0.21     &  43.77     &  7.01     & 0.95     &  22.2     &   -0.58    &  3.9    &  1.5     &  5.7     & 1.7\\
PG 1411+442    &         0.06        &       42.72	&       8.20    &	0.29 &	11.3 & -0.8 & 3.0 &  1.1 & 4.6 & 1.3 \\ 
PG 1415+451    &         0.54        &       43.99	&       8.14    &	0.17 &	11.8 & - & 3.3 & 1.2  & 4.9 & 2.2\\ 
PG 1416-129    &         0.14        &       44.65	&       9.19    &	0.01 &	136.3 & -0.5 & 6.2 &  2.3 & 8.5 &  1.4 \\  
PG 1425+267 &         2.73      &       44.21	&       9.90    &	0.03 & $3.7 \times 10^4$	& 0.05 & $1.2\times10^4$ & $4.4\times10^3$  & $5.4\times10^3$ & -0.6 \\ 
{\bf PG 1426+015}     &  0.51     &   44.58    &  9.15     & 0.04      &  21.0     &   -0.38    &  2.7    &  1.0     &  4.2    & 2.3 \\
PG 1427+480  		&         1.26        &       44.30	&       8.22    &	0.37 &	2.1 & - & 3.0 & 1.1  & 4.6 & 2.6\\ 
PG 1435-067  		&         0.02        &       44.28	&       8.50    &	0.08 &	6.9 & -0.4   & 2.1 & 0.8  & 3.3 &  1.1  \\ 
PG 1440+356  		&         1.46        &       44.54	&       7.60    &	0.83 &	23.9 &  -1.2 & 6.7 & 2.5 & 9.1 & 2.4   \\ 
PG 1444+407  		&         0.66        &       44.55	&       8.44    &	0.72 &	25.0 & -&  26.3 &  9.9 & 29.3 & 1.4 \\ 
{\bf PG 1448+273}     &  0.23     &   43.86    &  7.09     & 0.95      &  9.8     &   -0.80    &  3.1    &  1.2     &  4.7     & 1.9 \\
{\bf PG 1501+106}     &  0.20     &  43.65     &  8.64     & 0.02      &  4.7     &   -1.00    &  1.0    &  0.4     &  1.7     & 2.3 \\
PG 1512+370  &         0.28        &       45.03	&       9.53    &	0.06 &	$1.0 \times 10^5$ & +0.3  & $2.9 \times 10^4$ &  $1.1\times10^4$ & $1.2\times10^4$ &  -2.0 \\ 
PG 1519+226    &	  0.3555	&	44.24	&	8.07	&	0.39	&  62.7  & - & 0.8 &  0.3 & 1.5 &  2.7  \\ 
PG 1534+580    &         0.05        &       43.54	&       8.30    &	0.02 &	4.1 & -0.7  & 1.0 & 0.4  & 1.8 & 1.7  \\ 
PG 1535+547    &         0.03        &       $<$41.76	&       7.30    &	0.17  & 1.7	& - & 0.3 &  0.1 & 0.7 &  1.9\\ 
PG 1543+489    &        18.57        &       44.40	&       8.16    &	3.59 &	364.1 & - & 89.9 &  34.0 & 84.3 & 2.3 \\ 
PG 1545+210  &              -        &       45.14	&       9.47    &	0.03 &	$1.1 \times 10^5$ & -0.1 & $1.1\times10^4$ & $4.1\times10^3$  & $5.2\times10^3$ & -  \\ 
PG 1612+261    &         1.39        &       44.53	&       8.19    &	0.31 &	195.9 & -1.4  & 71.0 & 26.9  & 68.8 & 1.3  \\ 
PG 1613+658    &         3.57        &       44.89	&       9.32    &	0.05 &	100.0 & -1.2  & 17.0 & 6.4  & 20.2 & 2.3  \\
PG 1617+175    &         0.06        &       44.15	&       8.91    &	0.03 &	32.4 & - &  14.7 & 5.6  & 17.8 & 0.6 \\
PG 1626+554    &              -        &       44.48	&       8.63    &	0.04 &	6.8  & -0.1 & 1.07 & 0.4  & 1.9 &  - \\
PG 1700+518    &      8.82    &   42.90    &    8.61     & 1.39 &   $1.3\times10^3$ & -1.5  & 481.0 & 181.9 & 348.9 &  1.3 \\   
PG 1704+608    &      6.09    &   44.64    &   9.55      & 0.17 &  $3.6 \times 10^5$ & -1.0  & $1.0\times10^5$ & $3.9\times10^4$  & $3.5\times10^4$ & -1.2  \\   
PG 2112+059    &     1.87     &   43.87    &     9.18    & 0.53 &   388.4 & -0.96 & 137.6 &  52.0 & 121.4 & 1.1   \\   
{\bf PG 2130+099}     &  0.50     &   44.35    &  8.04     &  0.31     &  18.1     &   -0.88    &  4.7    &  1.8    &  6.7    & 2.0 \\
PG 2233+134    &     1.21     &   44.42    &    8.19     & 1.44 &   108.3 & -0.7  & 19.1 & 7.2  & 22.4 & 1.8  \\   
{\bf PG 2209+184}     &  0.16     &   43.94    &  8.89     &  0.006    &   $3.1\times10^3$    &  -     &    63.5  &   24.0    &  62.5     & 0.4\\
{\bf PG 2214+139}     &  0.10     &  42.63     &  8.68     &  0.03     &   2.3    &    -      &    0.5  &   0.2       &  1.0        & 2.3 \\
PG 2251+113    &      0.51    &   44.09    &     9.15    & 0.11 & $1.2\times10^5$  & -0.7$^{***}$ &  $3.5\times10^4$ &  $1.3\times10^4$ & $1.4\times10^4$&  -1.8 \\
{\bf PG 2304+042}     &   0.001    &  43.89     &  8.68     &  0.003   &   3.1    &  0.05     &   0.3   &   0.1   &  0.6     & 0.6 \\
PG 2308+098    &       -   &    45.35   &    9.76     & 0.04 & $1.2\times 10^5$ & +0.3 & $3.0\times10^4$ &  $1.1\times10^4$ & $1.2\times10^4$ & -  \\   
\hline
\end{tabular}}
}
{\flushleft \tiny
The sources in boldface are from Paper I. The spectral index, $L_{1400}$, $L_{151}$, $\overline{Q}$, and $q_{\mathrm{IR}}$ were recalculated for these sources using the methodology described in Section 4.2.
$^{a}$ $8-1000~\mu$m host galaxy IR luminosity from \citet{Lyu17} \\
$^{b}$ $0.2-20$ keV luminosity from \citet{LaorBehar08}\\
$^{c}$ BH mass from \citet{Shangguan18}\\ 
$^{d}$ Eddington Ratio ($L_{\mathrm{bol}}$/$L_{\mathrm{Edd}}$) estimated using $M_{BH}$ and $L_{\mathrm{bol}}$ from \citet{Lyu17} \\ 
$^{e}$ 5 GHz total luminosity from \citet{Kellermann89} at $18\arcsec$ resolution \\
$*$ Flux of -0.02 mJy reported in \citet{Kellermann89} \\
$**$ Flux of 0.0 mJy reported in \citet{Kellermann89} \\
$***$ The spectral index for this source, when calculated using the 5 GHz flux from \citet{Kellermann89}, yielded an ultra-steep value of approximately $-3$. Therefore, the spectral index adopted here is taken from \citet{Baghel24}. \\
}
\end{center}
\end{table*}

\section{Acknowledgments}
We thank the referee for their helpful suggestions that have improved this paper significantly. We acknowledge the insightful discussions with Agniva Roychowdhury at later stages of this project.
We acknowledge the support of the Department of Atomic Energy, Government of India, under the project 12-R\&D-TFR-5.02-0700. 
SS acknowledges funding from ANID through Fondecyt Postdoctorado (project code 3250762), Millenium Nucleus NCN23\_002 (TITANs), and Comite Mixto ESO-Chile. LCH was supported by the National Science Foundation of China (12233001) and the China Manned Space Program (CMS-CSST-2025-A09).
We thank the staff of the Giant Metrewave Radio Telescope (GMRT) that made these observations possible.
GMRT is run by the National Centre for Radio Astrophysics of the
Tata Institute of Fundamental Research. This research has made use of the NASA/IPAC Extragalactic Database (NED), which is operated by the Jet Propulsion Laboratory, California Institute of Technology, under contract with the National Aeronautics and Space Administration.

\begin{figure*}
\centering
\includegraphics[height=7.1cm,trim=25 10 0 10]{ 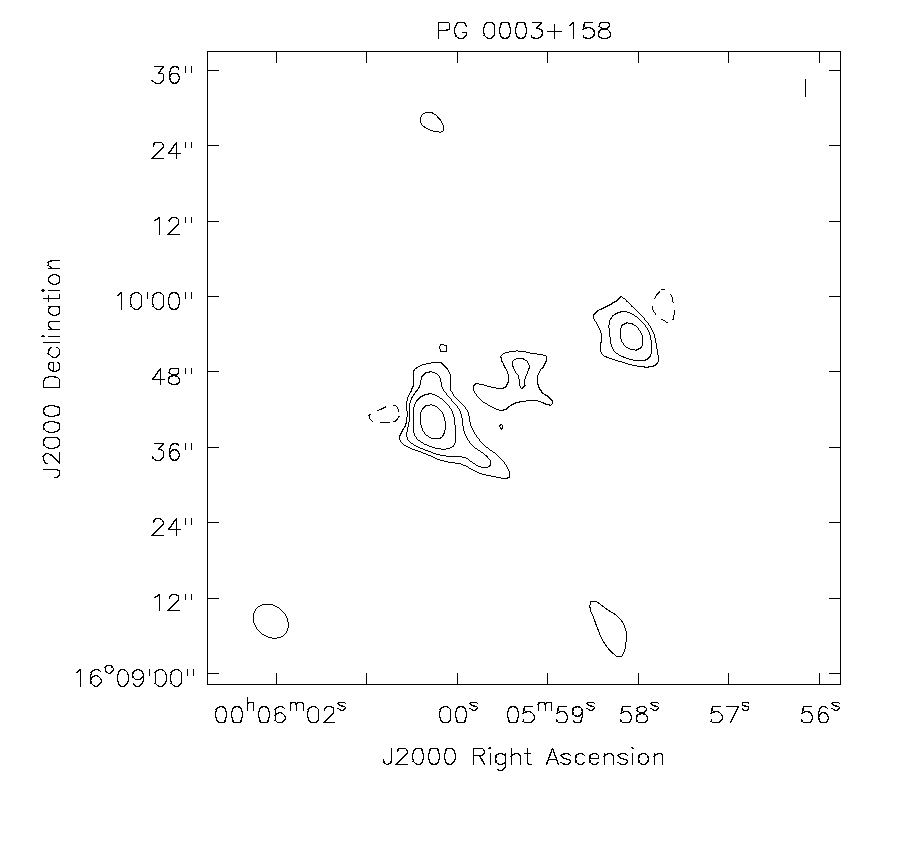}
\includegraphics[height=7.1cm,trim=25 10 0 10]{ 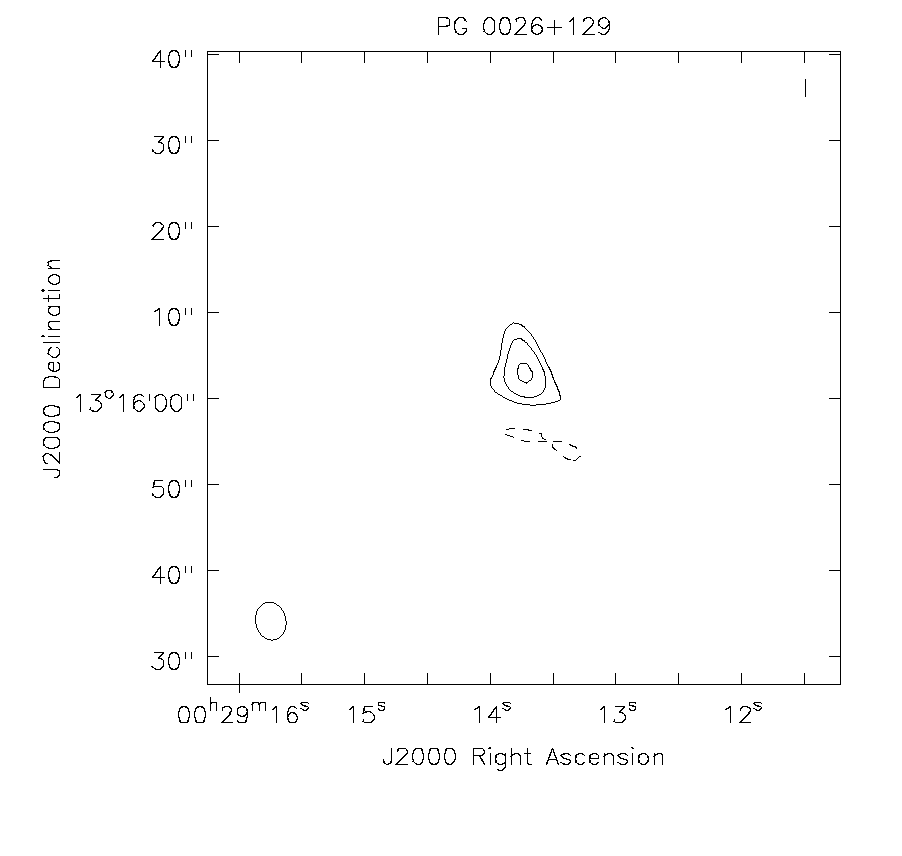}\\
\includegraphics[height=7.1cm,trim=25 10 0 10]{ 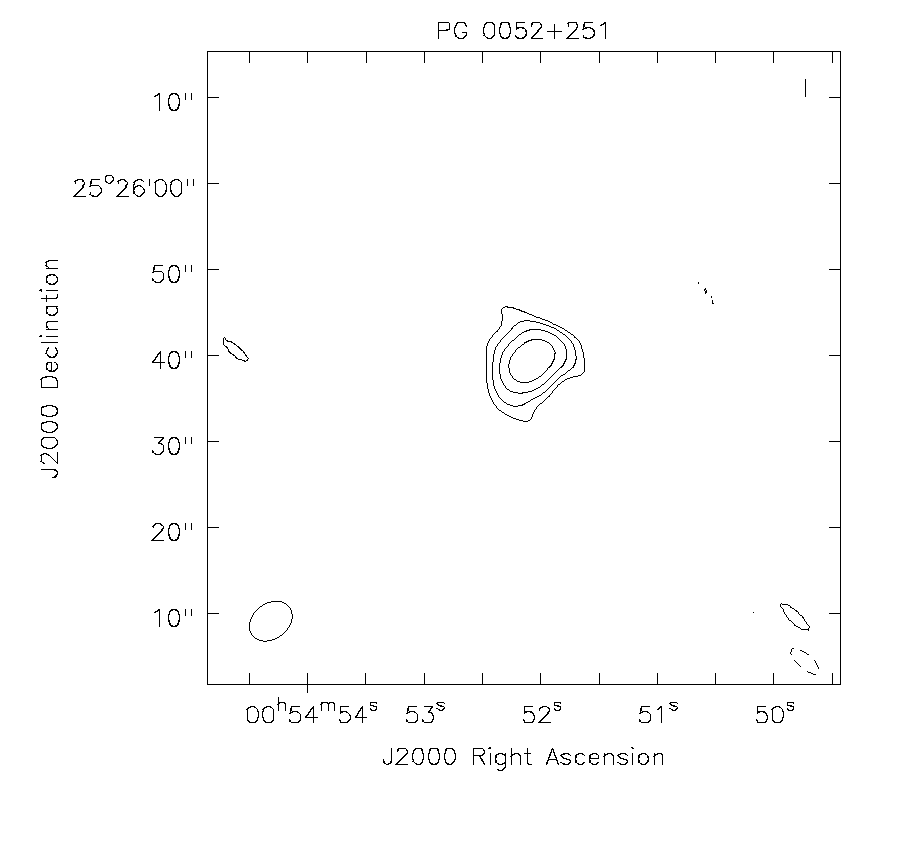}
\includegraphics[height=7.1cm,trim=25 10 0 10]{ 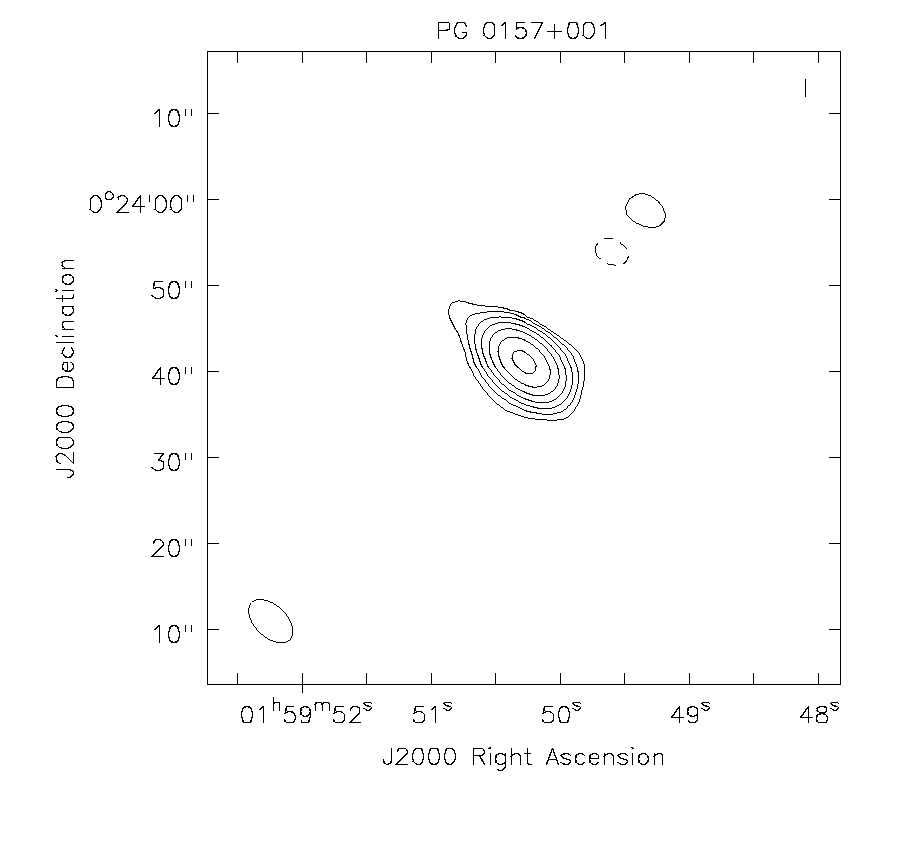}\\
\includegraphics[height=7.1cm,trim=25 10 0 10]{ 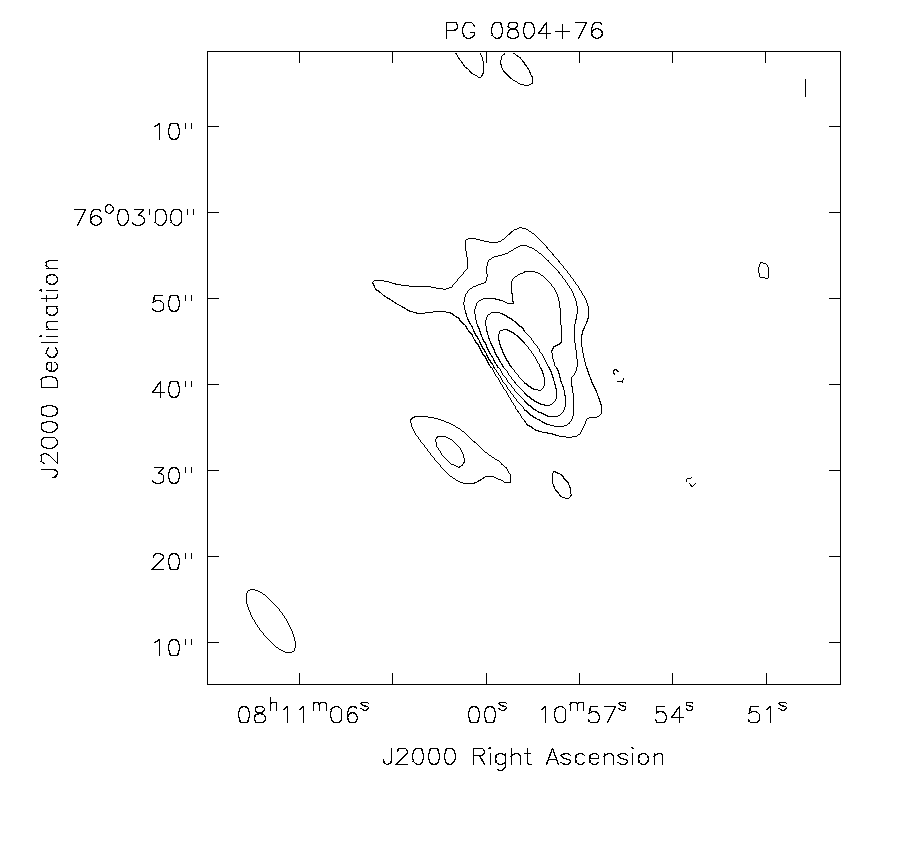}
\includegraphics[height=7.1cm,trim=25 10 0 10]{ 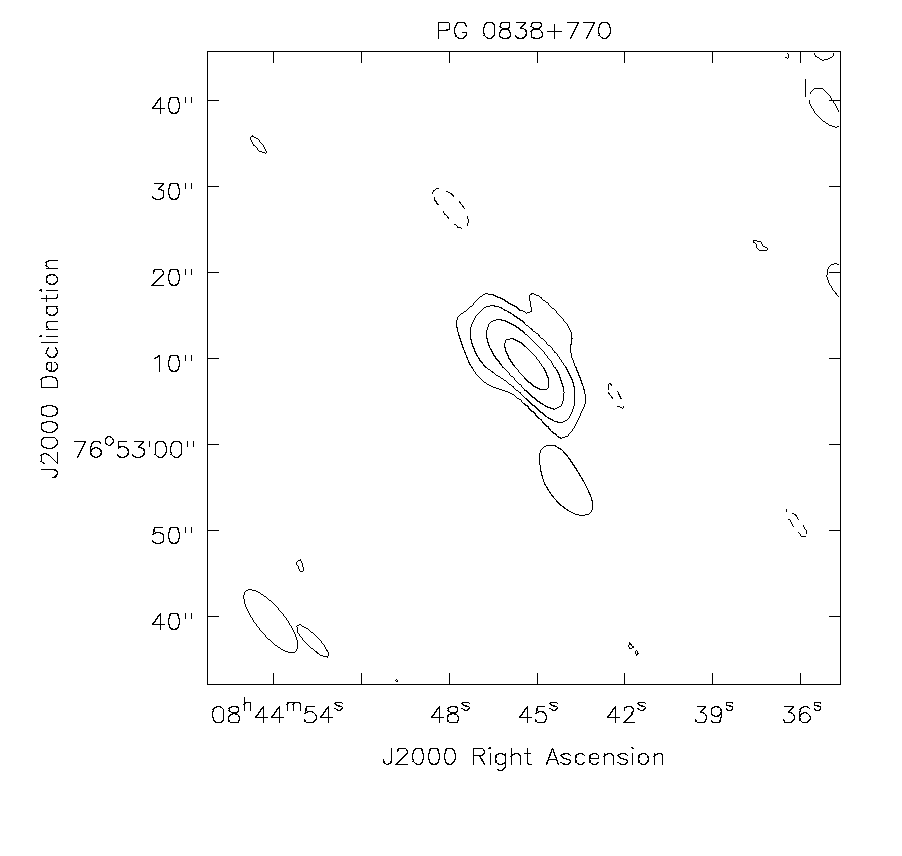}
\caption{uGMRT 685 MHz radio contour images of PG 0003+158 (RL), PG 0026+129, PG 0052+251, PG 0157+001, 
PG 0804+761, and PG 0838+770. The contour levels are $3\sigma \times (-1, 1, 2, 4, 8, 16, 32, 64, 128, 256, 516)$.}
\label{fig13}
\end{figure*}

\begin{figure*}
\centering
\includegraphics[height=7.1cm,trim=25 10 0 10]{ 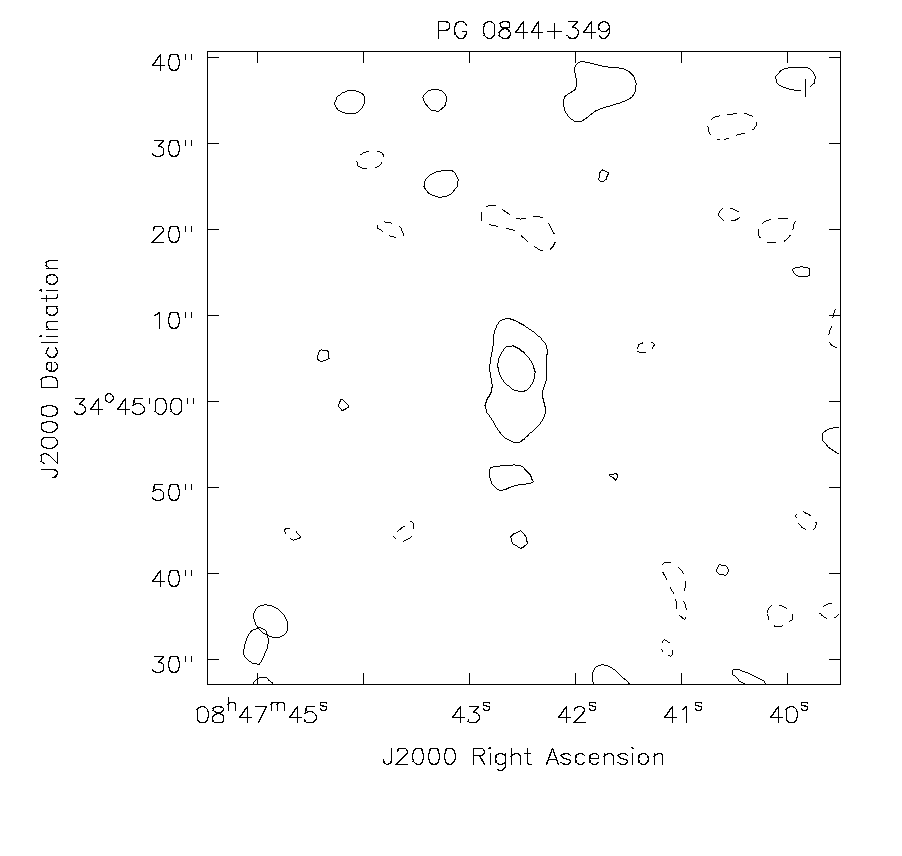}
\includegraphics[height=7.1cm,trim=25 10 0 10]{ 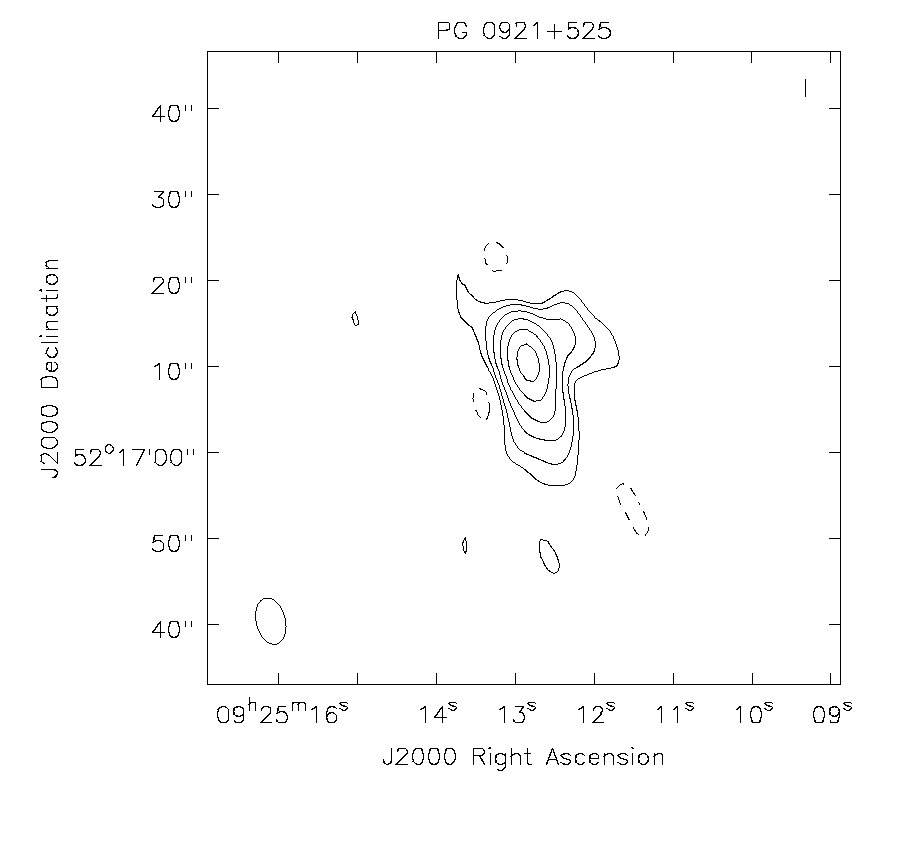}\\
\includegraphics[height=7.1cm,trim=25 10 0 10]{ 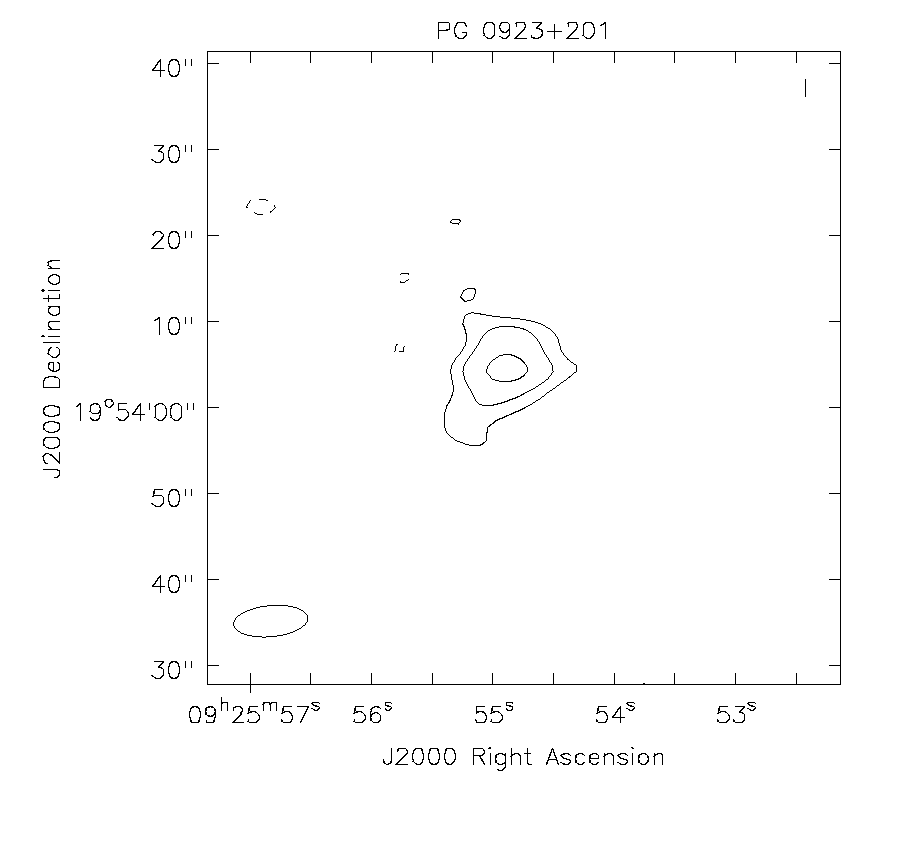}
\includegraphics[height=7.1cm,trim=25 10 0 10]{ 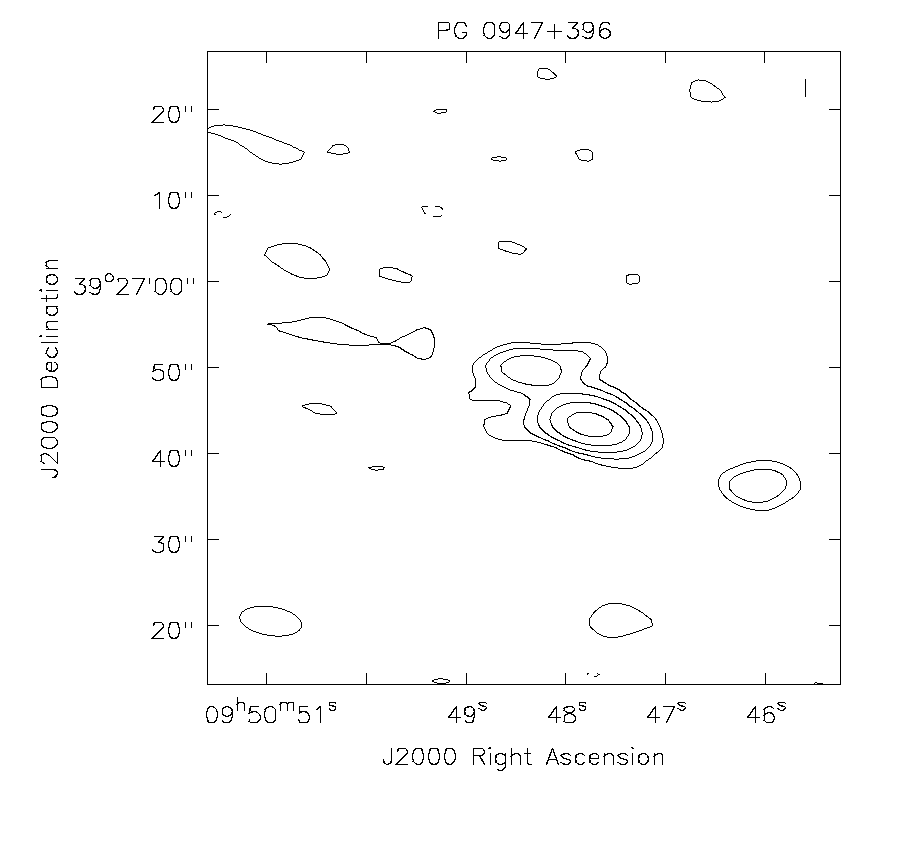}\\
\includegraphics[height=7.1cm,trim=25 10 0 10]{ 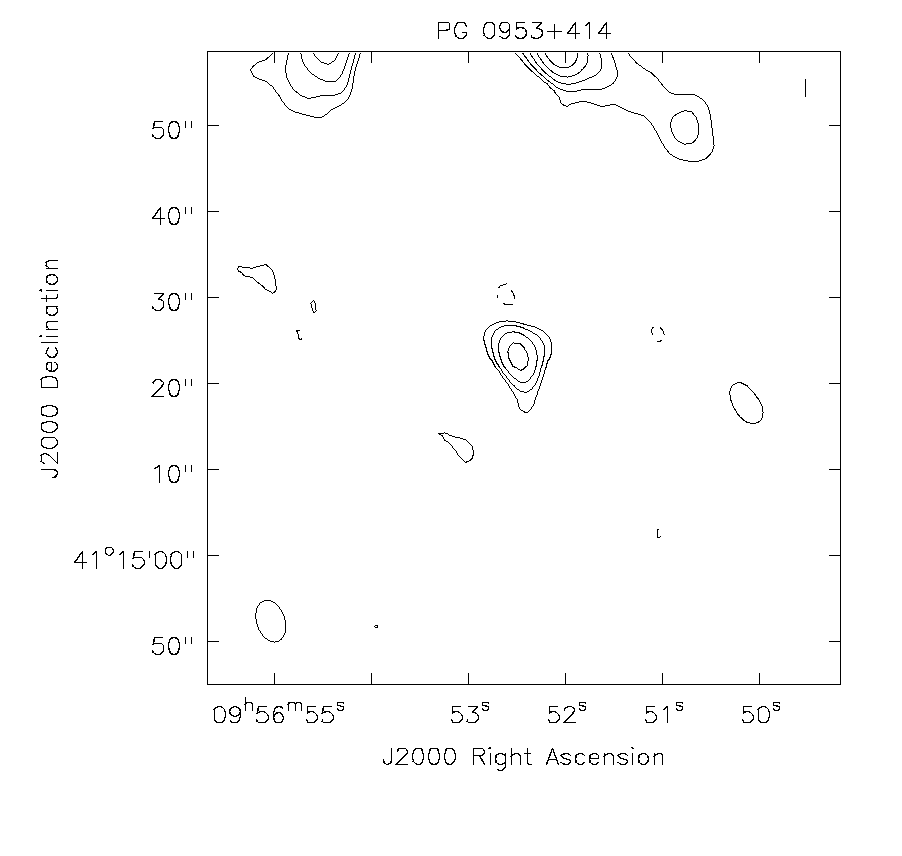}
\includegraphics[height=7.1cm,trim=25 10 0 10]{ 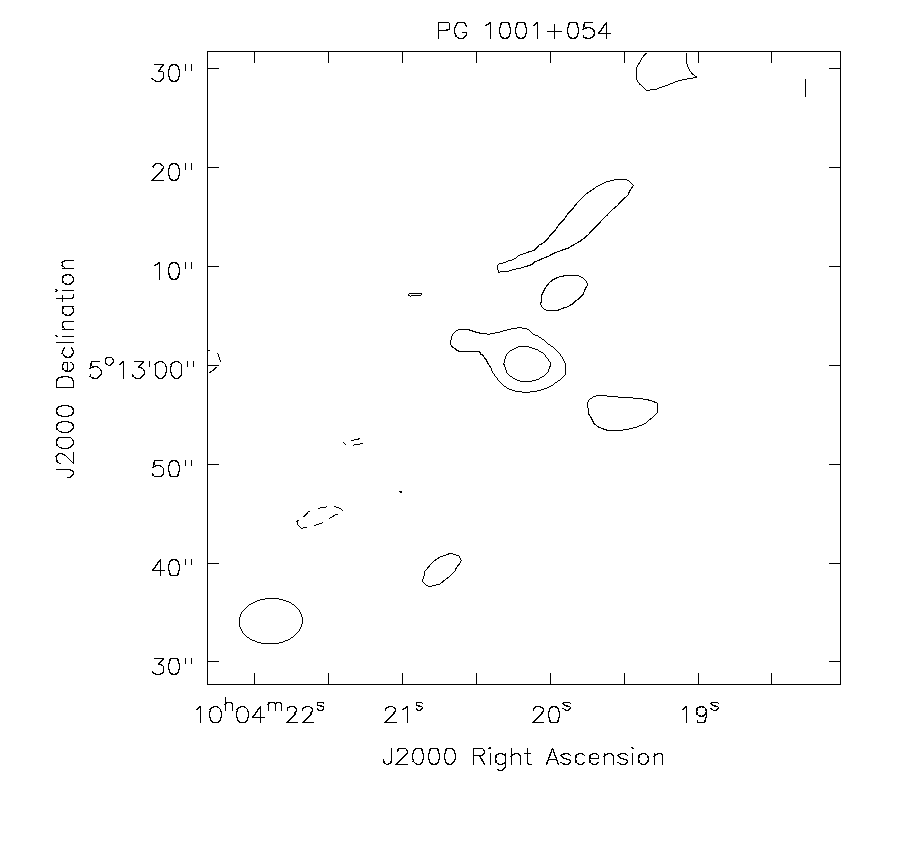}
\caption{uGMRT 685 MHz radio contour images of PG 0844+349, PG 0921+525, PG 0923+201, PG 0947+396, PG 0953+414, and PG 1001+054. The contour levels are $3\sigma \times (-1, 1, 2, 4, 8, 16, 32, 64, 128, 256, 516)$.}
\label{fig14}
\end{figure*}

\begin{figure*}
\centering
\includegraphics[height=7.1cm,trim=25 10 0 10]{ 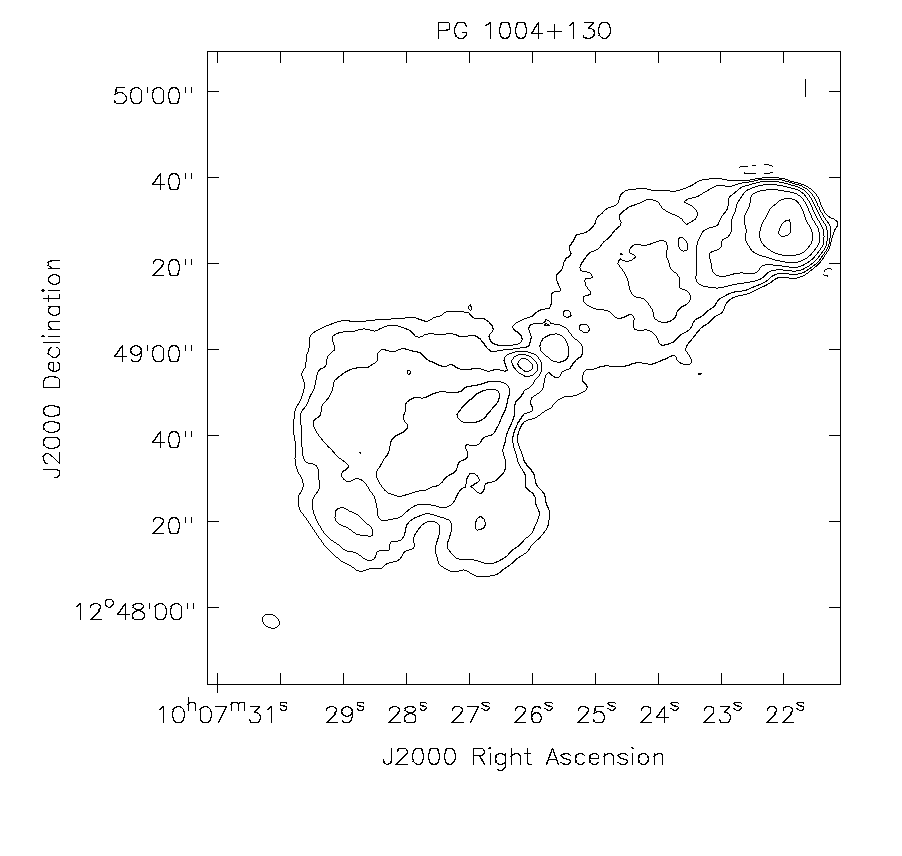}
\includegraphics[height=7.1cm,trim=25 10 0 10]{ 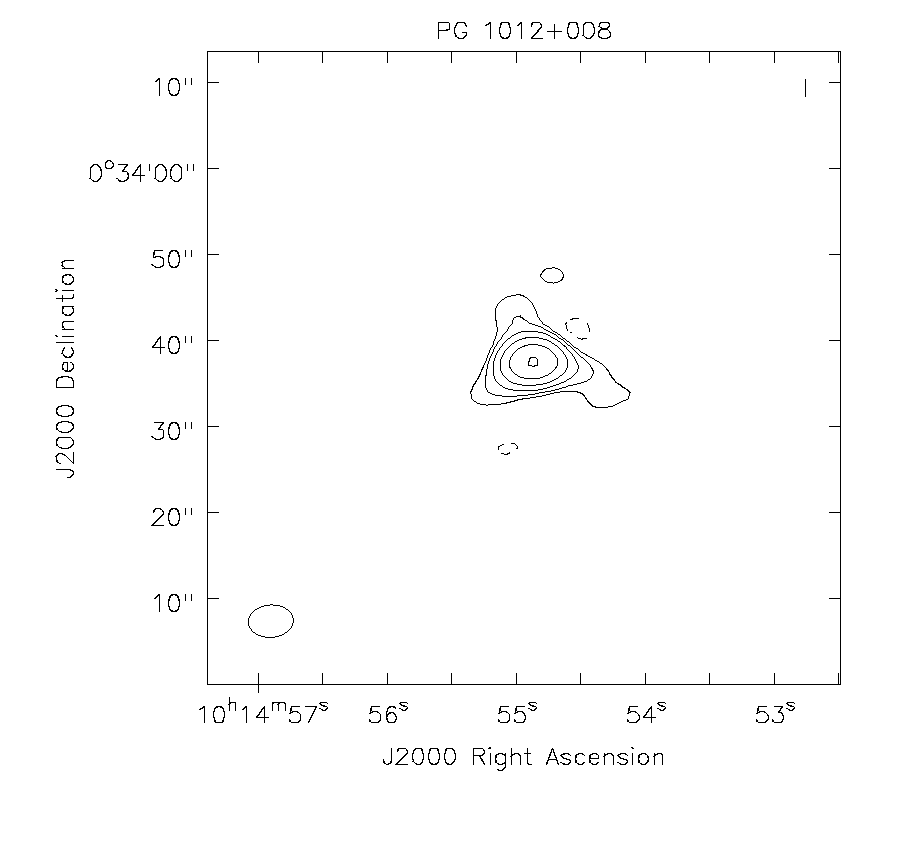}\\
\includegraphics[height=7.1cm,trim=25 10 0 10]{ 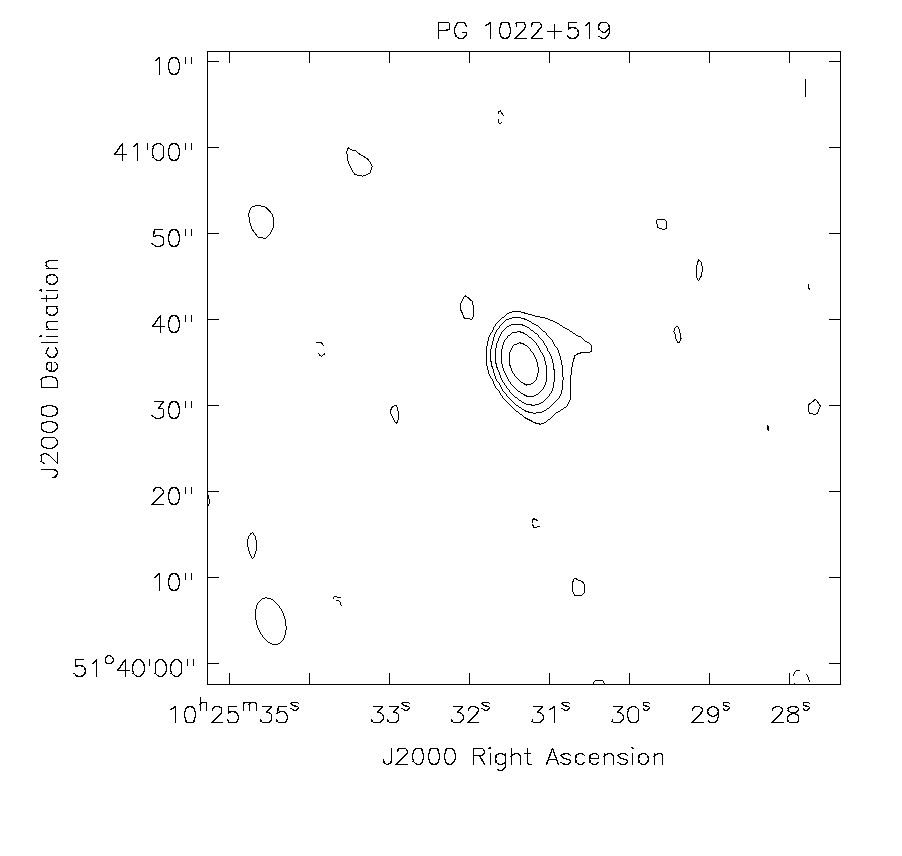}
\includegraphics[height=7.1cm,trim=25 10 0 10]{ 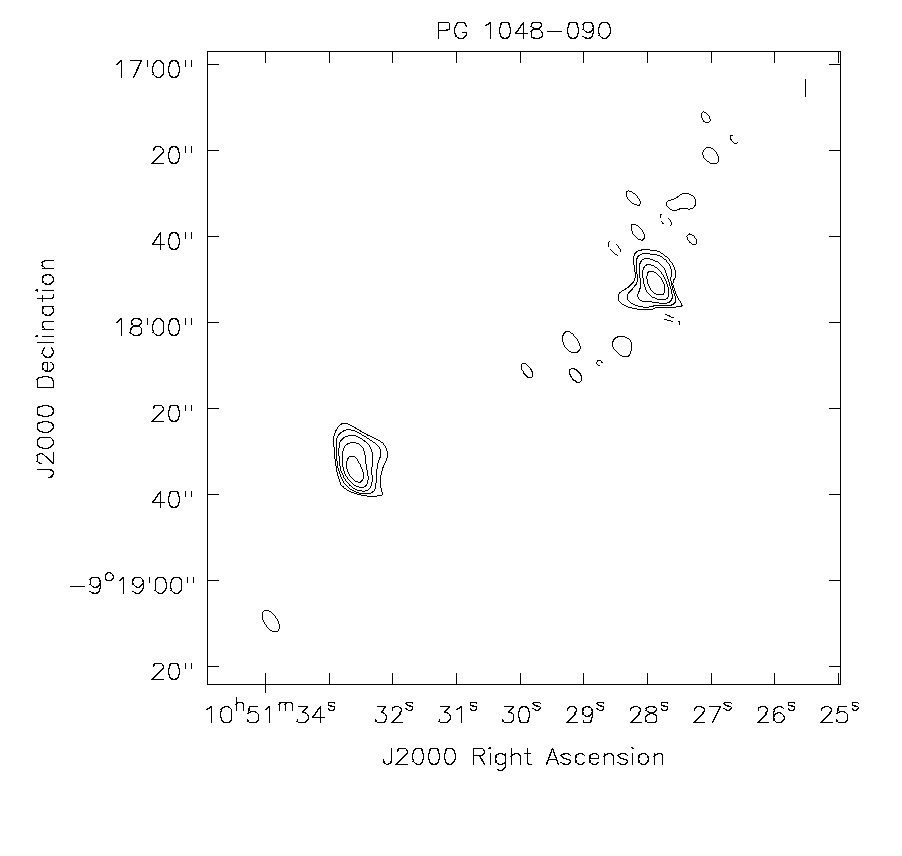}\\
\includegraphics[height=7.1cm,trim=25 10 0 10]{ 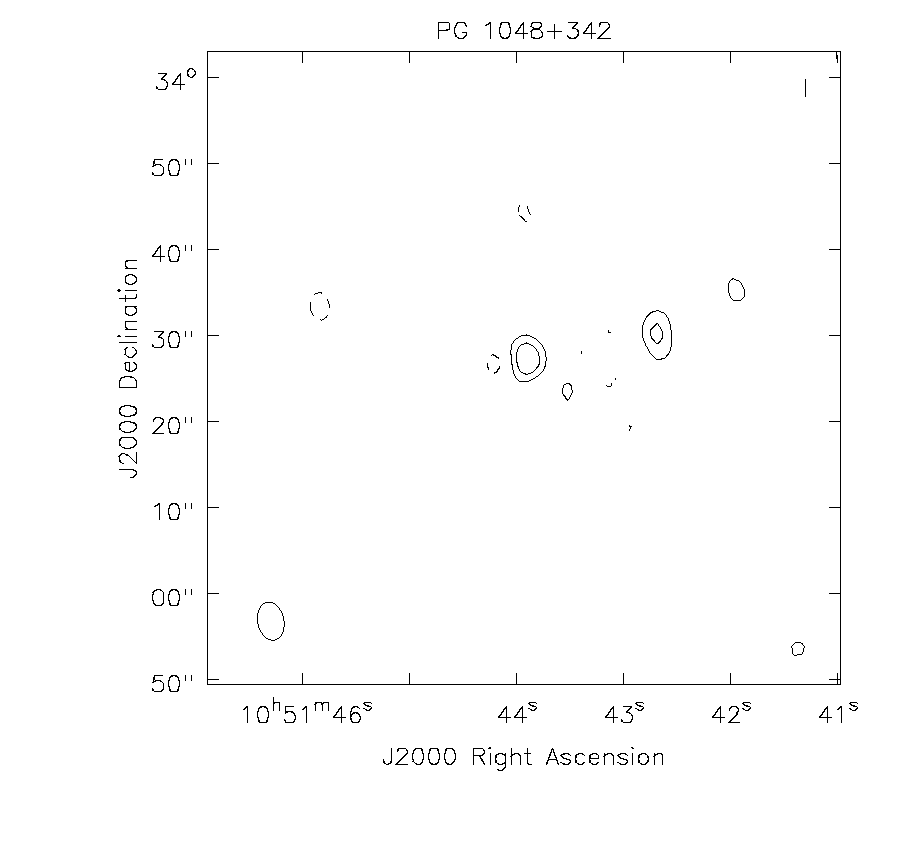}
\includegraphics[height=7.1cm,trim=25 10 0 10]{ 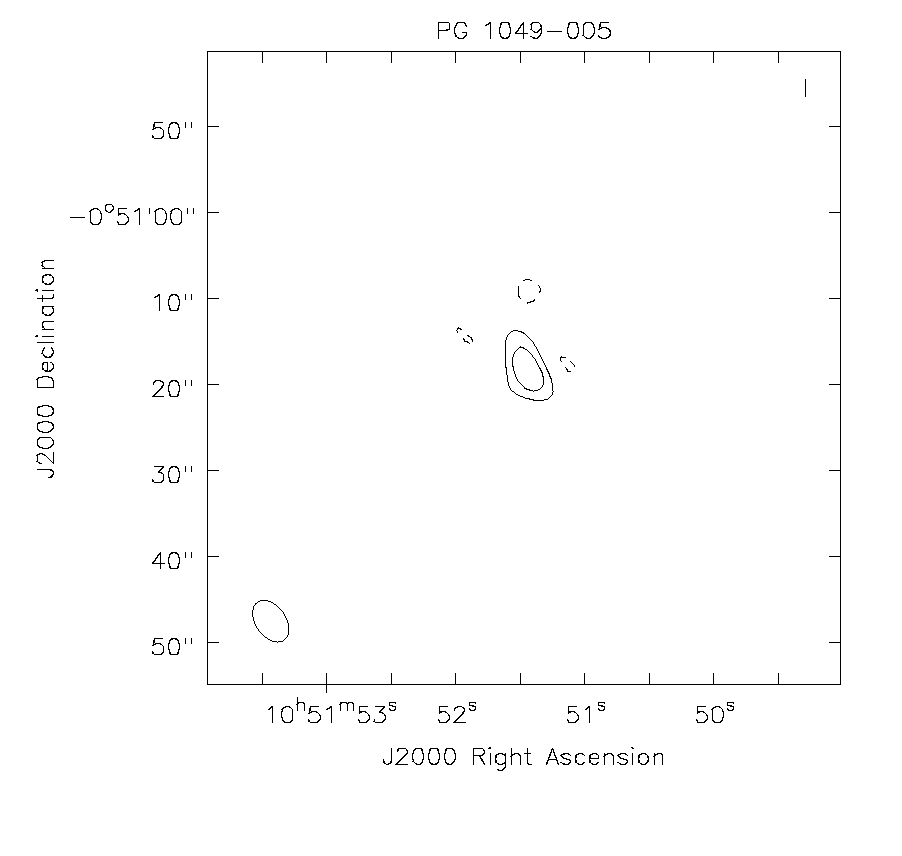}
\caption{uGMRT 685 MHz radio contour images of PG 1004+130 (RL), PG 1012+008, PG 1022+519, PG 1048-090 (RL), PG 1048+342, PG 1049-005. The contour levels are $3\sigma \times (-1, 1, 2, 4, 8, 16, 32, 64, 128, 256, 516)$.}
\label{fig15}
\end{figure*}

\begin{figure*}
\centering
\includegraphics[height=7.1cm,trim=25 10 0 10]{ 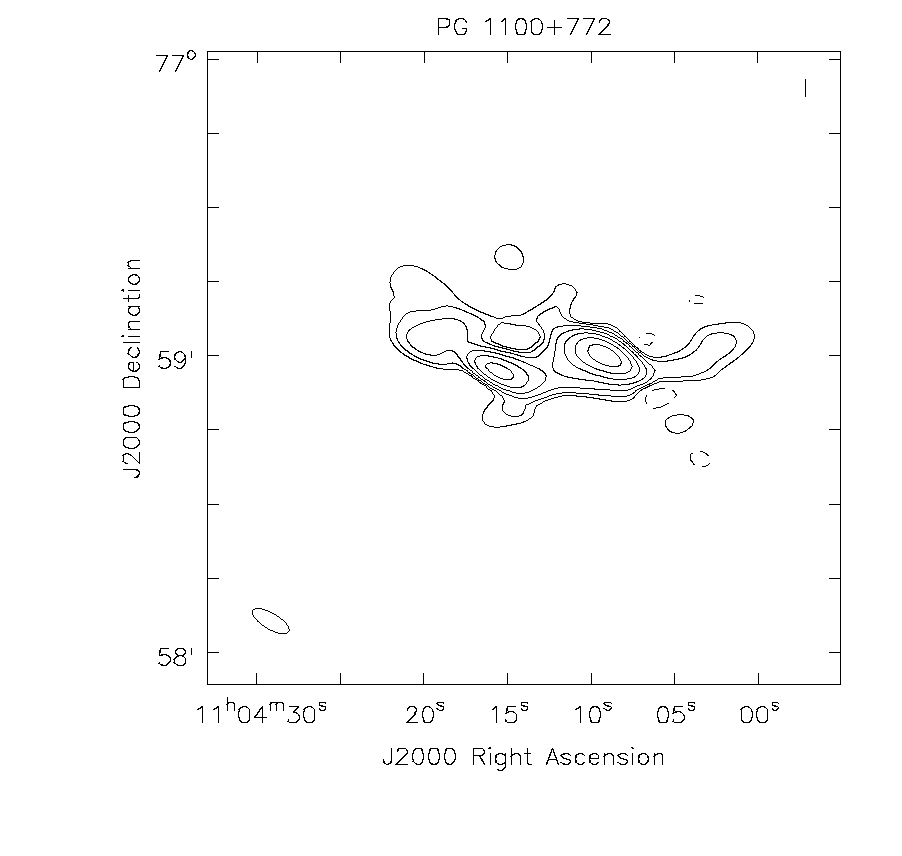}
\includegraphics[height=7.1cm,trim=25 10 0 10]{ 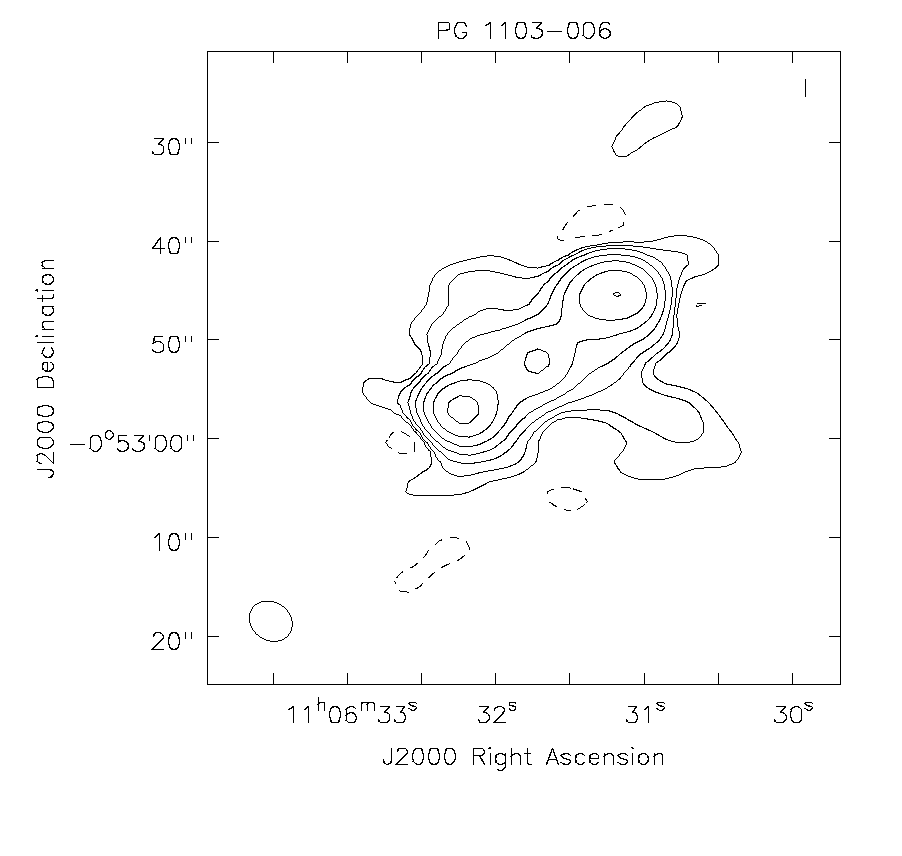}\\
\includegraphics[height=7.1cm,trim=25 10 0 10]{ 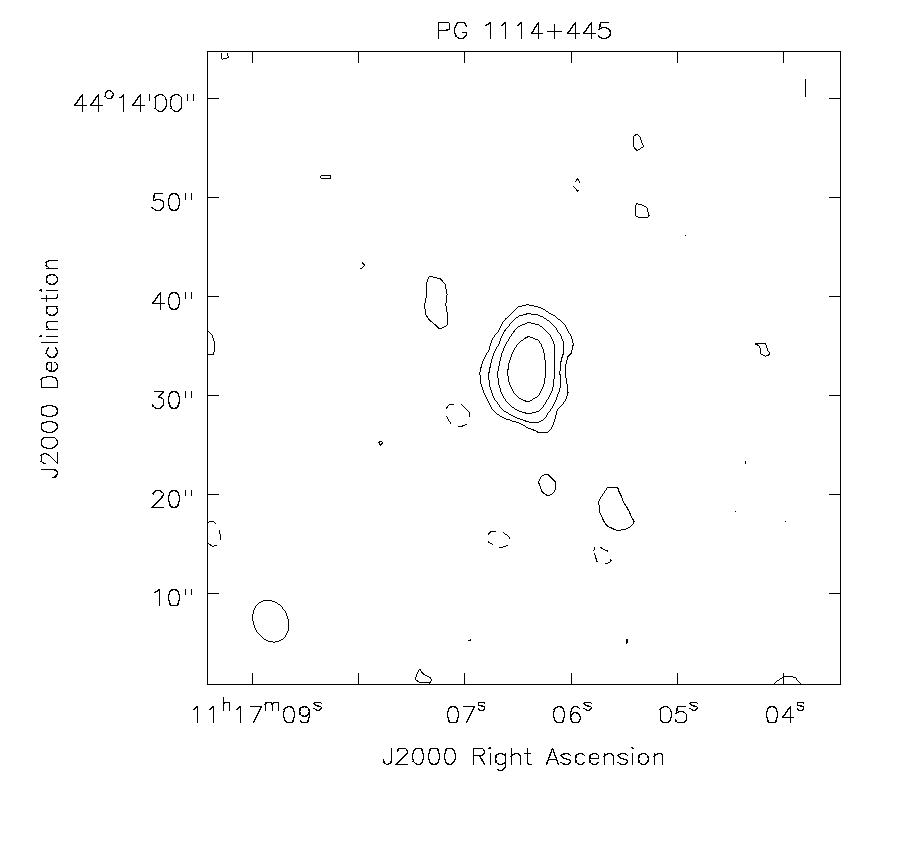}
\includegraphics[height=7.1cm,trim=25 10 0 10]{ 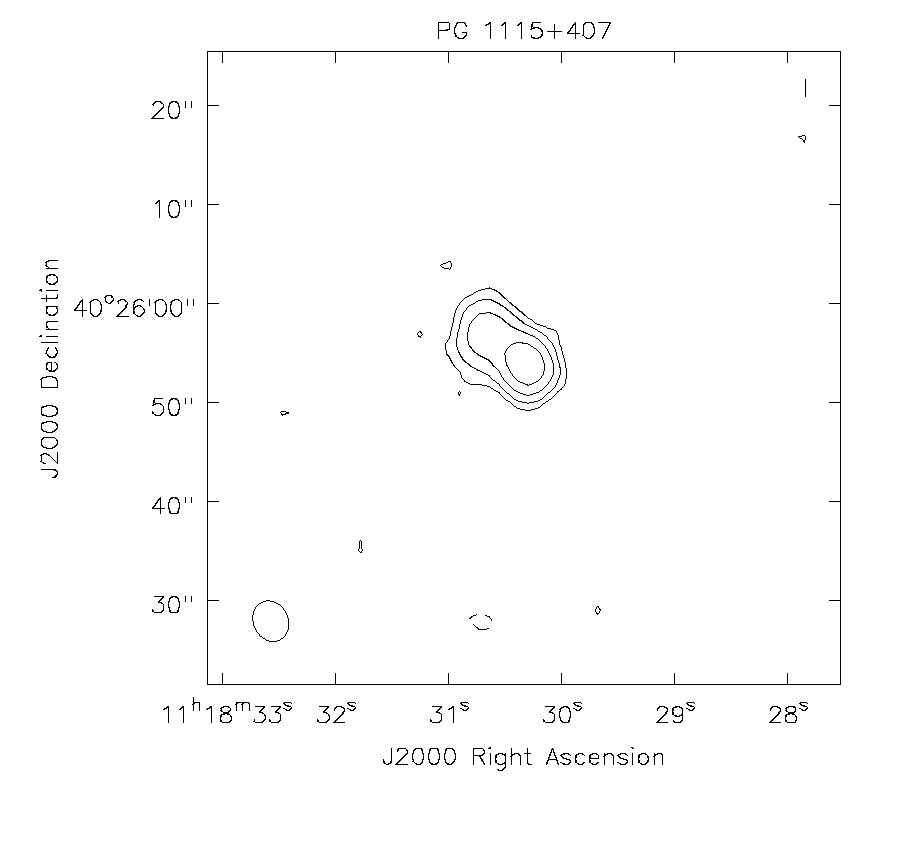}\\
\includegraphics[height=7.1cm,trim=25 10 0 10]{ 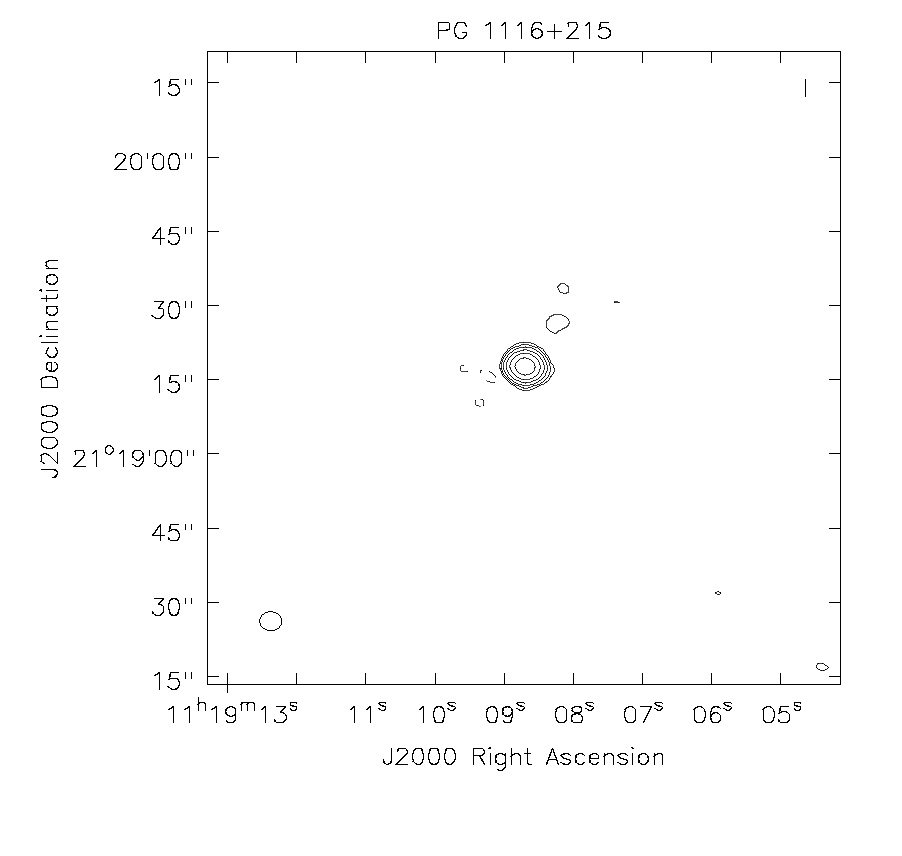}
\includegraphics[height=7.1cm,trim=25 10 0 10]{ 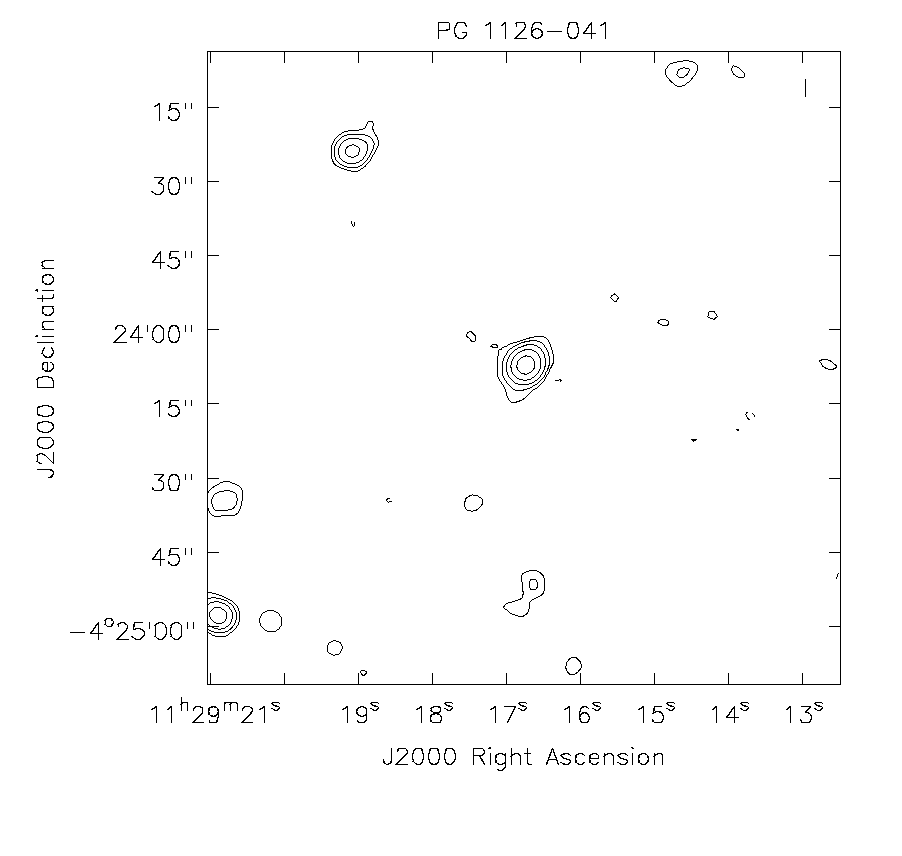}
\caption{\small uGMRT 685 MHz radio contour images of PG~1100+772 (RL), PG~1103-006 (RL), PG~1114+445, PG~1115+407, PG 1116+215, and PG 1126-041. The contour levels are $3\sigma \times (-1, 1, 2, 4, 8, 16, 32, 64, 128, 256, 516)$.}
\label{fig16}
\end{figure*}

\begin{figure*}
\centering
\includegraphics[height=7.1cm,trim=25 10 0 10]{ 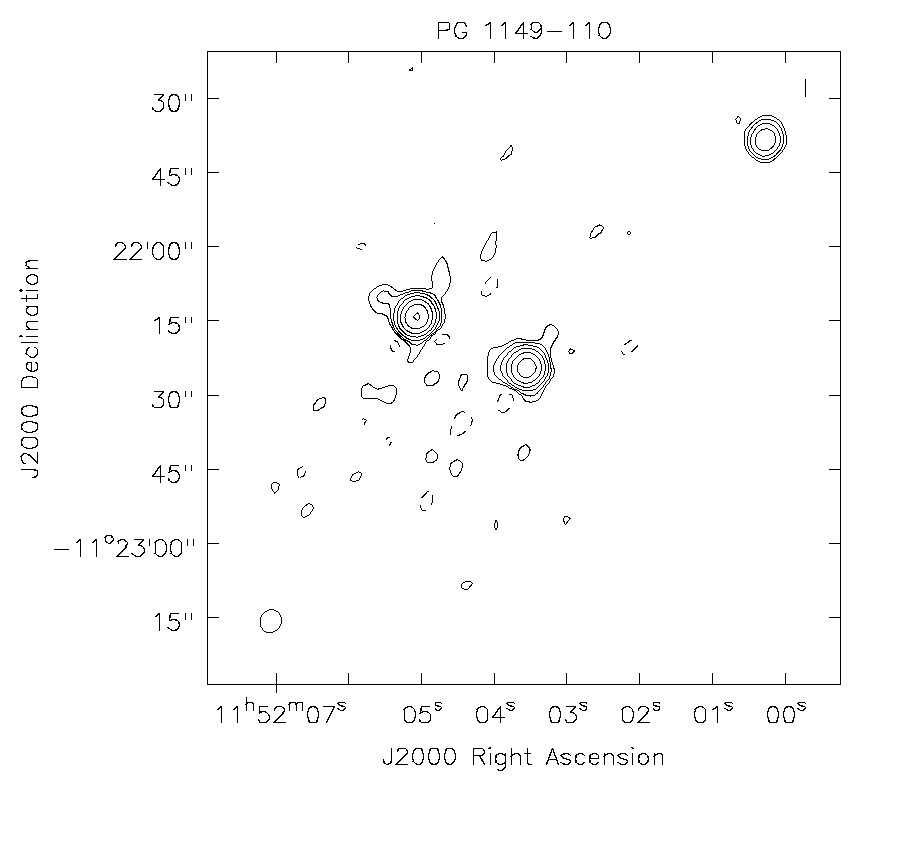}
\includegraphics[height=7.1cm,trim=25 10 0 10]{ 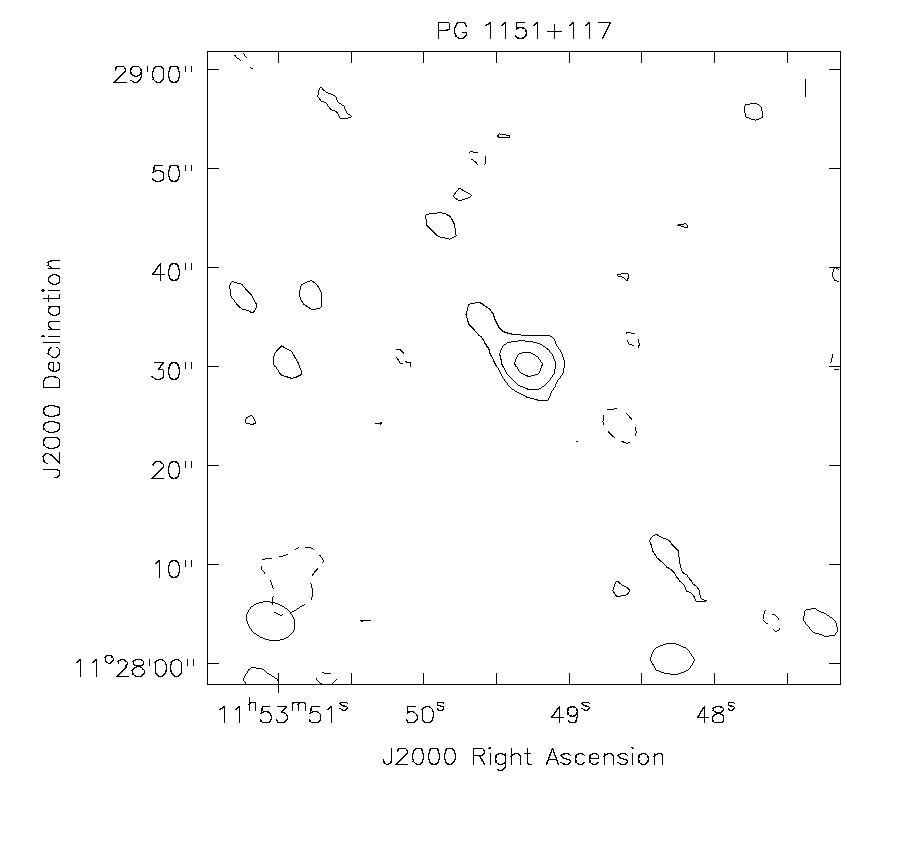}\\
\includegraphics[height=7.1cm,trim=25 10 0 10]{ 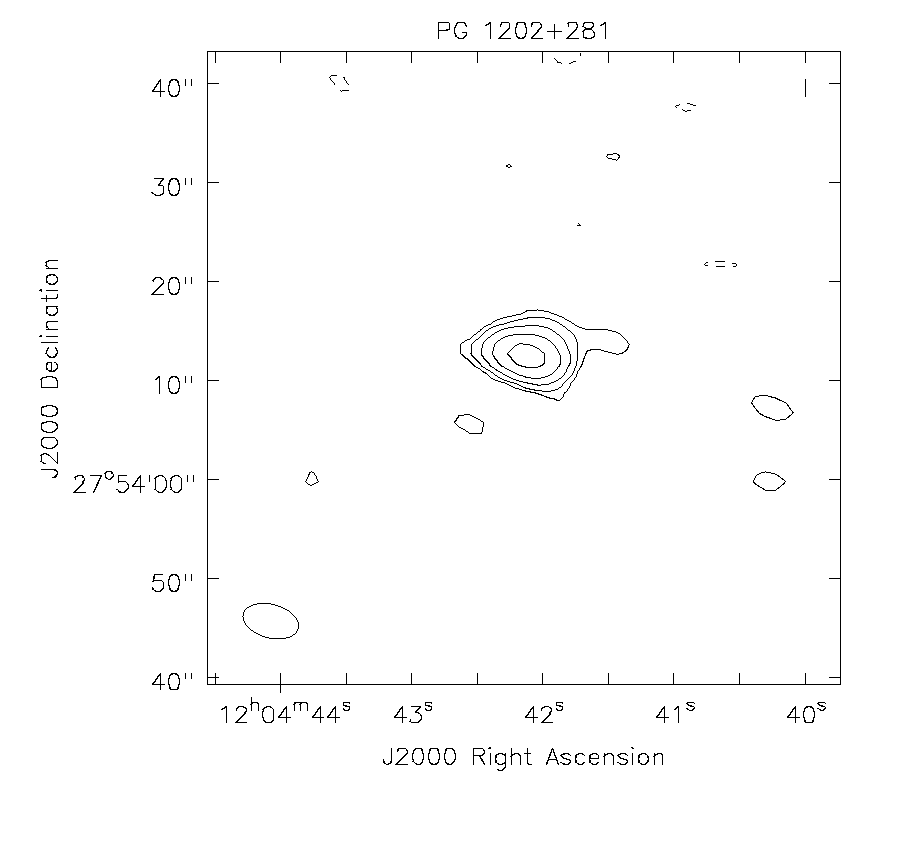}
\includegraphics[height=7.1cm,trim=25 10 0 10]{ 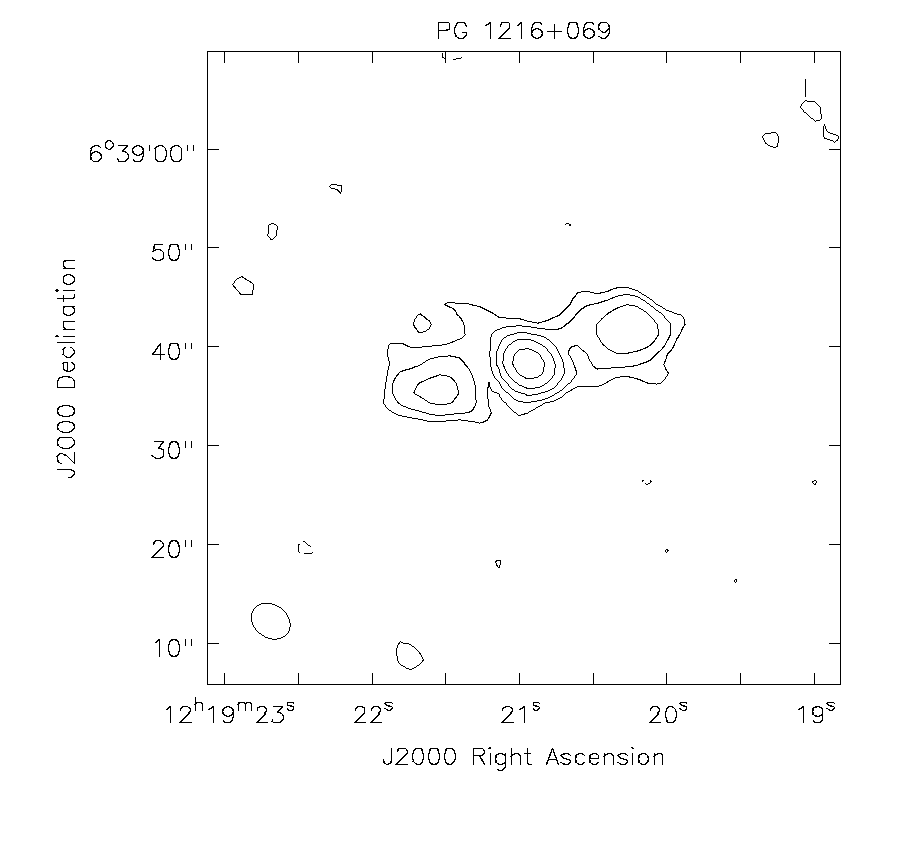}\\
\includegraphics[height=7.1cm,trim=25 10 0 10]{ 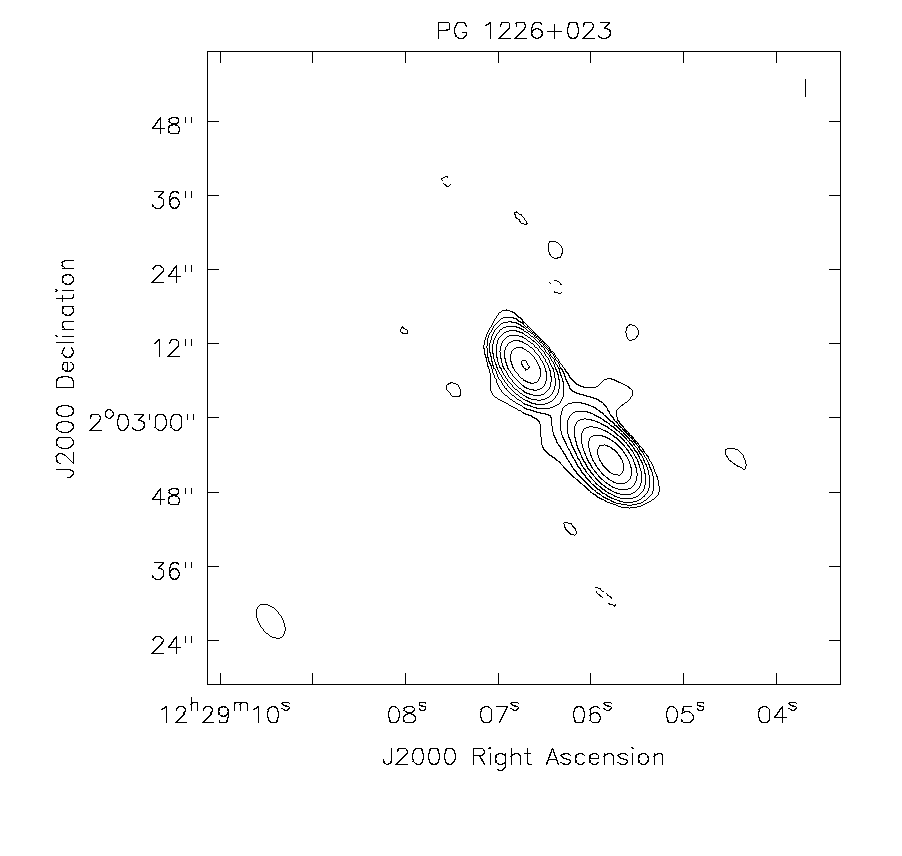}
\includegraphics[height=7.1cm,trim=25 10 0 10]{ 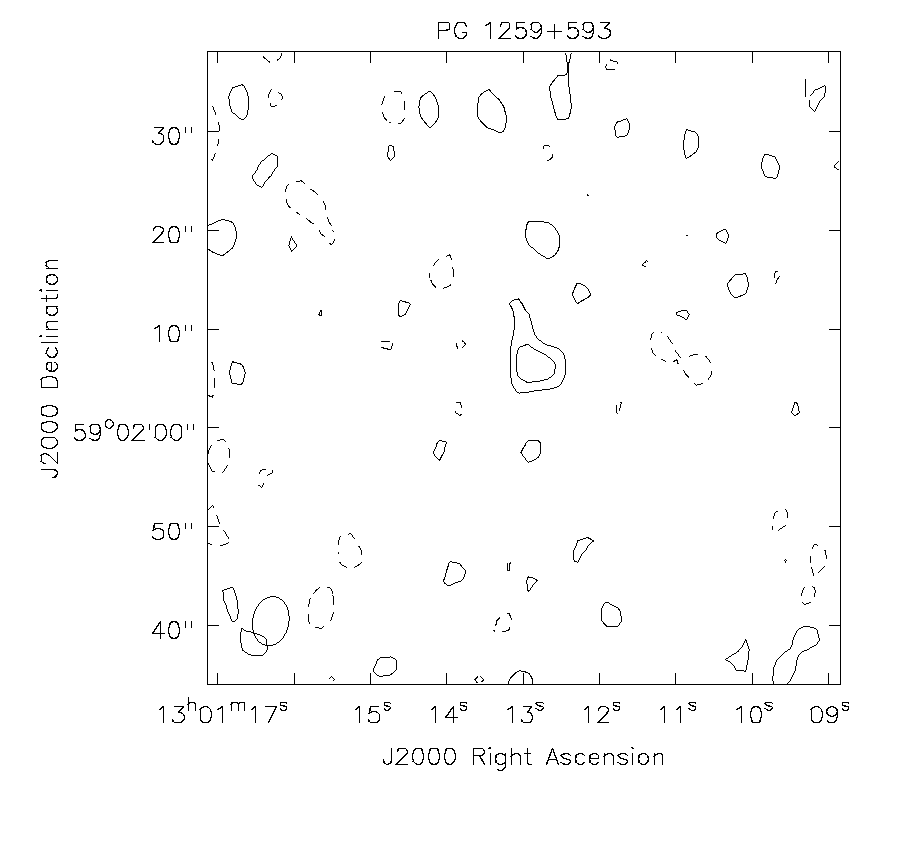}
\caption{\small uGMRT 685 MHz radio contour images of PG 1149-110, PG 1151+117, PG 1202+281, PG 1216+069, PG 1226+023 (RL), and PG 1259+593. The contour levels are $3\sigma \times (-1, 1, 2, 4, 8, 16, 32, 64, 128, 256, 516)$.}
\label{fig17}
\end{figure*}

\begin{figure*}
\centering
\includegraphics[height=7.1cm,trim=25 10 0 10]{ 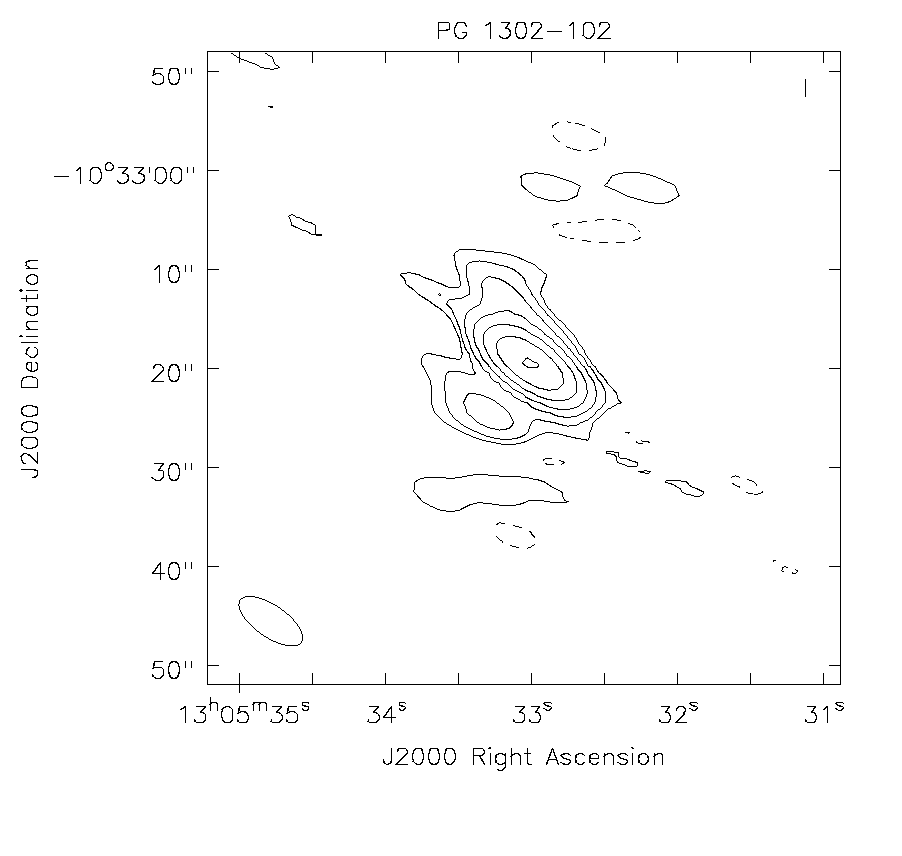}
\includegraphics[height=7.1cm,trim=25 10 0 10]{ 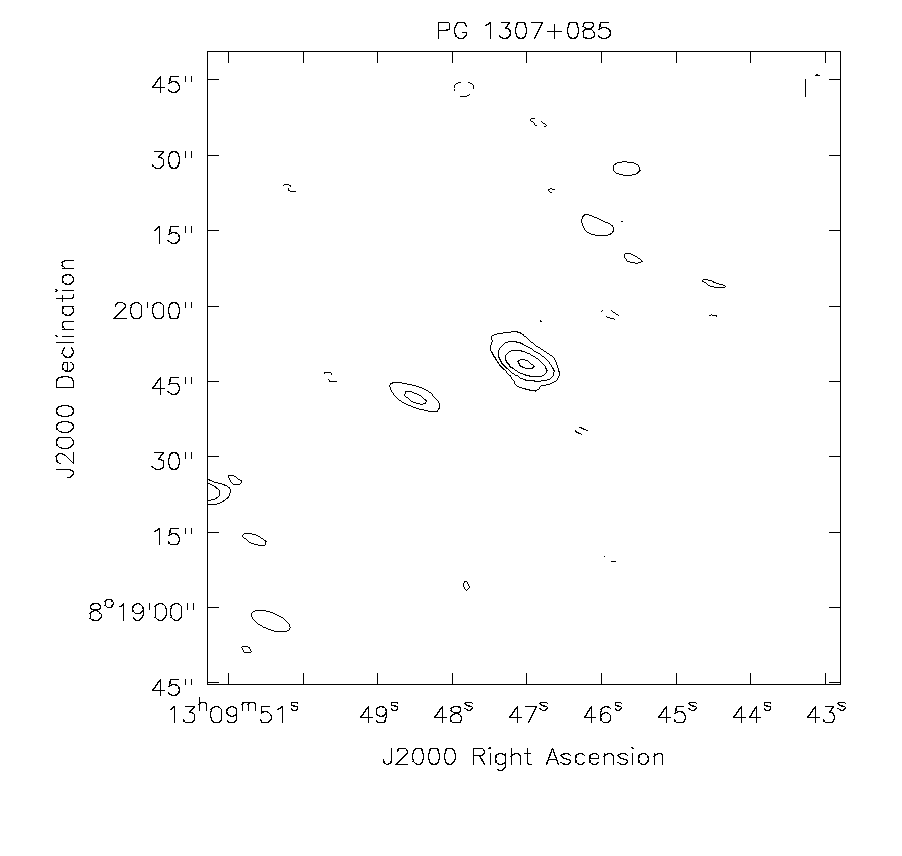}\\
\includegraphics[height=7.1cm,trim=25 10 0 10]{ 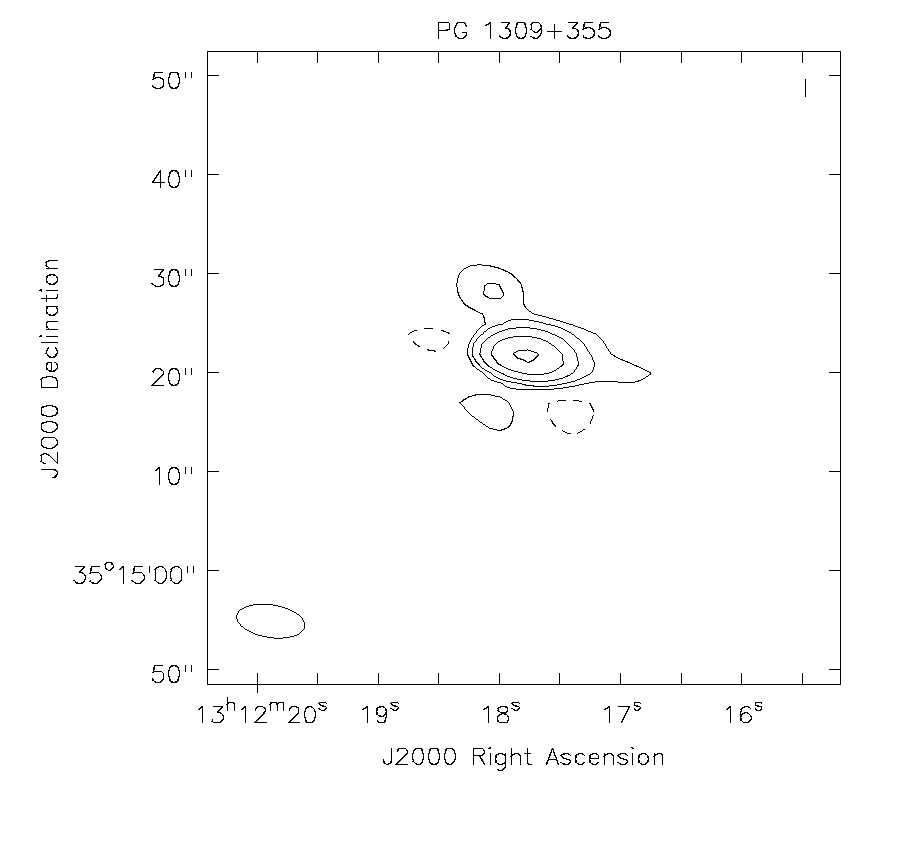}
\includegraphics[height=7.1cm,trim=25 10 0 10]{ 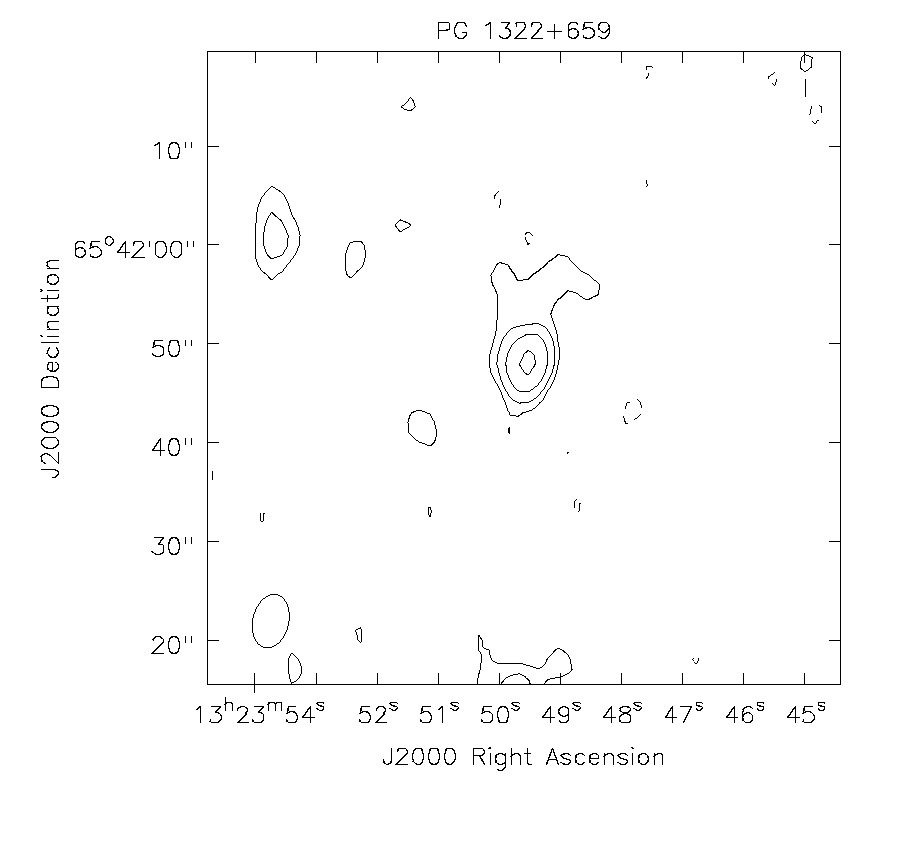}\\
\includegraphics[height=7.1cm,trim=25 10 0 10]{ 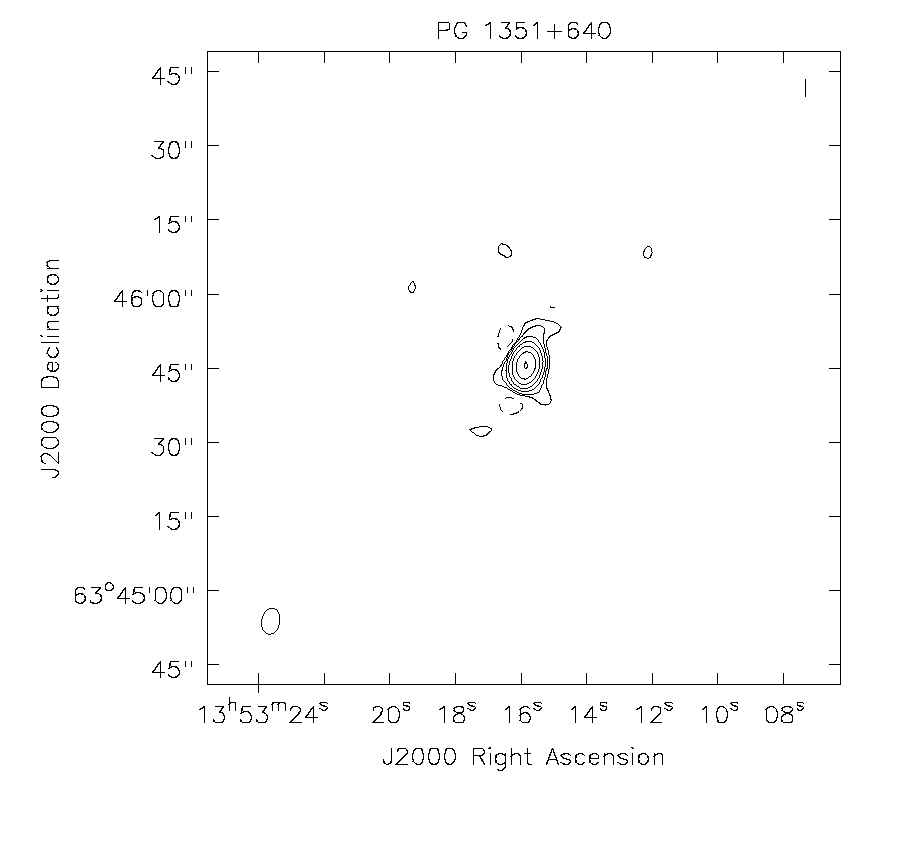}
\includegraphics[height=7.1cm,trim=25 10 0 10]{ 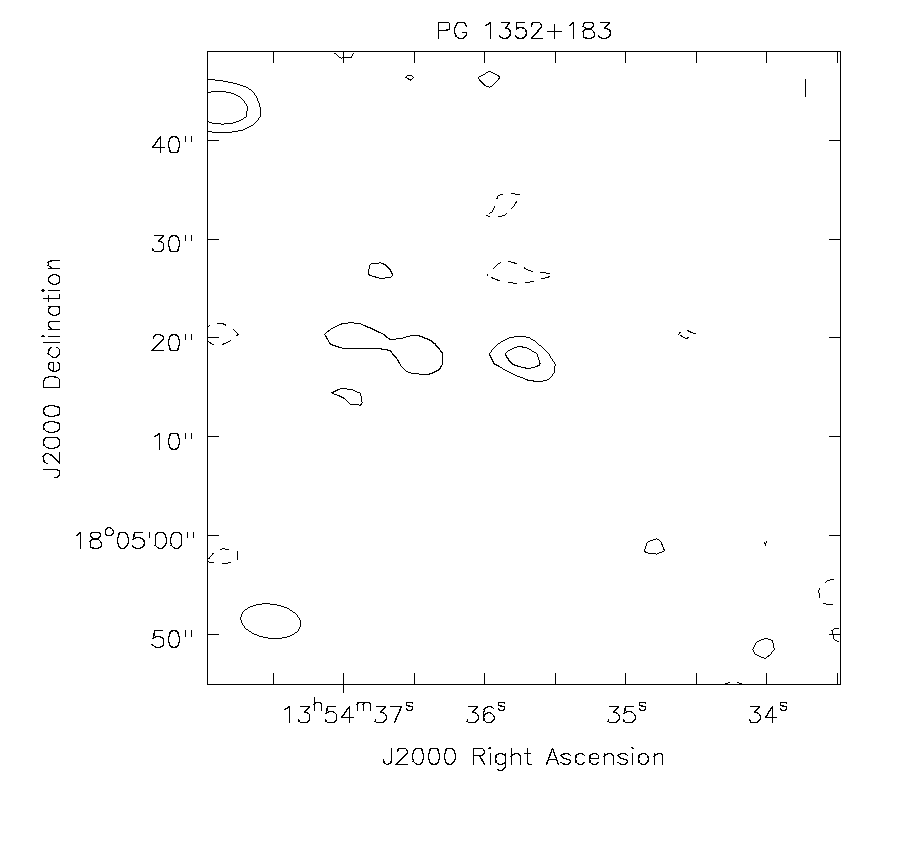}
\caption{\small uGMRT 685 MHz radio contour images of PG 1302-102 (RL), PG 1307+085, PG 1309+355 (RL), PG 1322+659, PG 1351+640, and PG 1352+183. The contour levels are $3\sigma \times (-1, 1, 2, 4, 8, 16, 32, 64, 128, 256, 516)$.}
\label{fig18}
\end{figure*}

\begin{figure*}
\centering
\includegraphics[height=7.1cm,trim=25 10 0 10]{ 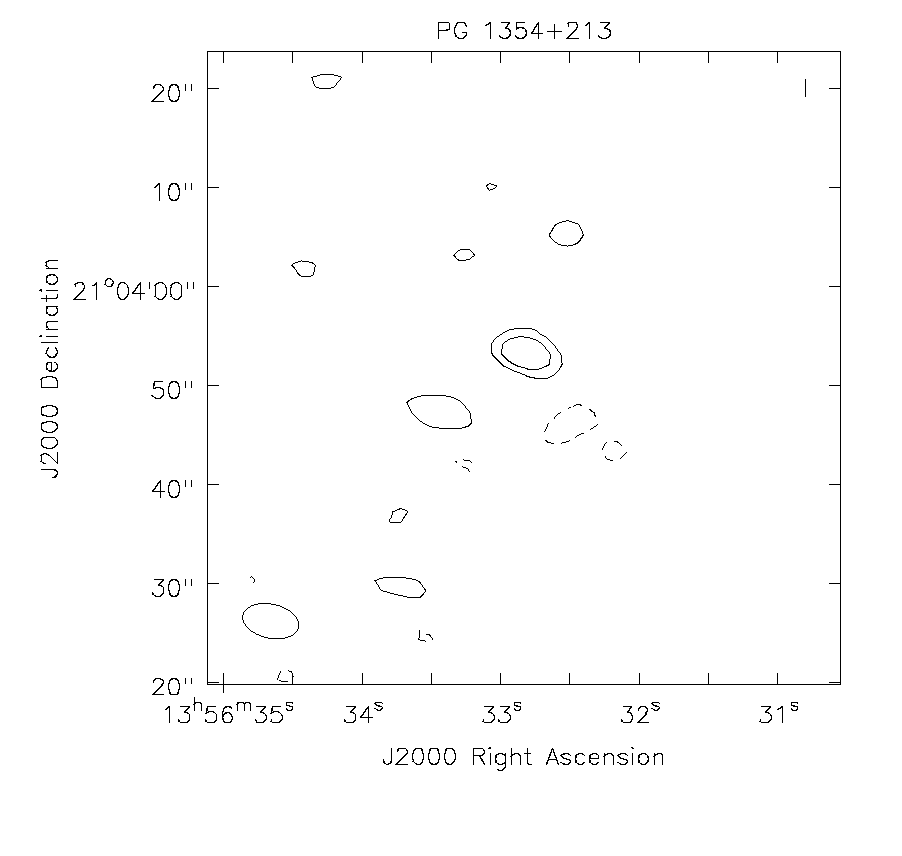}
\includegraphics[height=7.1cm,trim=25 10 0 10]{ 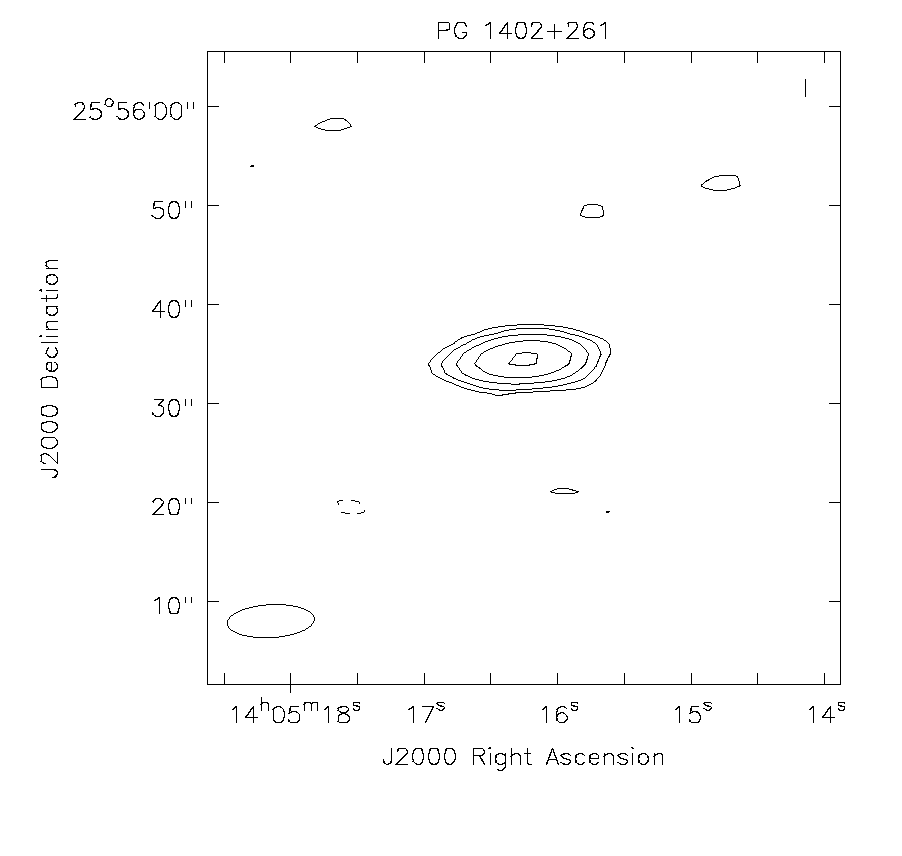}\\
\includegraphics[height=7.1cm,trim=25 10 0 10]{ 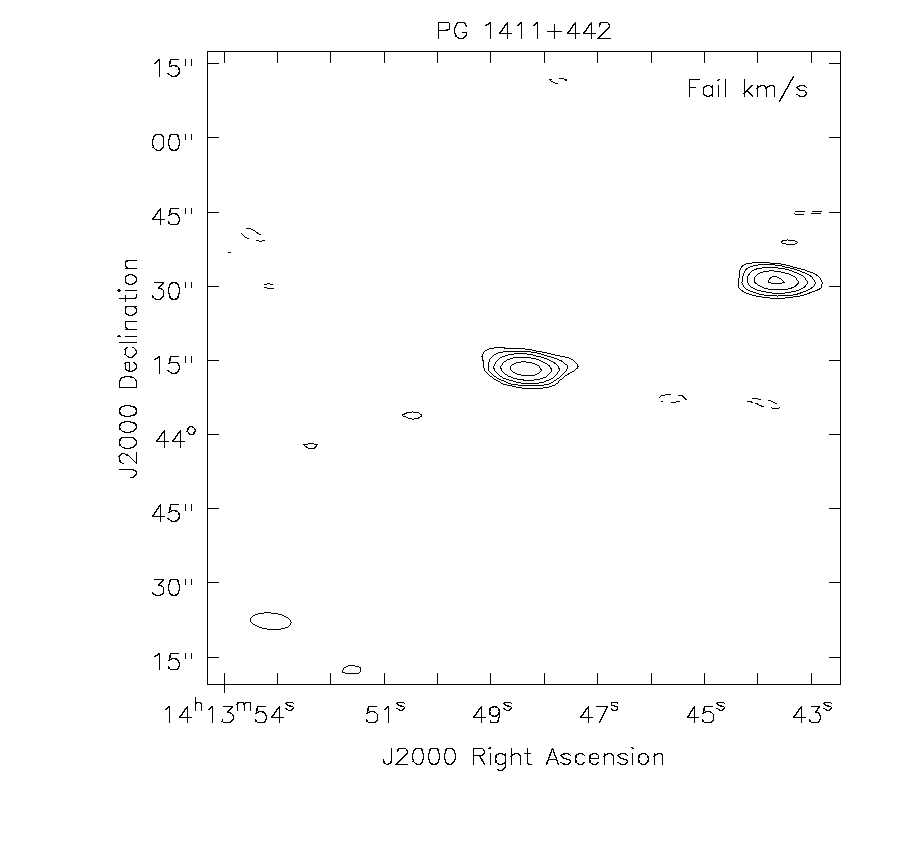}
\includegraphics[height=7.1cm,trim=25 10 0 10]{ 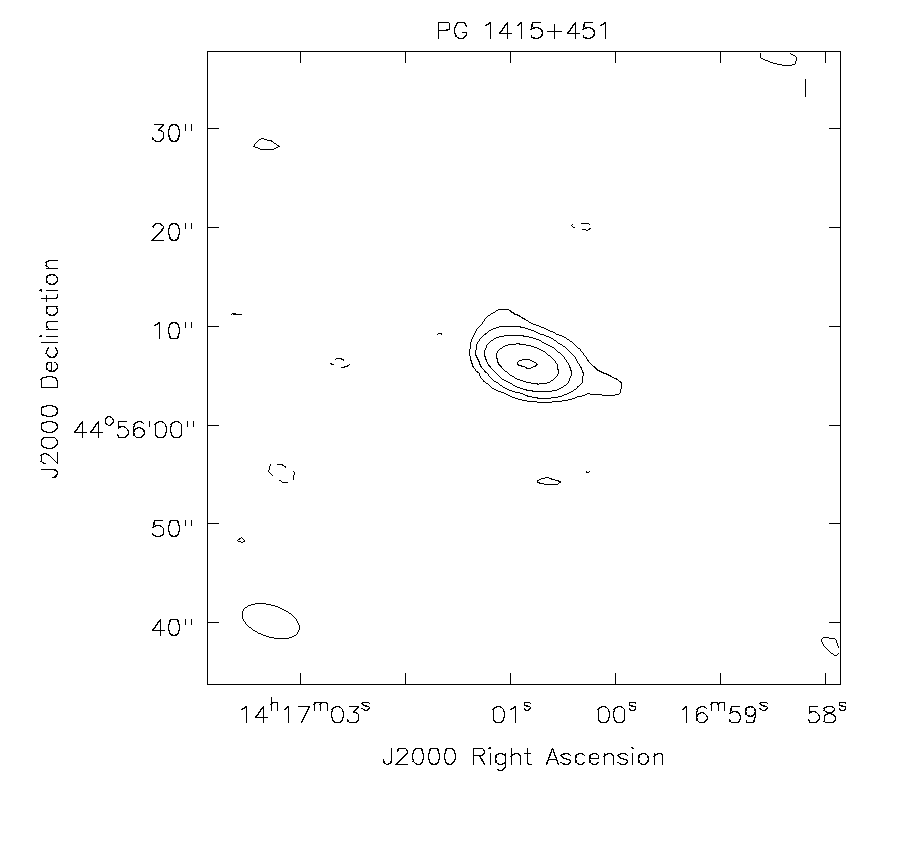}\\
\includegraphics[height=7.1cm,trim=25 10 0 10]{ 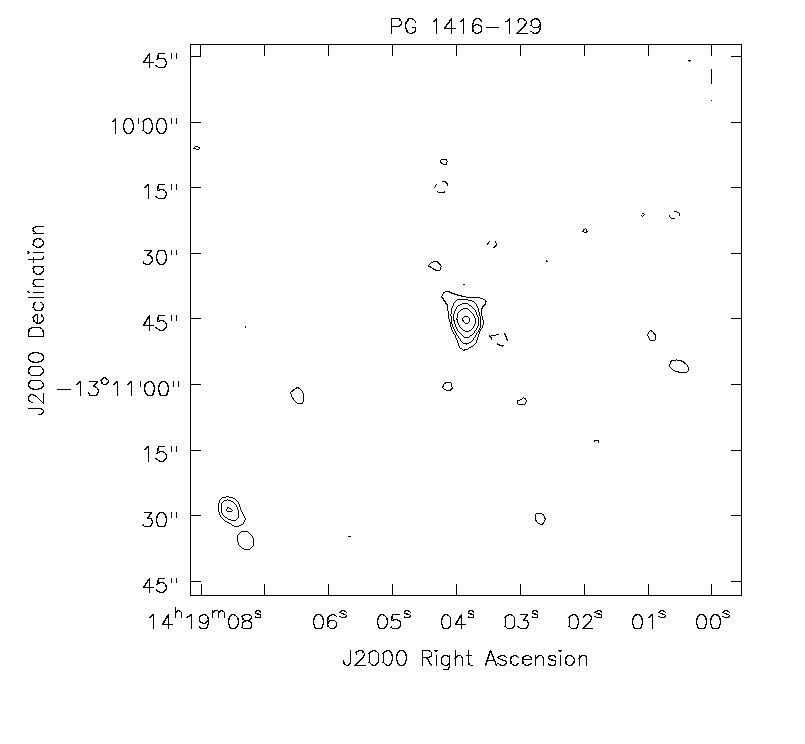}
\includegraphics[height=7.1cm,trim=25 10 0 10]{ 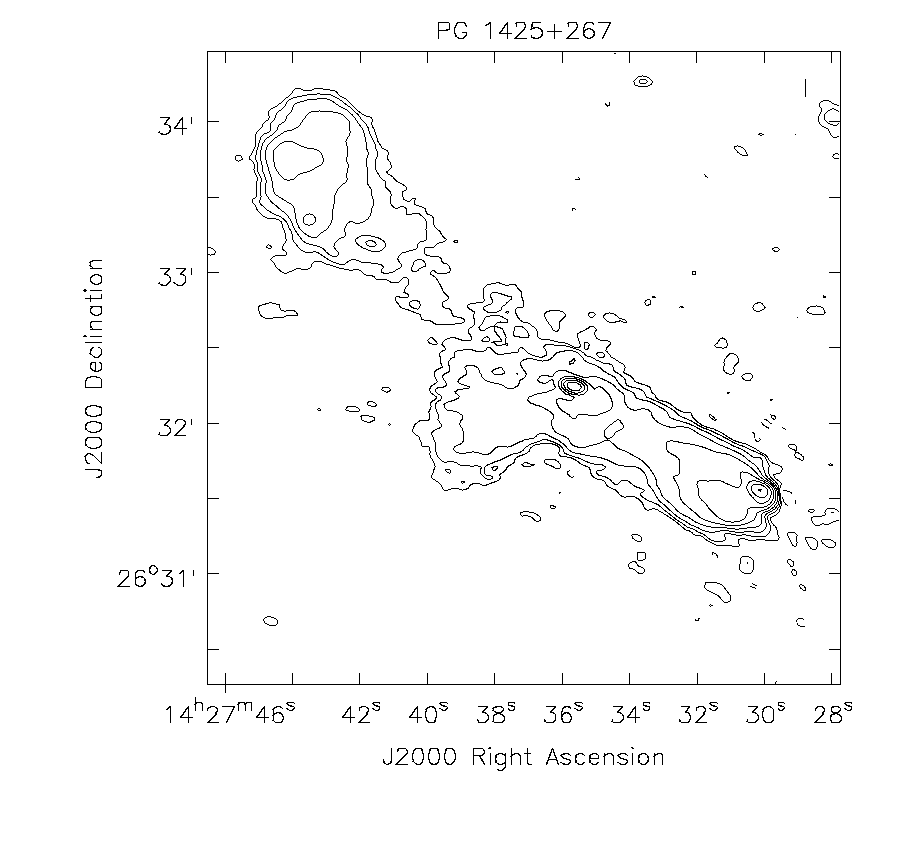}
\caption{\small uGMRT 685 MHz radio contour images of PG 1354+213, PG 1402+261, PG 1411+442, PG 1415+451, PG 1416-129, and PG 1425+267 (RL). The contour levels are $3\sigma \times (-1, 1, 2, 4, 8, 16, 32, 64, 128, 256, 516)$.}
\label{fig19}
\end{figure*}

\begin{figure*}
\centering
\includegraphics[height=7.1cm,trim=25 10 0 10]{ 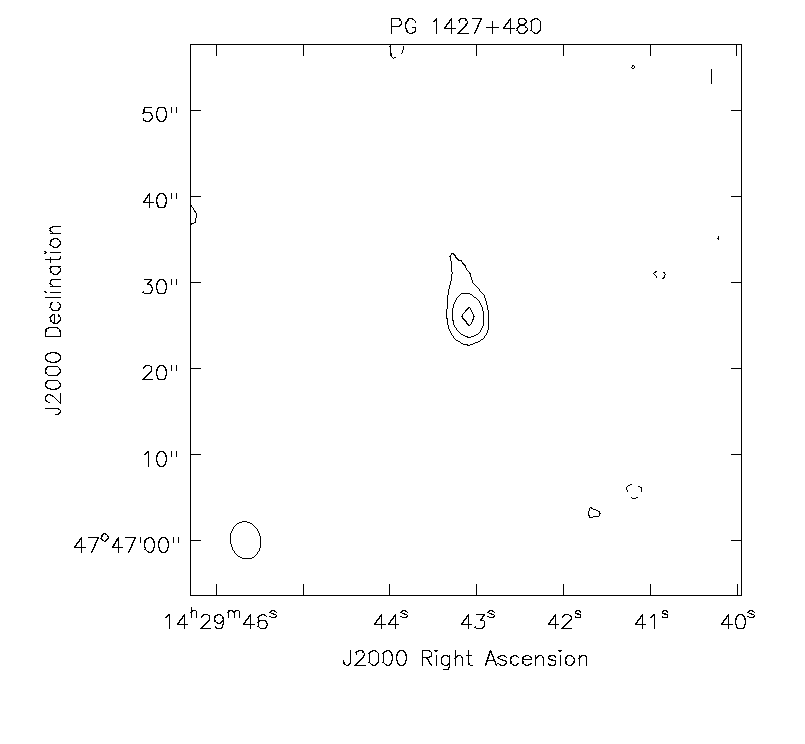}
\includegraphics[height=7.1cm,trim=25 10 0 10]{ 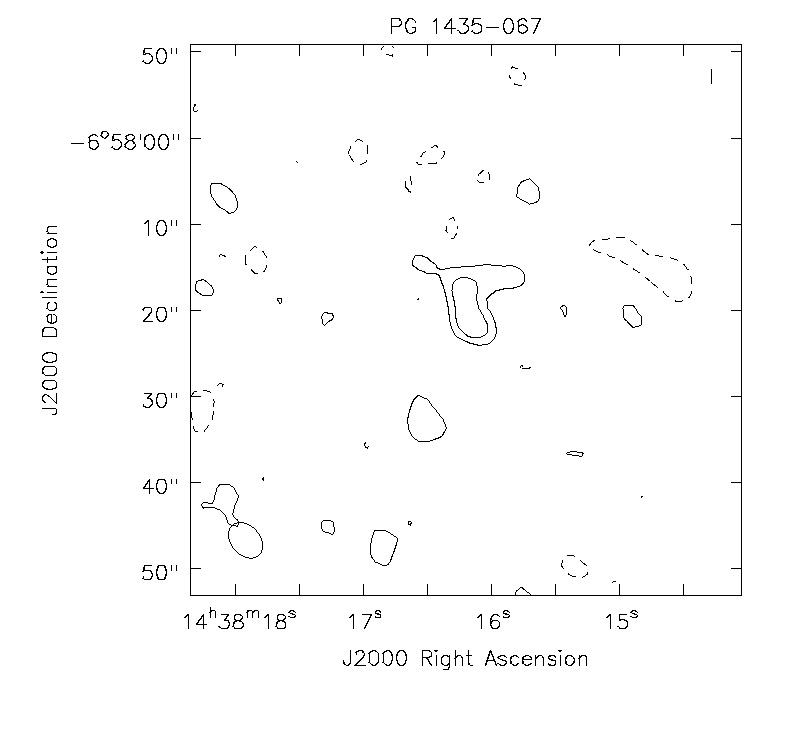}\\
\includegraphics[height=7.1cm,trim=25 10 0 10]{ 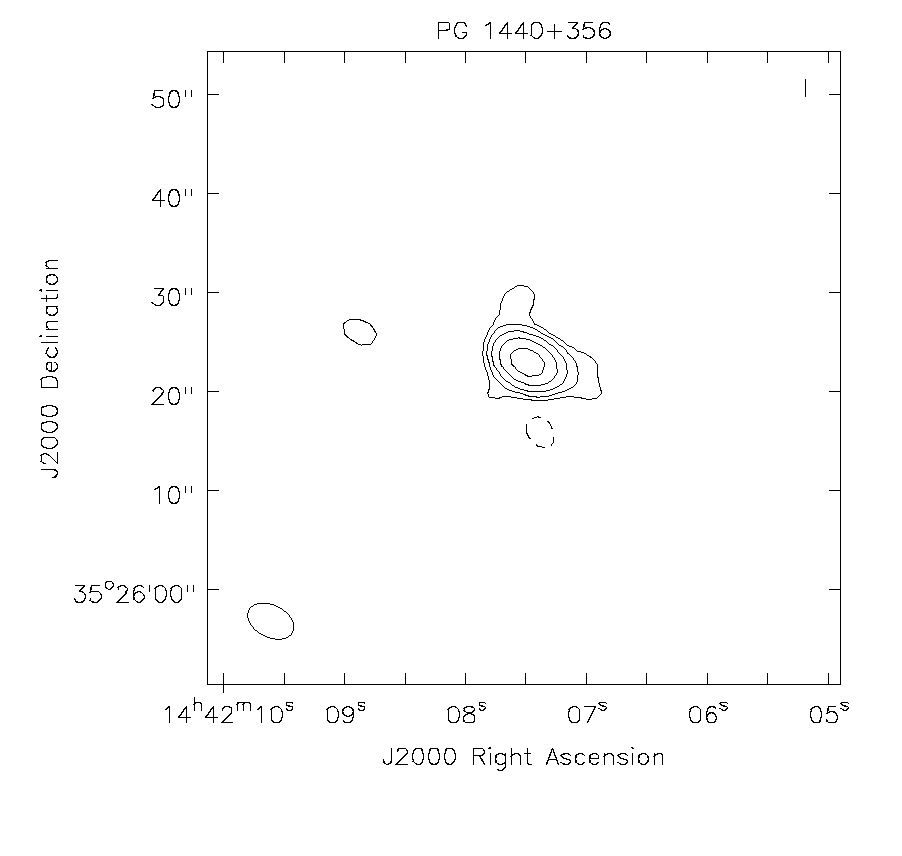}
\includegraphics[height=7.1cm,trim=25 10 0 10]{ 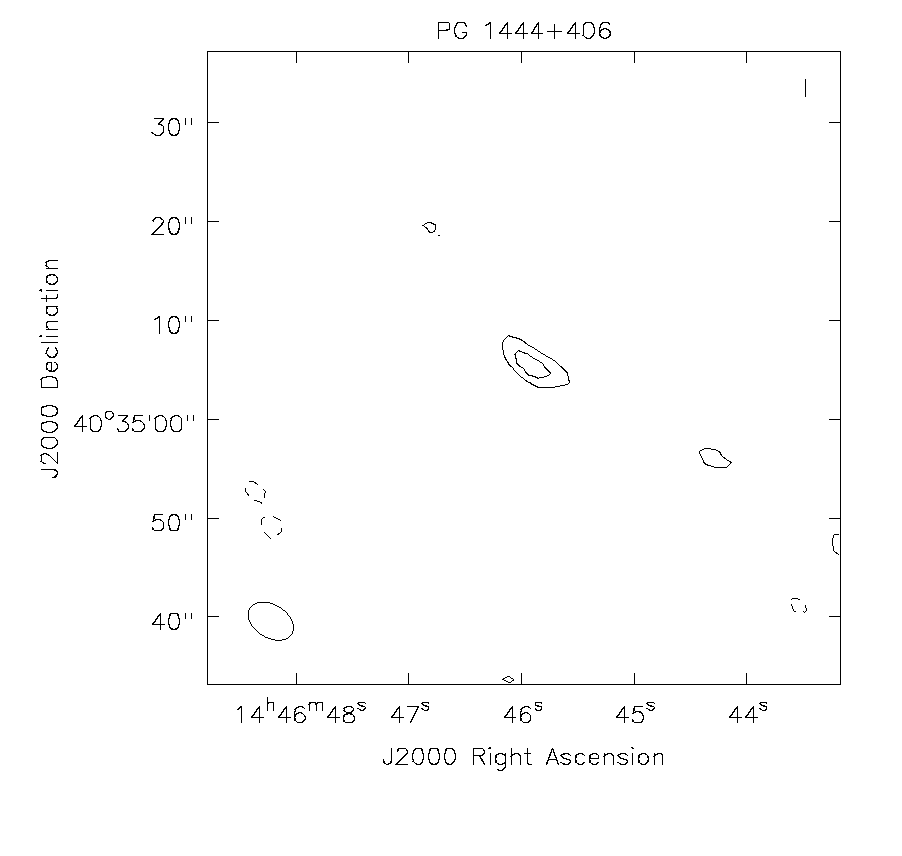}\\
\includegraphics[height=7.1cm,trim=25 10 0 10]{ 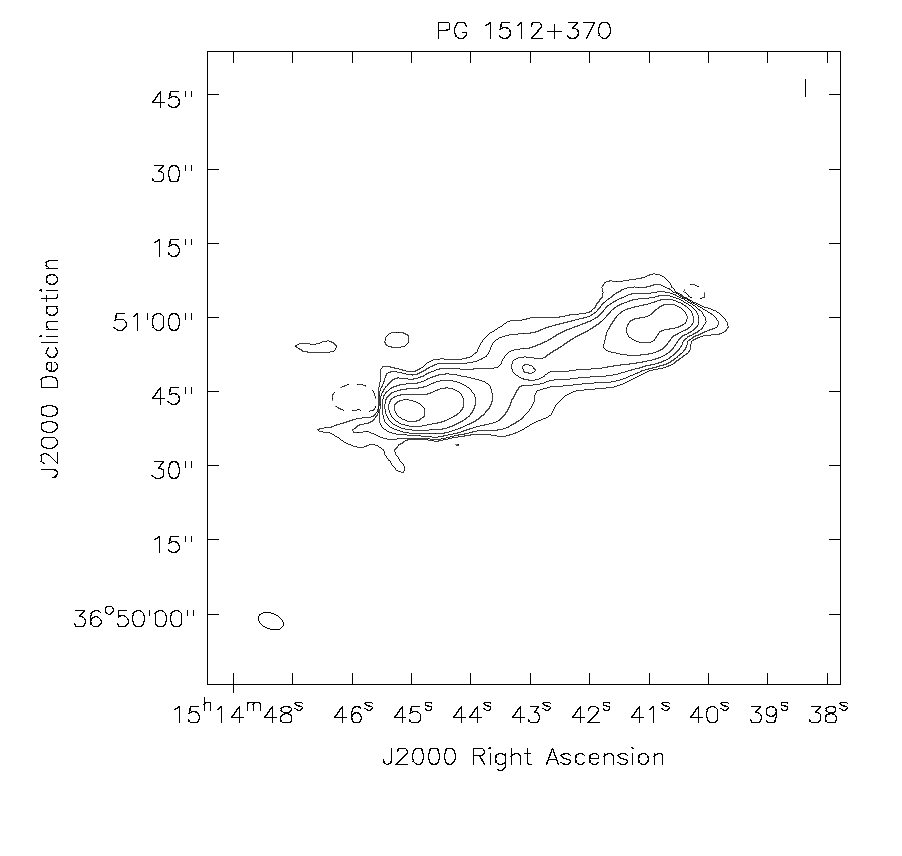}
\includegraphics[height=7.1cm,trim=25 10 0 10]{ 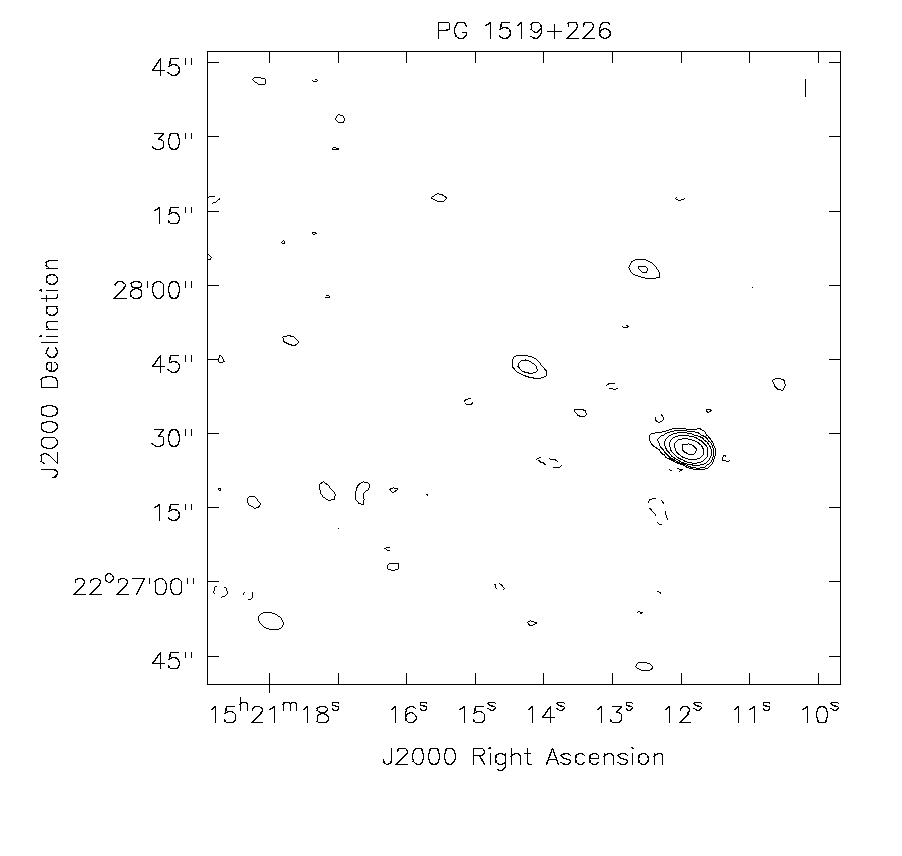}
\caption{\small uGMRT 685 MHz radio contour images of PG 1427+480, PG 1435-067, PG 1440+356, PG 1444+406, PG 1512+370 (RL), and PG 1519+226 (the source is the weak one at the centre of the image). The contour levels are $3\sigma \times (-1, 1, 2, 4, 8, 16, 32, 64, 128, 256, 516)$.}
\label{fig20}
\end{figure*}

\begin{figure*}
\centering
\includegraphics[height=7.1cm,trim=25 10 0 10]{ 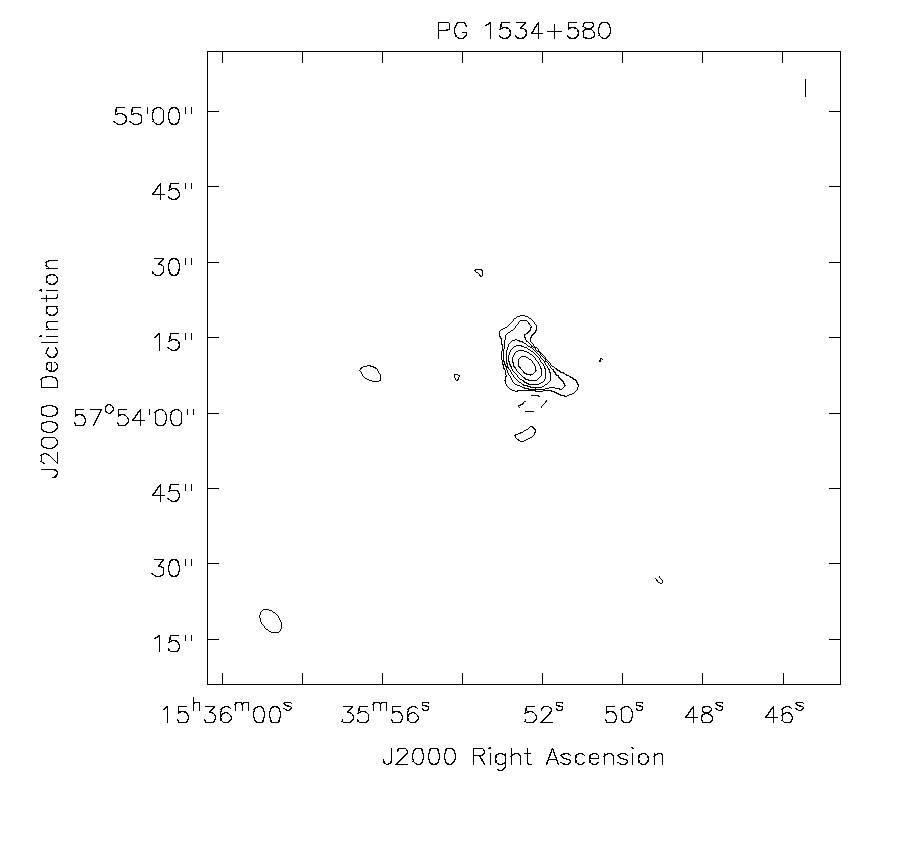}
\includegraphics[height=7.1cm,trim=25 10 0 10]{ 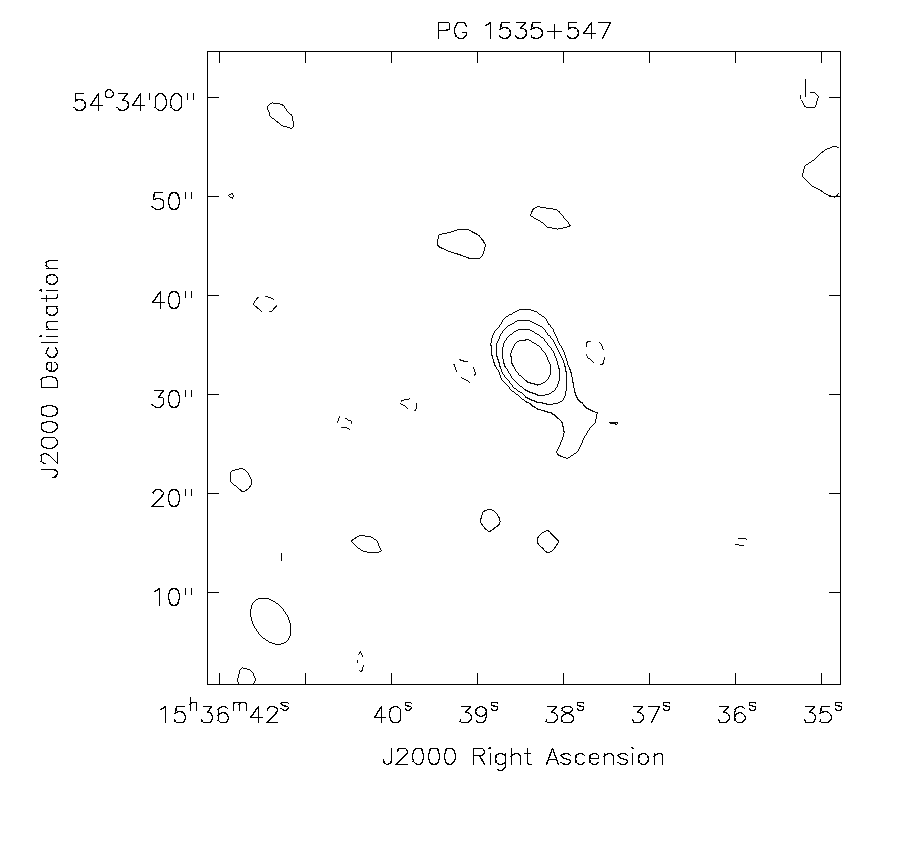}\\
\includegraphics[height=7.1cm,trim=25 10 0 10]{ 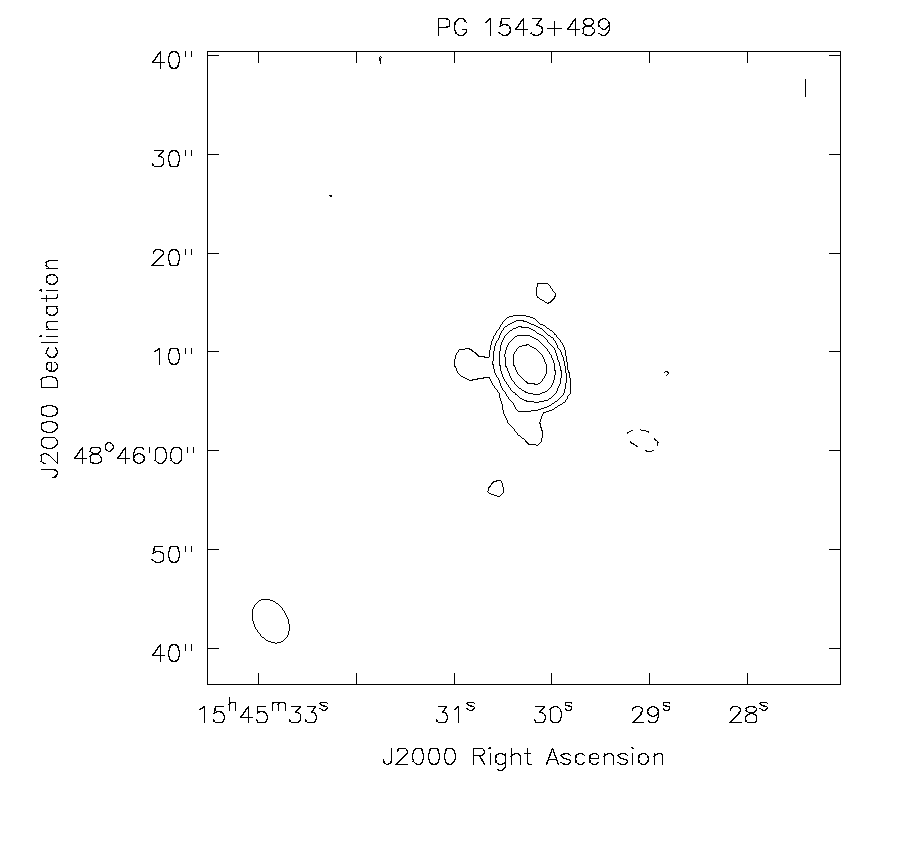}
\includegraphics[height=7.1cm,trim=25 10 0 10]{ 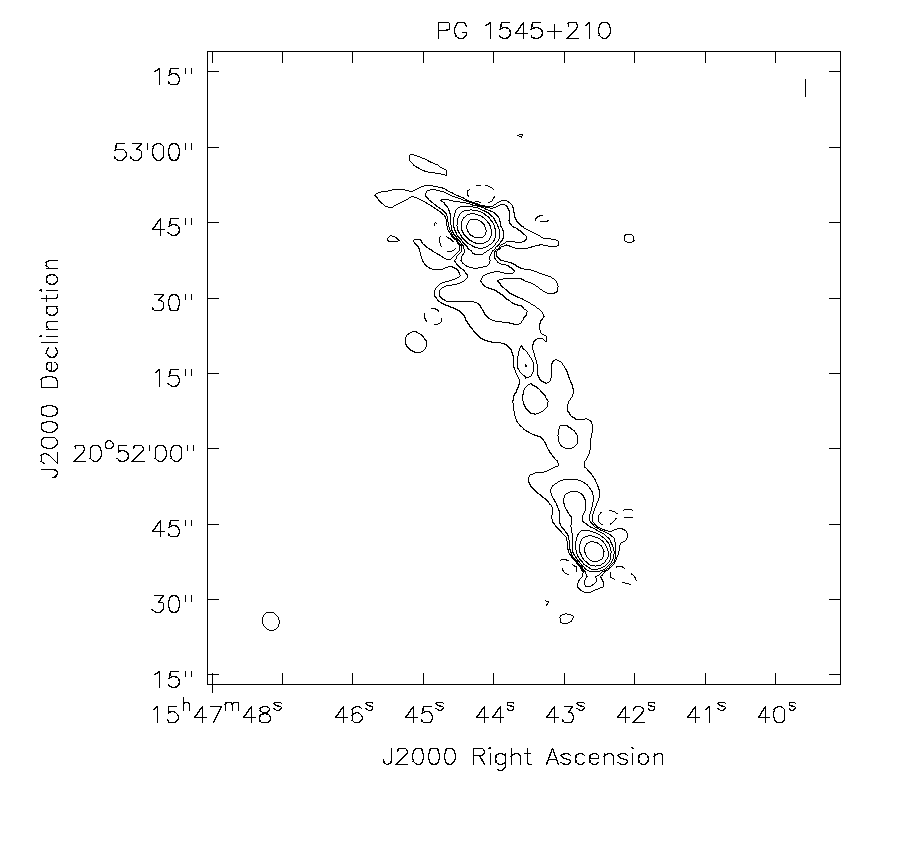}\\
\includegraphics[height=7.1cm,trim=25 10 0 10]{ 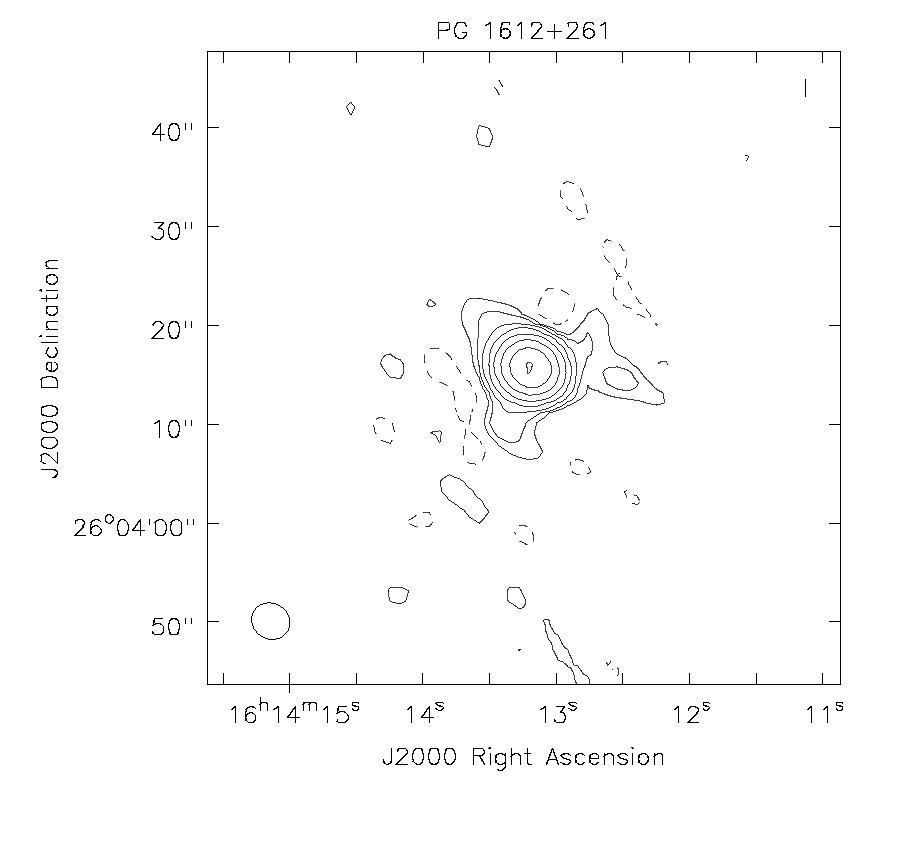}
\includegraphics[height=7.1cm,trim=25 10 0 10]{ 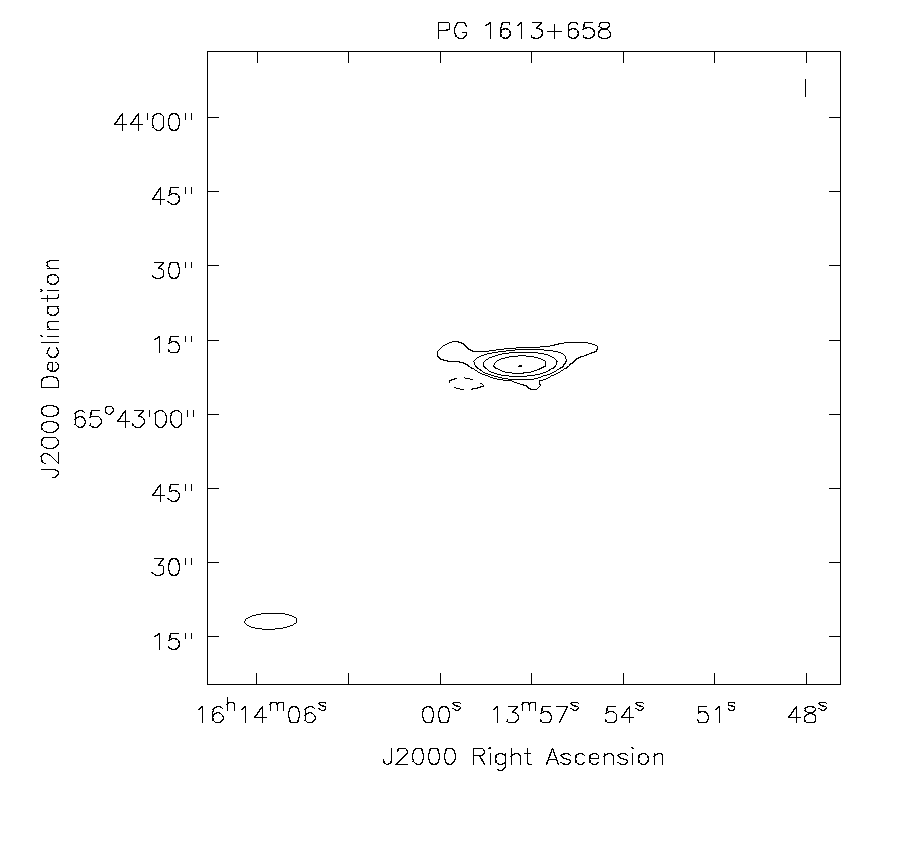}
\caption{\small uGMRT 685 MHz radio contour images of PG 1543+580, PG 1535+547, PG 1543+489, PG 1545+210 (RL), PG 1612+261, and PG 1613+658. The contour levels are $3\sigma \times (-1, 1, 2, 4, 8, 16, 32, 64, 128, 256, 516)$.}
\label{fig21}
\end{figure*}

\begin{figure*}
\centering
\includegraphics[height=7.1cm,trim=25 10 0 10]{ 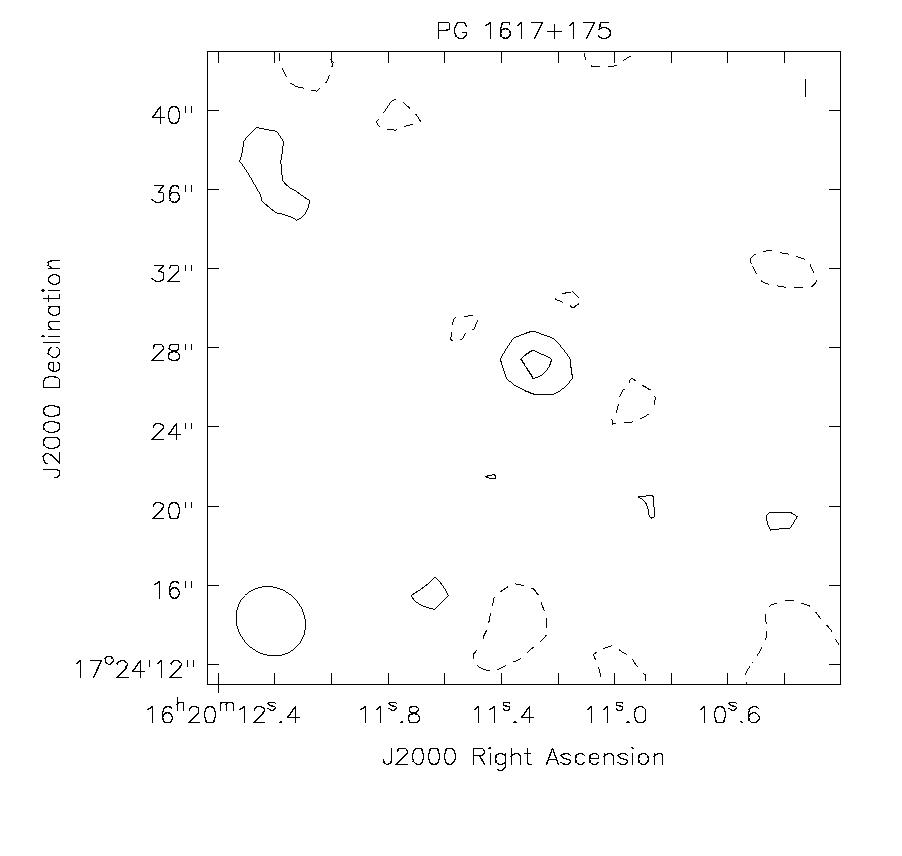}
\includegraphics[height=7.1cm,trim=25 10 0 10]{ 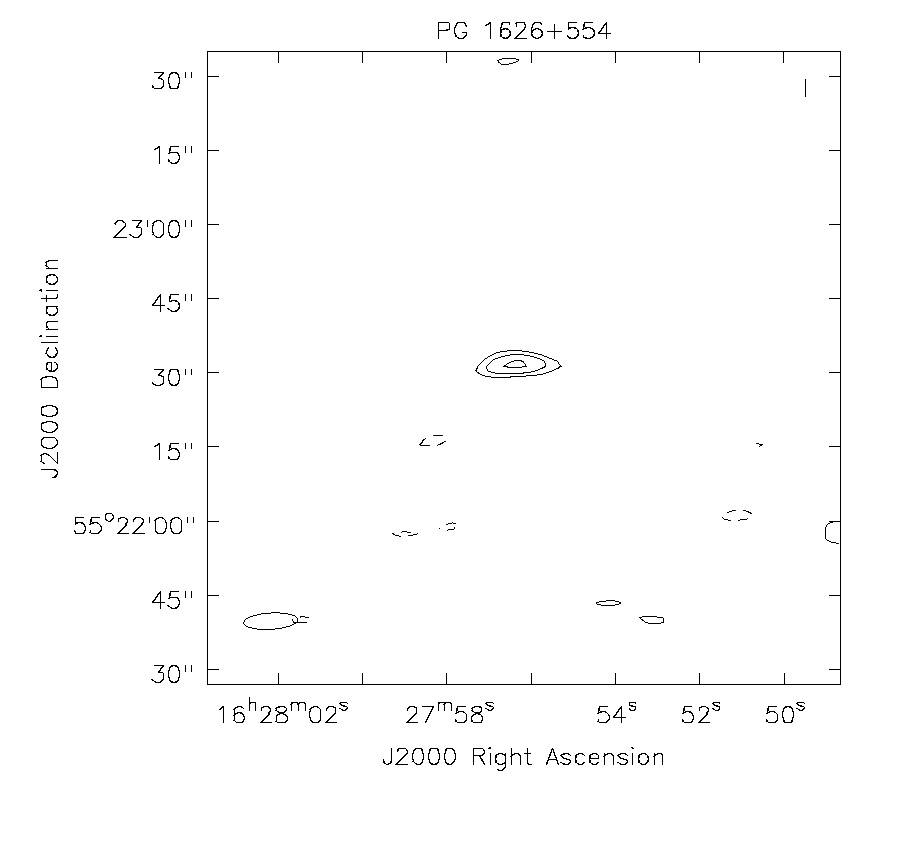}\\
\includegraphics[height=7.1cm,trim=25 10 0 10]{ 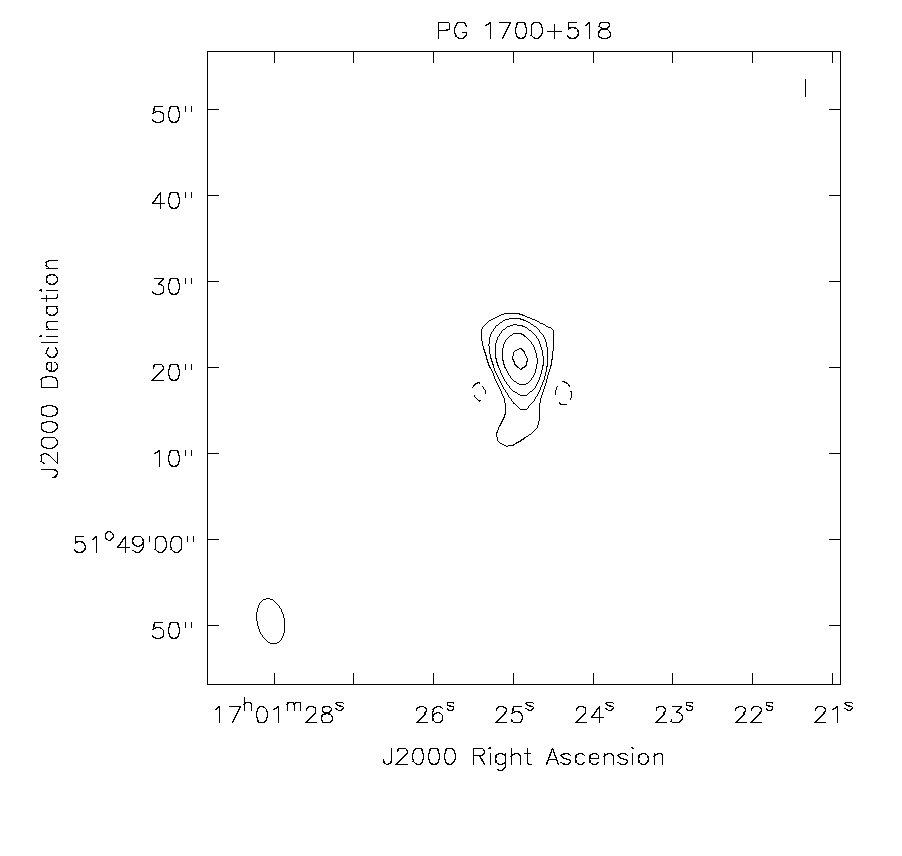}
\includegraphics[height=7.1cm,trim=25 10 0 10]{ 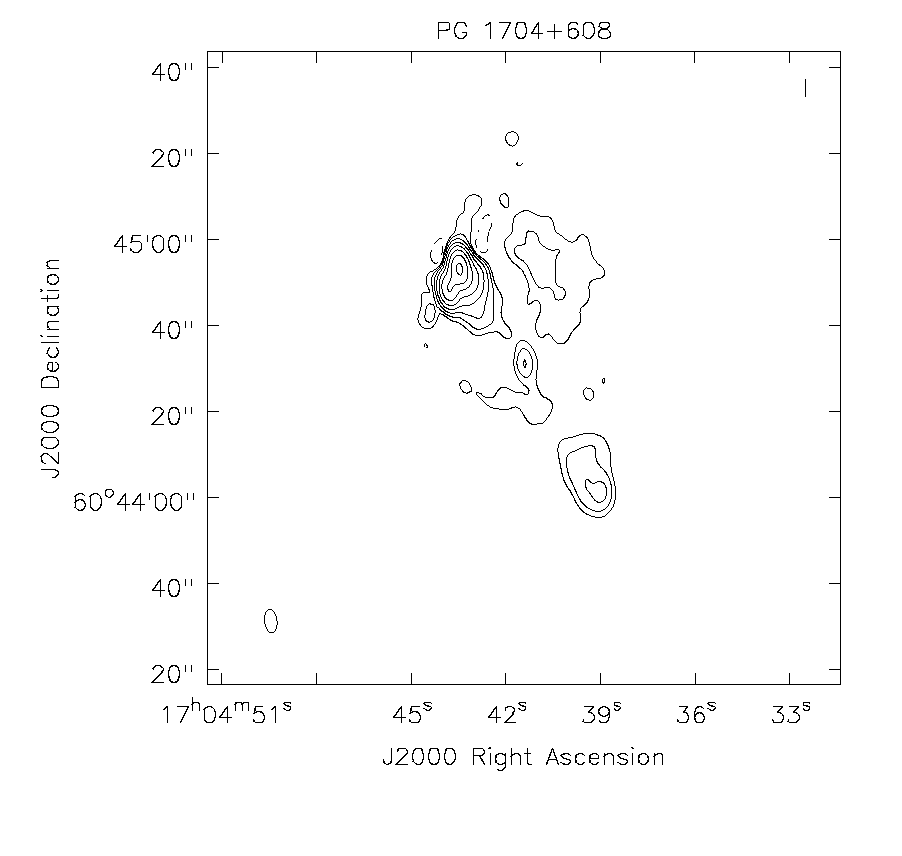}\\
\includegraphics[height=7.1cm,trim=25 10 0 10]{ 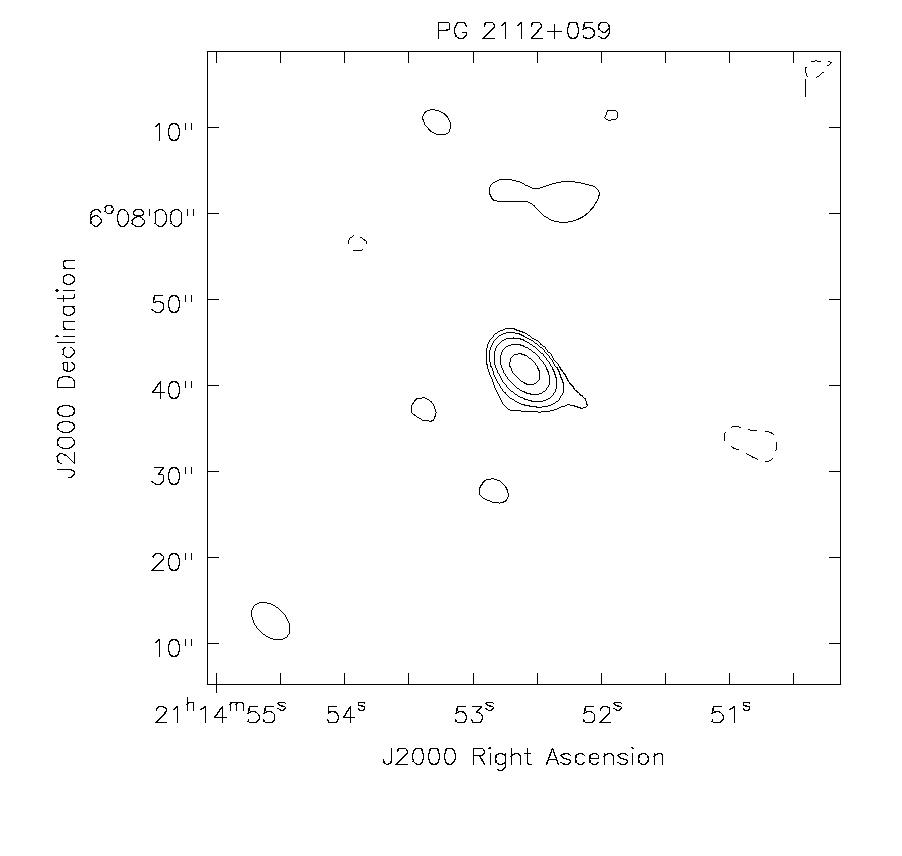}
\includegraphics[height=7.1cm,trim=25 10 0 10]{ 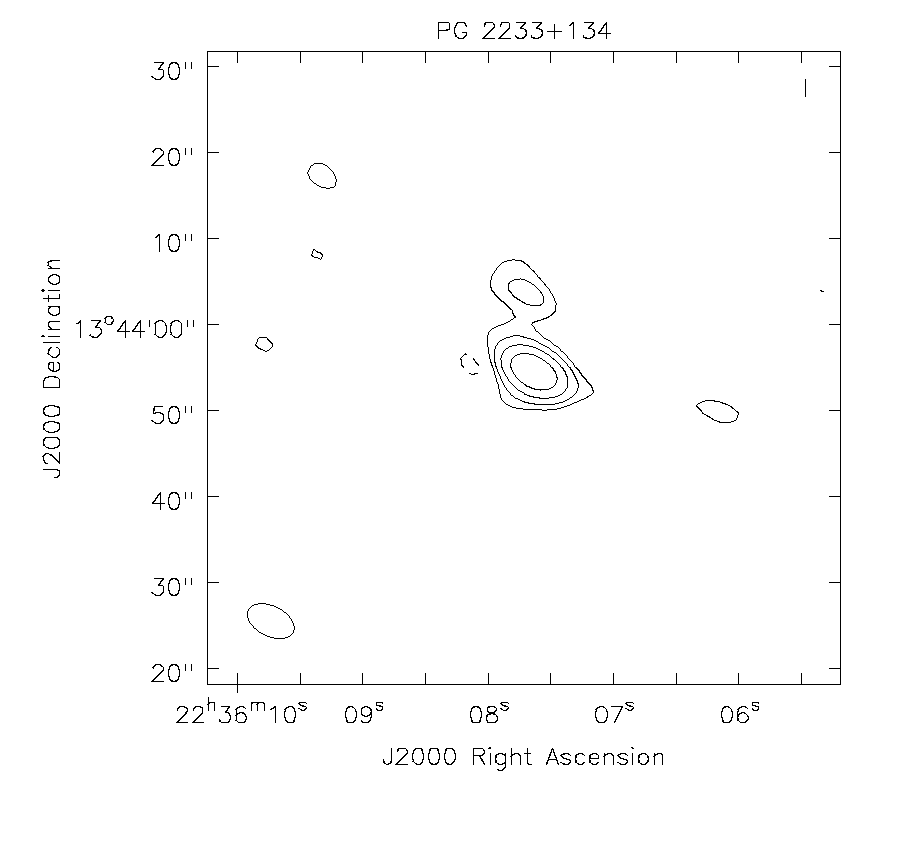}
\caption{\small uGMRT 685 MHz radio contour images of PG 1617+175, PG 1626-554, PG 1700+518, PG 1704+068 (RL), PG 2112+059, and PG 2233+134. The contour levels are $3\sigma \times (-1, 1, 2, 4, 8, 16, 32, 64, 128, 256, 516)$.}
\label{fig22}
\end{figure*}

\begin{figure*}
\centering
\includegraphics[height=7.1cm,trim=25 10 0 10]{ 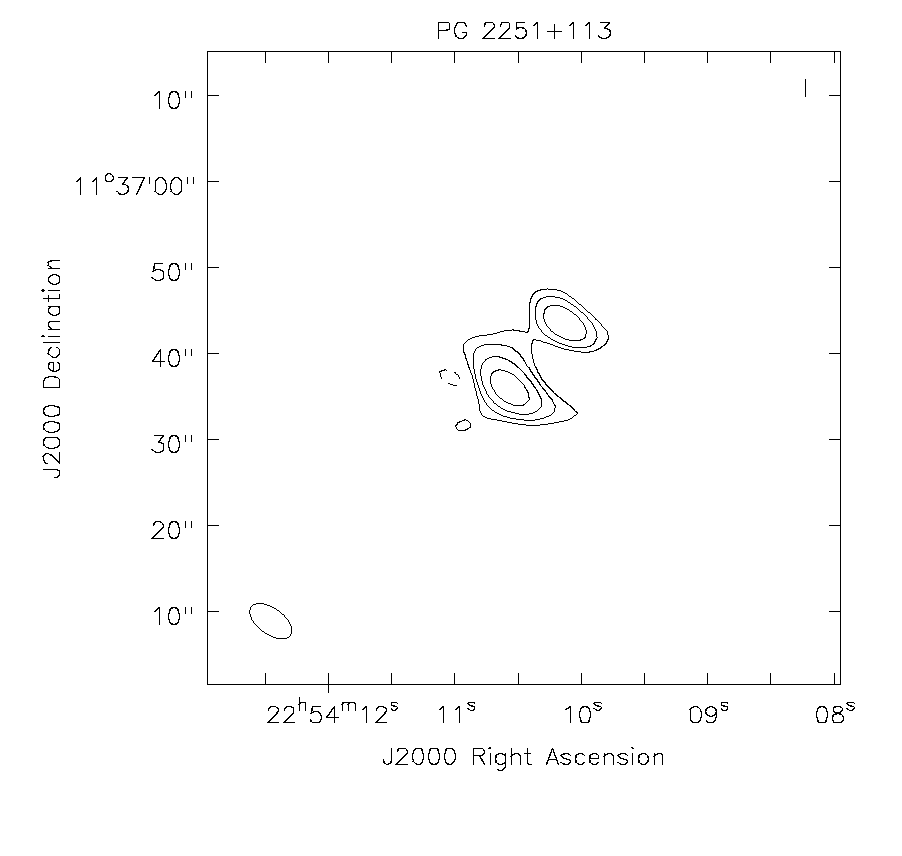}
\includegraphics[height=7.1cm,trim=25 10 0 10]{ 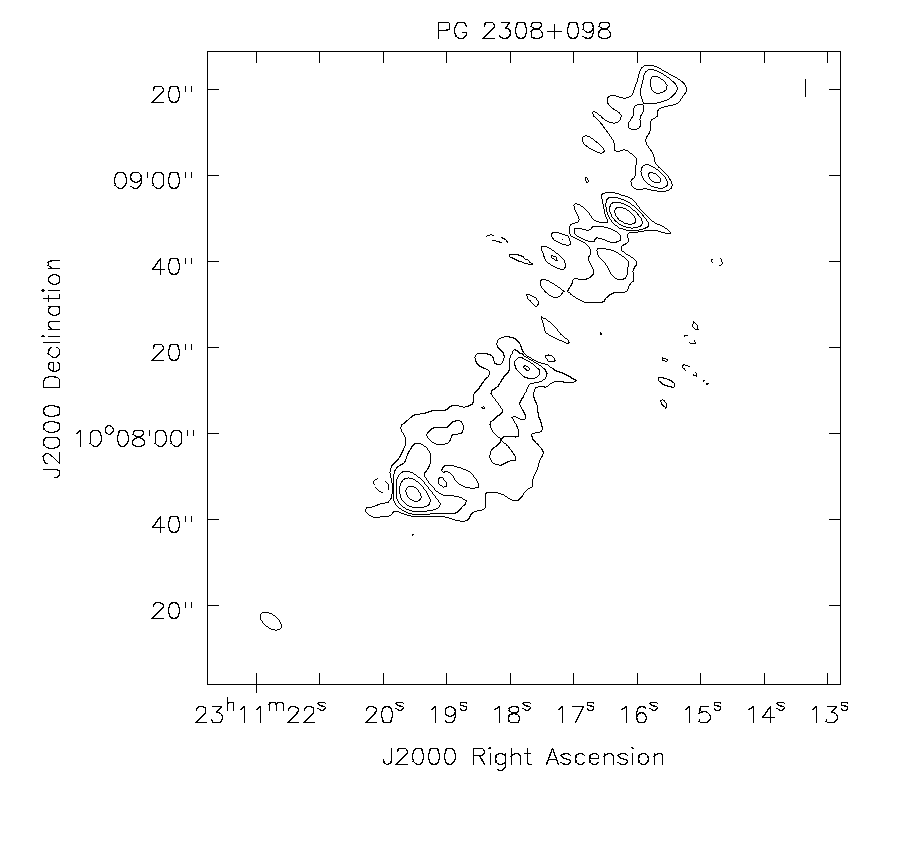}
\caption{\small uGMRT 685 MHz radio contour images of PG 2251+113 (RL) and PG 2308+098 (RL). The contour levels are $3\sigma \times (-1, 1, 2, 4, 8, 16, 32, 64, 128, 256, 516)$.}
\label{fig23}
\end{figure*}

\end{document}